\documentclass[12pt]{article}
\usepackage{latexsym}
\usepackage{plain}

\usepackage[german,english]{babel}
\usepackage[T1]{fontenc}
\usepackage[latin1]{inputenc}

\usepackage{amsfonts}
\usepackage{amsmath}
\usepackage{mathrsfs}
\usepackage{graphicx}
\usepackage{amssymb}
\usepackage{array}
\usepackage{rotating}

\numberwithin{equation}{section}

\begin{document}

\def\beq{\begin{equation}}
\def\eeq{\end{equation}}

\def\cmp{\complement}
\def\vnth{\varnothing}
\def\uvnth{\underline\vnth}

\def\cA{{\mathcal A}}
\def\tB{{\tt B}}
\def\cC{{\tt C}}
\def\cD{{\mathcal G}}
\def\td{{\tt d}}

\def\tte{{\tt e}}
\def\tF{{\tt F}}
\def\tG{{\tt G}}
\def\tg{{\tt g}}
\def\ti{{\tt i}}
\def\tI{{\tt I}}
\def\tj{{\tt j}}
\def\tn{{\tt n}}

\def\tO{{\tt O}}
\def\tP{{\tt P}}
\def\tq{{\tt q}}
\def\ttr{{\tt r}}
\def\tP{{\tt P}}
\def\tR{{\tt R}}
\def\tS{{\tt S}}
\def\tT{{\tt T}}
\def\ttg{{\tt g}}
\def\ttG{{\tt G}}
\def\bttg{\overline{\tg}}
\def\tu{{\tt u}}
\def\tU{{\tt U}}
\def\tv{{\tt v}}
\def\tV{{\tt V}}
\def\tw{{\tt w}}
\def\tX{{\mathcal X}}
\def\tx{{\tt x}}
\def\ty{{\tt y}}
\def\tz{{\tt z}}
\def\tA{{\tt A}}
\def\tE{{\tt E}}

\def\bgam{{\mbox{\boldmath$\gamma$}}}
\def\ubgam{{\underline\bgam}}

\def\Gam{\Gamma}
\def\OGam{\overline\Gam}
\def\uGam{\underline\Gam}
\def\wGam{\widetilde\Gam}

\def\uphi{\underline\phi}
\def\bphi{{\mbox{\boldmath$\phi$}}}
\def\ubphi{\underline\bphi}
\def\boeta{{\mbox{\boldmath$\eta$}}}
\def\oboeta{\overline\boeta}
\def\ups{\upsilon}

\def\Om{\Omega}
\def\om{\omega}
\def\oom{\overline\omega}
\def\bttg{\mbox{\boldmath${\tt g}$}}
\def\btau{\mbox{\boldmath${\tau}$}}
\def\bom{{\mbox{\boldmath$\omega$}}}
\def\obom{\overline\bom}
\def\0bom{{\bom}^0}
\def\0obom{{\obom}^0}
\def\nbom{{\bom}_n}
\def\0nbom{{\bom}_{n,0}}
\def\n*bom{{\bom}^*_{(n)}}
\def\wt{\widetilde}
\def\wtbom{\widetilde\bom}
\def\whbom{\widehat\bom}
\def\oom{\overline\om}
\def\wtom{\widetilde\om}
\def\bOm{\mbox{\boldmath${\Om}$}}
\def\obOm{\overline\bOm}
\def\whbOm{\widehat\bOm}
\def\wtbOm{\widetilde\bOm}

\def\Lam{\Lambda}
\def\uLam{\underline\Lam}
\def\oLam{\overline\Lam}
\def\lam{\lambda}

\def\Ups{\Upsilon}
\def\utheta{\underline\theta}
\def\ovr{\overline r}

\def\oG{\overline G}
\def\oL{\overline L}

\def\bbC{\mathbb C}
\def\bbE{\mathbb E}
\def\bbP{\mathbb P}
\def\fB{\mathfrak B}
\def\fG{\mathfrak G}
\def\fW{\mathfrak W}
\def\bbQ{\mathbb Q}

\def\bfe{\mathbf e}
\def\bi{\mathbf i}
\def\bj{\mathbf j}
\def\bn{\mathbf n}
\def\bt{\mathbf t}
\def\bu{\mathbf u}
\def\bw{\mathbf w}
\def\bX{\mathbf X}
\def\ubX{\underline\bX}
\def\ubY{\underline\bY}
\def\ubZ{\underline\bZ}
\def\bx{\mathbf x}
\def\ubx{\underline\bx}
\def\bY{\mathbf Y}
\def\by{\mathbf y}
\def\bZ{\mathbf Z}
\def\bz{\mathbf z}

\def\cl{\centerline}

\def\cA{\mathcal A}
\def\cB{\mathcal B}
\def\cC{\mathcal C}
\def\tD{{\tt D}}
\def\cE{\mathcal E}
\def\cF{\mathcal F}
\def\cG{\mathcal G}
\def\cH{\mathcal H}
\def\bbH{\mathbb H}
\def\cK{\mathcal K}
\def\cL{\mathcal L}
\def\cN{\mathcal N}
\def\tS{{\tt S}}
\def\cT{\mathcal T}
\def\cV{\mathcal V}
\def\cW{\mathcal W}
\def\ocH{\overline\cH}
\def\ocW{\overline\cW}
\def\bbB{\mathbb B}
\def\bbR{\mathbb R}
\def\bbS{\mathbb S}
\def\bbT{\mathbb T}
\def\bbU{\mathbb U}
\def\bbZ{\mathbb Z}
\def\ba{\mathbf a}
\def\bg{\mathbf g}
\def\bX{\mathbf X}
\def\bx{\mathbf x}
\def\wtbx{\widetilde\bx}
\def\ui{{\underline i}}

\def\oA{{\overline A}}
\def\uA{{A}}
\def\uuA{{\underline{A_{}}}}

\def\uk{{\underline k}}
\def\ux{{\underline x}}
\def\wtux{\widetilde\ux}
\def\uX{{\underline X}}
\def\by{\mathbf y}
\def\uy{\underline y}
\def\bY{\mathbf Y}
\def\uY{\underline Y}

\def\uj{{\underline j}}
\def\unn{\underline{n_{}}}
\def\unp{\underline p}
\def\ovp{\overline p}
\def\bx{\mathbf x}
\def\ox{\overline x}
\def\obx{\overline\bx}
\def\uz{\underline z}
\def\bz{\mathbf z}
\def\uv{\underline v}
\def\dist{\textrm{dist}}
\def\diy{\displaystyle}
\def\ov{\overline}
\def\u0{{\underline 0}}

\def\oomega{\overline\omega}
\def\oUpsilon{\overline\Upsilon}
\def\wtomega{\widetilde\omega}
\def\wtz{\widetilde z}
\def\wtheta{\widetilde\theta}
\def\wtalpha{\widetilde\alpha}
\def\wh{\widehat}
\def\oV{\overline {\mathcal V}}

\def\bI{\mathbf I}
\def\bN{\mathbf N}
\def\bbN{\mathbf N}
\def\bP{\mathbf P}
\def\bV{\mathbf V}
\def\oW{\overline W}
\def\ofW{\overline\fW}
\def\LT{{\mathbb{LT}}}
\def\mucr{{\mu_{cr}}}

\def\rA{{\rm A}}
\def\rB{{\rm B}}
\def\rb{{\rm b}}

\def\urB{\underline\rB}
\def\rc{{\rm c}}
\def\rC{{\rm C}}
\def\rd{{\rm d}}
\def\rD{{\rm D}}
\def\rd{{\rm d}}
\def\re{{\rm e}}
\def\rE{{\rm E}}
\def\rF{{\rm F}}
\def\rI{{\rm I}}
\def\rK{{\rm K}}
\def\rL{{\tt L}}

\def\rn{{\rm n}}
\def\rN{{\rm N}}

\def\rP{{\rm P}}
\def\rO{{\rm O}}
\def\rQ{{\rm Q}}
\def\rr{{\rm r}}
\def\rR{{\rm R}}

\def\rs{{\rm s}}
\def\rS{{\rm S}}
\def\rT{{\rm T}}
\def\rU{{\tt U}}

\def\rV{{\rm V}}
\def\rv{{\rm v}}
\def\rw{{\rm w}}
\def\rW{{\rm W}}

\def\rx{{\rm x}}
\def\ry{{\rm y}}
\def\rtr{\rm{tr}}

\def\oa{b}
\def\ua{a}
\def\uua{{\ua}}

\def\uk{\underline k}
\def\un{\underline n}
\def\ux{\underline x}
\def\uy{\underline y}
\def\wtux{\widetilde\ux}
\def\uX{\underline X}

\def\oJ{\overline J}
\def\oP{\overline P}
\def\utC{{\underline\cC}}
\def\utE{{\underline\rE}}
\def\urB{{\underline\rB}}
\def\urC{{\underline\rC}}
\def\urD{{\underline\rD}}
\def\urE{{\underline\rE}}

\def\b!{{\bf !}}

\def\a {\alpha} 
\def\b {\beta} 
\def\g {\gamma}
\def\tg {\tilde \gamma} 
\def\d {\delta} 
\def\f {\varphi}
\def\ff{\bar\varphi} 
\def\s {\sigma}
\def\bs {\bar \sigma}
\def\t {\tau} 
\def\o {\bo}
\def\T {\Theta} 
\def\L {\Lambda} 
\def\l {\lambda}
\def\G {\bG}
\def\e {\varepsilon}
\def\r{\varrho}
\def\O {\Omega} 
\def\S{\Sigma} 
\def\D{\Delta}
\def\k{\kappa}
\def\vt{\nu}
\def\p{\partial}

\def\cAi{\cA(\bbR^d\| i)}

\def\cCi{\cC(\bbZ^d\| i)}
\def\cDi{\cD(\bbZ^d\| i)}

\def\BCCs{{contour collections}}
\def\BCC{{contour collection}}
\def\BC{{contour}}
\def\BCs{{contours}}
\def\basic{}
\def\LBCs{{large contours}}
\def\LBC{{large contour}}
\def\SBC{{small contour}}
\def\SBCs{{small contours}}
\def\BL{{boundary layer}}
\def\BLs{{boundary layers}}

\def\St{\mathbb S}
\def\Dm{\mathbb D}
\def\sh{\sharp}

\def\teLam{{^{2R}\Lam}}

\def\sX{{\mathscr X}} \def\fX{{\mathfrak X}} \def\rX{{\rm X}}

\title{A classical WR model\\ with $q$ particle types}
\author{ A. Mazel$^{1}$ \and Yu. Suhov~$^{\,2 ,3,4}$ \and I. Stuhl~$^{3,5}$}
\vspace{1mm}

\maketitle {\footnotesize
\noindent $^{1}$ 
E-mail: alik@speakeasy.org

\noindent $^{\,2}$ Statistical Laboratory, DPMMS, University of Cambridge, UK;\\
Institute of Mathematics and Statistics, University of S\~ao Paulo, Brazil\\
E-mail: yms@statslab.cam.ac.uk

\noindent $^3$ Institute of Mathematics
and Statistics,  University of S\~ao Paulo, Brazil

\noindent $^4$ Institute for Information Transmission Problems, Moscow, Russia

\noindent $^5$ University of Debrecen, Hungary\\
E-mail: izabella@ime.usp.br}

\begin{abstract}
A version of the Widom--Rowlinson model is considered, where 
particles of $q$ types coexist, subject to pairwise hard-core exclusions. For $q\leq 4$, 
in the case of large equal fugacities, we give a complete description of the pure phase
picture based on the theory of dominant ground states. 
\vskip.1truecm


\end{abstract}

\section{Introduction}

{\bf 1.1. Preliminaries.} The Widom--Rowlinson 
(WR) model in Euclidean space $\bbR^d$ with two types of particles was invented in \cite{WR}. It was considered in numerous papers among which we mention \cite{BKL, CCK, GH, LL, R} as most closely related to the current study. In this paper we analyze a $q$-type version of the WR model with equal fugacities and arbitrary hard-core exclusion diameters between particles of different types. Particles of the same type do not interact with each other. Formally, a particle of type $i \in \{1,\ldots, q\}$ located at point $x \in \bbR^d$ and a particle of different type $j \in \{1,\ldots, q\}$ located at point $x' \in \bbR^d$ interact via the potential 
\beq\Phi_{ij}(x,x')=
\begin{cases}
0,&|x-x'|>D(i,j),\\ 
+\infty ,&|x-x'|\leq D(i,j).
\end{cases} \label{1.01}\eeq
representing a hard-core exclusion of diameter $D(i,j)=D(j,i)\in (0,\infty )$. As was said, we assume identical fugacities:
\beq \label{1.02}z_1=\ldots =z_q=z>0\eeq 
in addition, we require a triangular-type inequality 
\beq  D(i,j)<D(i,l)+D(j,l) \label{1.03}\eeq
for any pairwise distinct triple $i,j,l\in\{1,\ldots ,q\}$. This property guarantees that two particles of types $i$ and $j$ 
do not interact with each other as soon as they are separated by a layer of particles of type $l$, provided that 
this layer has a finite thickness and high enough density of particles.

Under assumptions \eqref{1.01}, \eqref{1.02} and \eqref{1.03}, for $q\leq 4$ and for large values of $z$,  
we specify the set of ergodic infinite-volume Gibbs/Dobrushin--Lanford--Ruelle (DLR) measures (pure phases) for each 
collection   
\beq\label{1.03A}\Dm=\big\{D(i,j),\,1\leq i<j\leq q\big\}\eeq
of hard-core diameters; see Theorem 1.1.
 
The ergodic DLR measures are constructed in the infinite-volume (thermodynamic) limit  
$\Lam\nearrow\bbR^d$ from Gibbs distributions in a bounded `box` $\Lam$ with boundary conditions 
(possibly randomized) outside $\Lam$ 
induced by particles of a given type $i\in\{1,\ldots q\}$. Particle configurations specifying boundary conditions are 
taken to be dense enough in the following sense:  given a partition of $\bbR^d$ into a grid of cubic cells of a fixed size 
smaller than $\ua =\min\;[D(i,j):\;1\leq i<j\leq q]$, the boundary condition configuration 
is required to have at least one particle of type $i$ inside each cell located outside $\Lam$. Depending on collection $
\Dm$, some particle types turn out to be `stable` meaning that the corresponding boundary conditions generate a pure 
phase. The remaining particle types are `unstable` as they generate only convex combinations of the pure phases 
produced by stable types. See Theorem 1.2 where these convex combinations are specified explicitly.

Methodologically, the present work is related to the Pirogov--Sinai theory (PST), \cite{PS1, PS2, Si, Z}. In Ref  
 \cite{BKL} an extension of the PST  for the WR model has been proposed where, for a given 
collection $\Dm$, the vector $\bz=(z_1,\ldots ,z_q)$ of fugacities $z_1, \ldots, z_q>0$ is varied to achieve a pure phase 
coexistence between different subsets of particle types. Given a subset of particle types, ${\mathbb Q}\subset\{1,\ldots ,q\}$, a 
hypersurface $\bbH=\bbH ({\mathbb Q})$ of dimension $q+1-\sh{\mathbb Q}$ in the fugacity orthant $\bbR^q_+$ has been  
constructed such that  for $\bz$ lying in this hyper-surface the only pure phases are the ones generated by the particle 
types $i\in{\mathbb Q}$. (Here and below, $\sh$ stands for the cardinality of a given (finite) set.) As is customary in the 
PST, these hypersurfaces are obtained by equalizing sums 
of absolutely convergent series for the free energies of the coexisting pure phases. The series are constructed using a 
rather involved perturbation theory. Consequently, given a point $\bz\in\bbR^q_+$ (e.g., $\bz =(z,\ldots z)$ as in Eqn 
\eqref{1.02}), it is hard to conclude in which hypersurface $\bbH$ it lies. Solving such an `inverse problem` requires 
an additional effort which frequently 
employs a so-called dominant ground states analysis; see, e.g., \cite{BS1, BS2, MS1, MS2}.

The current paper presents a solution to the aforementioned inverse problem for the case of equal (large) fugacities.
Contrary to \cite{BKL} we consider fugacities fixed and, in a sense,  vary hard-core diameter collection $\Dm$. In this
set-up, for $q\leq 4$ and the fugacity $z$ large enough it is possible to specify phase coexistence regions explicitly in
terms of linear inequalities between hard-core diameters. Our analysis is still based upon a perturbation theory and
carried out in the spirit of \cite{BS1, BS2, MS1, MS2}. It turns out that for $q\leq 4$, only few leading terms of 
the perturbation series are need to be analyzed in detail; the results can be expressed via linear inequalities 
between hard-core diameters. For $q > 4$ one needs to analyze higher order terms which leads to rather 
cumbersome equations. Cf. Remark 1.2 below. 

It is instructive to compare our results with those obtained for a symmetric (continuous) version of the WR model 
where the fugacities $z_i\equiv z$, $1\leq i\leq q$, and the diameters $D(i,j)\equiv D>0$, $1\leq i<j\leq q$. 
Historically, a  symmetric version of the WR model (with $q=2$) was the first one to be analyzed in \cite{R} through 
the so-called Peierls argument. A further progress for symmetric models was achieved in \cite{LL} and later on in 
\cite{GH}, for an arbitrary $q$. In particular, in Ref \cite{GH} some interaction potentials 
were added, within and between particle types, which obey the symmetry of the model 
and have a more generic form than a hard-core exclusion. The existence of at least $q$ 
distinct pure phases for $z$ 
large enough has been proven in \cite{GH}, based on a clever stochastic dominance type of an argument. Other 
applications of stochastic dominance and the FKG-inequality type of arguments to the WR model can be found in 
\cite{CCK, GHM}. 

In contrast to \cite{GH, BKL}, our set-up is partially symmetric (as the fugacities $z_i$ are equal) and 
partially without a symmetry (as the hard-core exclusion diameter collection $\Dm$ is arbitrary). Given a collection 
$\Dm$, our Theorems~1.1~and~1.2 calculate the set of pure phases which appear to be the same for all $z > z_0(\Dm)$. Such 
independence of $z$ is a consequence of a symmetry because several pure phases are observed only if 
there exists a permutation of the set  $\{1,\ldots ,q\}$ mapping $\Dm$ into itself and the corresponding stable particle 
types into each other (unstable particle types are not necessarily symmetric under this map). To establish 
Theorems~1.1~and~1.2 we employ the machinery of the PST which is designed to treat cases without any symmetry 
as in \cite{BKL}. We do not know any `simpler` inequalities or stochastic dominance-based arguments which can 
produce the same results. For some specific cases of $\Dm$ one can establish that for any bounded box $\Lam$ 
and any value of fugacity $z$ the 
partition function with stable boundary conditions is not less than the corresponding partition function with unstable 
boundary conditions. This could slightly simplify the proof of Theorem~1.1. We do not use such arguments as they 
cannot be applied to an arbitrary $\Dm$. We again refer the reader to Remark 1.2 where some particular models are 
discussed.
\bigskip

{\bf 1.2. A formal description of the model. The results.} The definition of the phase space and the standard sigma-algebras of its subsets 
for a $q$-type WR model can be found in \cite{GH}. The set-up adopted in \cite{GH} provides a way to describe the DLR measures for this model.  

In this paper we use an alternative approach, starting with 
$q$ independent Poisson processes of rate $z>0$ in $\bbR^d$, with samples
$\bX_1$, $\ldots$, $\bX_q$. We write $\ubX =(\bX_1,\ldots ,\bX_q)$, where $\bX_i$ identifies a type $i$ particle configuration
and $\ubX$ is a multi-type particle configuration. Given a bounded 
open subset $\Lam\subset\bbR^d$, we set $\bX_i^\Lam=\bX_i\cap\Lam$ and
$\ubX^\Lam =(\bX^\Lam_1,\ldots ,\bX^\Lam_q)$. Then $\bX_1^\Lam$, $\ldots$, $\bX_q^\Lam$  are finite 
subsets in $\Lam$ such that 
\begin{description}
\item[(i)] $\bX_1^\Lam$, $\ldots$, $\bX_q^\Lam$ are independent, 
\item[(ii)] $\forall$ $1\leq i\leq q$, 
\beq\bbP (\sh\bX_i^\Lam =s)=\diy\frac{e^{-z \ups (\Lam )}z^s\ups (\Lam )^s}{s!}, \;\;s=0,1,\ldots ,
 \label{1.04}\eeq
\item[(iii)] $\forall$ $1\leq i\leq q$, the  distribution of points $X$ from 
$\bX^\Lam_ i$, conditional on $\sh\bX_i^\Lam =s$, 
is obtained by symmetrizing a sample of $s$ independent uniform random points in $\Lam$.
\end{description}
Here and below, $\ups (\;\cdot\;)$ stands
for the Lebesgue volume and $\bbP$ for the underlying (joint) Poisson processes distribution.

In a number of  forthcoming definitions, we refer a cubic box 
\beq\Lam =[-L,L]^d,\;\hbox{ where $\;\diy\frac{L}{R}\;$ is a positive integer.} \label{1.05}\eeq  
Here $R\leq \ua =\min\;[D(i,j):\;1\leq i<j\leq q]$ is a `coarse-graining` constant which will be chosen later. Such
a box is naturally divided into cubic cells of size $R$; such a cell will be denoted by $\Ups$. The choice of 
$R$ will not affect the forthcoming
constructions, and we will omit this symbol from the notations.
Let $\mu_\Lam =\mu_{\Lam ,\Dm ,z}$ denote the conditional probability distribution 
\beq\mu_\Lam (\,\cdot\,):=\bbP\Big(\;\cdot\;\big|\;\cA (\Lam)\;\Big)=\frac{\bbP
\Big(\;\cdot\;\cap\;\cA (\Lam)\;\Big)}{{\bbP}(\cA (\Lam ))} \label{1.06}\eeq
where $\cA (\Lam )=\cA (\Dm,\Lam)$ stands for the event 
\beq\begin{array}{r}\cA (\Lam ):=\Big\{\ubX =(\bX_1,\ldots ,\bX_q):\;
|x-x'|>\;D(i,j),\qquad{}\\ \:\forall\;x\in\bX^\Lam_i,\;x'\in\bX^\Lam_j,
\;\;1\leq i<j\leq q\;\Big\}\,.\end{array} \label{1.07}\eeq
Then $\mu_{\Lam}$ is called the Gibbs distribution in $\Lam$ corresponding to the 
potential \eqref{1.01}, fugacity $z$ and an empty boundary condition outside $\Lam$.
Multi-type particle configurations $\ubX$ belonging to $\cA (\Lam )$ are called {\it admissible 
in} $\Lam$. (The adjective multi-type will be omitted.)

Next, consider an event $\cA$ defined by
\beq\begin{array}{r}\cA :=\Big\{\ubX=\{\bX_1,\ldots ,\bX_q\}:\;|x-x'|>\;D(i,j),\qquad\quad{}\\ \forall\;\;x\in\bX_i
\hbox{ and }\;x'\in\bX_j,\;\;
\forall\;\;1\leq i<j\leq q\;\Big\}.\end{array} \label{1.08}\eeq
A configuration $\ubX=\{\bX_1,\ldots ,\bX_q\}\in\cA$ is called {\it admissible}. 
Further, given $\ubY\in\cA$ and denoting $\Lam^\cmp = \bbR^d \setminus \Lam$, set:
\beq
\cA\left(\Lam |\,\ubY^{\Lam^\cmp}\right):=\Big\{\ubX =(\bX_1,\ldots ,\bX_q):\;\ubX^\Lam\vee\ubY^{\Lam^\cmp}\in\cA\Big\}.
 \label{1.09}\eeq
Here, as before, $\ubX^{\Lam}=(\bX_1^{\Lam},\ldots ,\bX_q^{\Lam})$, 
$\bX_j^{\Lam}=\bX_i\cap\Lam$, and $\ubX^\Lam\vee\ubY^{\Lam^\cmp}$ denotes the 
concatenated configuration: 
\beq\ubX^\Lam\vee\ubY^{\Lam^\cmp}=(\bX^\Lam_1\cup\bY^{\Lam^\cmp}_1,\ldots ,
\bX^\Lam_q\cup\bY^{\Lam^\cmp}_q).\label{1.10}\eeq
The Gibbs distribution $\mu_{\Lam}\left(\;\cdot \;| \ubY^{\Lam^\cmp}\right)$ in $\Lam$ 
with the boundary condition $\ubY^{\Lam^\cmp}$ is determined by
\beq\mu_\Lam\left(\;\cdot \;| \ubY^{\Lam^\cmp}\right)
:={\bbP}\left(\;\cdot \;\Big|\cA\left(\Lam |\ubY^{\Lam^\cmp}\right)\right)=\frac{{\bbP}\left(\;\cdot\;\cap
\cA\left(\Lam |\ubY^{\Lam^\cmp}\right)\right)}{{\bbP}\left(
\cA\left(\Lam |\ubY^{\Lam^\cmp}\right)\right)}. \label{1.11}\eeq

Next, it is convenient to introduce a Gibbs distribution $\mu_{\Lam}(\;\cdot\;\|\,i)$ with a randomized 
boundary condition of type $i$ outside $\Lam$.
To this end, we pass from $\Lam$ to the `twice-extended' box $\;\teLam$:
\beq\label{1.12A}\teLam=[-2R-L,L+2R]^d.\eeq
Cube $\teLam$ is obtained by adding a layer of width $2R$ to $\Lam$; equivalently,  
$\teLam$ is the union of cells belonging 
to cubes of linear size $5R$ centered at $\Ups\subset\Lam$. Given $i\in\{1,\ldots ,q\}$, set:
\beq\begin{array}{rl}
\cA (\Lam\|\,i):=&\Big\{\ubX =(\bX_1,\ldots ,\bX_q)\in\cA\left(\,\teLam\right):\\
&\;\;\,\hbox{$\bX_i\cap\Ups\neq\varnothing$ for each cell $\Ups\subset\,\teLam
\setminus\Lam$}\Big\}\,.\end{array} \label{1.13}\eeq
Further, set:
\beq\mu_{\Lam}(\;\cdot\;\|\,i):=\bbP\Big(\;\cdot\;\big|\;\cA (\Lam\|\,i)\;\Big)=\frac{\bbP
\Big(\;\cdot\;\cap\;\cA (\Lam\|\,i)\;\Big)}{{\bbP}(\cA (\Lam\|\,i))}. \label{1.12}\eeq
We note that the choice of width $2R$ is not occasional: it reflects a convenient screening 
property discussed after Definition 2.1 in the next section. Furthermore, 
our choice of $R$ guarantees that for $\ubX\in\cA (\Lam\|\,i)$ the intersection $\bX_j\cap\Ups
=\varnothing$ for all $\Ups\subset\,\teLam\setminus\Lam$ and $j\neq i$.

As was indicated above, we study so-called infinite-volume Gibbs/DLR probability measures, corresponding with hard-core exclusion potentials $\Phi_{ij}$
as in \eqref{1.01}, \eqref{1.03} and with symmetric fugacities equal to $z$ as in \eqref{1.02}. 
The definition of such a measure $\mu =\mu_{\Dm ,z}$ follows the DLR 
(Dobrushin--Lanford--Ruelle) equation. Given 
$\Lam$ as in Eqn \eqref{1.05}, let $\fX(\Lam )$ and $\fX(\Lam^\cmp)$ denote the sigma-algebras generated 
by  restrictions $\ubX^\Lam$ and $\ubX^{\Lam^\cmp}$, respectively.
Then the restriction of the conditional distribution $\mu (\,\cdot\,|\fX (\Lam^\cmp) )$ to $\fX^\Lam$  
coincides with $\mu_\Lam\left(\;\cdot \;| \ubY^{\Lam^\cmp}\right)$ for $\mu$-a.a. $\ubY$. In other words, for any $\cB\in\fX^\Lam$
\beq\left[\mu\left(\;\cB\;|\fX (\Lam^\cmp )\right)\right](\ubY)=\mu_\Lam\left(\;\cB\;|\;\ubY^{\Lam^\cmp}\right).
\label{1.14}\eeq

Examples of DLR-measures are
(weak) limiting points for measures $\mu_\Lam(\,\cdot\,)$, $\mu_\Lam\left(\;\cdot\;|\ubY^{\Lam^\cmp}\right)$ 
or  $\mu_{\Lam}\big(\;\cdot\;\|\,i\big)$ when 
$\Lam$ is taken as in Eqn \eqref{1.05} and $\Lam\nearrow\bbR^d$  (we will also use the notation
 $L\to\infty$).  
In fact, the set of 
DLR measures is the closure of the convex hull of  
limit points of probability distributions $\mu_\Lam\left(\;\cdot\;|\ubY^{\Lam^\cmp}\right)$ 
with arbitrary (admissible) boundary conditions $\ubY^{\Lam^\cmp}$. 

A shift-invariant and ergodic DLR measure is interpreted as a pure phase. 
In this paper, for $q=2,3,4$, we propose the following algorithm of identifying the 
pure phases. Let us list the distinct values of the hard-core diameters 
in an increasing order:
\beq\ua=:a(1)<\ldots <a(k):=\oa \label{1.17}\eeq
where $1\leq k\leq q(q-1)/2$.

For each $j=1,\ldots ,q$ we form a $k$-dimensional 
incidence-occupation number vector $\unn (j)$ with non-negative integer 
entries: 
\beq\unn (j)=\big(n(j,1),\ldots,n(j,k)\big). \label{1.18}\eeq
Here 
$n(j,l)$ stands for the number of types $i\in\{1,\ldots ,q\}\setminus\{j\}$ such that  $D(j,i)=a(l)$, 
$1\leq l\leq k$. 
Then consider the collection $\St$ of types $j$ for which the vector $\unn (j)$ is lexicographically
maximal:
\beq\St =\St(\Dm )
:=\left\{j:\;\unn (j')\buildrel{\rm{lex}}\over \preceq\unn (j)\;\forall\;1\leq j'\leq q\right\}
. \label{1.19}\eeq
Mnemonically, to form set $\St$ we first select the types $j\in\{1,\ldots ,q\}$ which have the maximal 
number $n(j,1)$ of  incident `edges' $(j,j')$ with $D(j,j^\prime)=a(1)$ (i.e., the maximal number of types $j'$ 
that repel type $j$ with the shortest hard-core diameter $a(1)$). From these we select those types $j$
which have the maximal number $n(j,2)$ (representing edges $(j,j')$ with $D(j,j^\prime )=a(2)$), and so on. 
Pictorially speaking, types $i\in\St$ are `most tolerant` to other types $j$ in terms of  a `direct` repulsion. Cf.
Remark 1.3 below.

Geometrically, for $q=4$, particle types $j$ can be placed at the vertices of a tetrahedron, and 
edges $j\leftrightarrow l$
can be `painted` in different colors: e.g., green when $D(j,l)=a(1)$, blue when $D(j,l)=a(2)$, etc. 
(We do not mean here that $D(i,j)$ is equal to the length of an edge.) In this context,
vector $\unn (j)$ takes into account the multiplicities of single-edge paths of different colors 
beginning at $j$.

Values $i\in\St$ are called {\it stable particle types} or {\it stable types} for short.
\bigskip

{\bf Theorem 1.1.} {\sl Assume that collection $\Dm$ satisfies the triangular property \eqref{1.03}.
Then, for $d\geq 2$ and $q=2,3,4$, there exists $z_0\in (0,\infty )$ (depending on 
the collection $\Dm$) such that $\forall$ $z\geq z_0$: 

{\rm{(I)}} $\forall$ $i\in\St$ there exists a pure phase $\mu (\;\cdot\;\|\,i)$ (a shift-invariant, ergodic
DLR-measure, with exponential mixing) obtained as 
the limit 
\beq\lim\limits_{\Lam\nearrow\bbR^d}\mu_{\Lam}(\;\cdot\;\|\,i)=\mu (\;\cdot\;\|\,i). \label{1.20}\eeq
Measures $\mu (\;\cdot\;\|\,i)$ are distinct for different $i\in\St$.

{\rm{(II)}} Every shift-periodic DLR-measure is a convex linear 
combination of measures $\mu (\;\cdot\;\|\,i)$, $i\in\St$.}
\bigskip

{\bf Remark 1.1.} Under an additional 
assumption that $k=1$ (i.e., when $D(i,j)\equiv \ua$), assertion (I) has been proved for any value of $q\geq 2$ in \cite{RL} and generalized to a wider class if symmetric interactions in \cite{GH}.

The only place using the triangular property is Eqn \eqref{2.33}.
$\qquad\blacktriangle$
\bigskip

{\bf Theorem 1.2.} {\sl Take $q=4$ and adopt the same assumptions as in Theorem~1.1.
Then, for $d\geq 2$ and $\forall$ $z\geq z_0$, the following assertions hold true.

{\rm{(I)}} Suppose that  $\sh\St =1$ and, without loss of generality, $\St =\{1\}$ (i.e., $1$ is a unique 
stable particle type). Then 
\beq\lim_{\Lam\nearrow\bbR^d}\mu_\Lam(\;\cdot\;\|\,j)=\mu (\;\cdot\;\|\,1)\;\;\forall\;j\in\{1,2,3,4\}. \label{1.21}\eeq
Moreover, any family of Gibbs distributions $\mu_\Lam\left(\;\cdot\;|\ubY^{\Lam^\cmp} \right)$ with 
boundary conditions $\ubY^{\Lam^\cmp}$ induced by  $\ubY\in\cA$ converges to $\mu (\;\cdot\;\|\,i)$ as $\Lam\nearrow\bbR^d$.

{\rm{(IIa)}} Suppose that $\sh\St =2$ and $\St =\{1,2\}$ (i.e., there are two stable types, 
$1$ and $2$). Suppose $D(1,3)=D(2,3)$.
Then $D(1,4)=D(2,4)$, and vice versa. In this case, 
\beq\lim_{\Lam\nearrow\bbR^d}\mu_\Lam (\;\cdot\;\|\,3)=\lim_{\Lam\nearrow\bbR^d}\mu_\Lam (\;\cdot\;\|\,4)
=\frac{1}{2}\mu (\;\cdot\;\|\,1)+\frac{1}{2}\mu (\;\cdot\;\|\,2). \label{1.22}\eeq

{\rm{(IIb)}} Suppose again that $\sh\St =2$ and $\St =\{1,2\}$. Suppose  $D(1,3)\neq D(2,3)$ 
then $D(1,4)\neq D(2,4)$, and vice versa. In this case
$D(1,4)=D(2,3)$ and $D(1,3)=D(2,4)$. Furthermore,  
\beq\lim_{\Lam\nearrow\bbR^d}\mu_\Lam(\;\cdot\;\|\,3)=\begin{cases}\mu (\;\cdot\;\|\,1),&\hbox{if }
D(1,3)<D(2,3),\\
\mu (\;\cdot\;\|\,2),&\hbox{if }
D(2,3)<D(1,3),\end{cases}  \label{1.23}\eeq
and, similarly,
\beq\lim_{\Lam\nearrow\bbR^d}\mu_\Lam (\;\cdot\;\|\,4)=\begin{cases}\mu (\;\cdot\;\|\,1),&\hbox{if }
D(1,4)<D(2,4),\\
\mu (\;\cdot\;\|\,2),&\hbox{if }
D(2,4)<D(1,4).\end{cases}  \label{1.24}\eeq

{\rm{(III)}} Suppose $\sh\St =3$ and $\St =\{1,2,3\}$ (i.e., there are three stable types, 
$1$, $2$ and $3$). Then 
\beq\lim_{\Lam\nearrow\bbR^d}\mu_\Lam(\;\cdot\;\|\,4)= \frac{1}{3}\mu (\;\cdot\;\|\,1)+\frac{1}{3}\mu (\;\cdot\;\|\,2)
+\frac{1}{3}\mu (\;\cdot\;\|\,3). \label{1.25}\eeq

{\rm{(IV)}} In all cases ($\sh\St=1,2,3,4$),
\beq\lim_{\Lam\nearrow\bbR^d}\mu_\Lam(\,\cdot\,) =\frac{1}{\sh\St}\sum\limits_{i\in\St}\;\mu (\;\cdot\;\|\,i)
. \label{1.26}\eeq}
\bigskip

{\bf Remark 1.2.} Assertions (IIa), (IIb), (III) and (IV) have a transparent geometrical meaning. 
Viz., conditions specifying case (IIa) mean that unstable types $3$ and $4$ are positioned
symmetrically relative to types $1$ and $2$ forming set $\St$. Next, cases (III) and (IV) possess 
a maximum geometric symmetry which leads to uniform coefficients. In contrast, in case (IIb) 
inequalities in \eqref{1.23} and \eqref{1.24} indicate that the unstable type generates a
phase corresponding to a `closer' type from $\St$. 

{\bf Remark 1.3.} For $q=5$, the particle types can be placed at the vertices of a simplex in 
${\bbR}^4$ (a pentagram). Here,  the algorithm of specifying the set $\St$ is based on the 
analysis of both single-edge and  two-edge paths  issued from a given vertex $i$. The number of 
edges in a path needed to be taken into account increases with $q$ (in general, non-monotonically). 
Formally, for $q\geq 5$ it is still true that any type $i$ which creates a pure phase lies in $\St$. However, 
not every $i\in\St$ yields a pure phase; the specification of the subset $\St^{\rm{st}}\subset\St$ 
corresponding to pure phases remains an open question.
%
%
$\qquad\blacktriangle$
\bigskip

The subsequent sections are dedicated to the proofs. Formally, we assume that $q=4$;
for $q=2,3$ the proofs simplify. 
The proof of assertion (I) of Theorem 1.1 is completed 
in sub-Section 4.4 and that of assertion (II) in  sub-Section 5.5. The proof of Theorem 1.2 
is completed in sub-Sections 5.1 - 5.4. 
\bigskip

\section{Contours: definitions and main facts}

The aim of this section is to introduce a representation of the WR  model under consideration in
terms of `contours`. To this end, we first perform a coarse graining procedure reducing the original
model to a model on a cubic lattice $\bbZ^d$. We then proceed with definitions of `rarefied` and 
`crystalline` partition functions that are standard concepts employed in the PST; see Eqns \eqref{2.34}
and \eqref{2.35}. The crystalline partition function plays a role of (and introduced as) a statistical 
weight of an external contour. As usual, key properties are factorizations into products; cf.
Eqns  \eqref{2.33},  \eqref{2.34} and \eqref{2.36}.

\bigskip

{\bf 2.1. Hard-core exclusion diameters.} Given set $\Dm$ of hard-core exclusion diameters we 
select a constant $R>0$ which is used as a discretization scale. This constant is taken to be small 
enough such that 
\beq  \min\left[\ua, \min_{1\leq l_1<l_2\leq k} |a(l_1) - a(l_2)|\right] > 10 Rd;\label{2.01}\eeq  
the number 10 is selected for definiteness only and is not optimal. Rescaling of $\bbR^d$ 
with magnification $1/R$ maps our original model into a similar model with rescaled values of $D(i,j)$ 
and $z$. Accordingly, the previous requirement takes the form
\beq  \min\left[\ua, \min_{1\leq l_1<l_2\leq k} |a(l_1) - a(l_2)|\right] > 10d. \label{2.02}\eeq
Without loss of generality we assume now that $R=1$ and \eqref{2.02} is satisfied. 
\bigskip

{\bf 2.2. The \basic contour definition.} Our next aim is to translate 
the original model given in $\bbR^d$ into a lattice model living in $\bbZ^d$ (cf. \cite{R, BKL}).
A peculiarity of the WR model is that it is convenient to introduce contours without discussing
(formal) Hamiltonians.

For this purpose we set:
\beq\begin{array}{l}
\bbB (\ty,S):=\left\{\tx=(x^1,\ldots ,x^d) \in \bbR^d:\;\sum\limits_{i = 1}^d(x^i - y^i)^2\leq S^2\right\}, \\ 
\bbB (S):=\bbB (0,S)\\
\Ups(\ty):=\left\{\tx=(x^1,\ldots ,x^d) \in \bbR^d:\;|x^i - y^i| < {\diy\frac{1}{2}}\quad  i=1,\ldots,d\right\},\end{array} \label{2.03}\eeq
In other words, $\bbB (\ty,S)$ and $\bbB (S)$ in \eqref{2.03} are  closed balls of radius $S$, and 
$\Ups (\ty)$ is a unit cube (cell)
centered at a point $\ty =(y^1,\ldots ,y^d) \in \bbZ^d$ with integer coordinates. 
For most of the time,  the reference to the center $\ty$ will be not 
important; accordingly, the argument  $\ty$ will be omitted. From now on, $\Lam$ will be the closure
of a union of a finitely many unit cells $\Ups (\ty )$ (an example of which is the cube \eqref{1.05}); we will 
sometimes refer to such $\Lam$ as a box. All definitions given in Section 1 remain valid for non-cubic boxes.

Depending on the context we will sometimes identify a cell $\Ups$  with its center $\ty$ and write 
$\Ups\in\bbZ^d$; similarly, a box
$\Lam$ is identified with the collection of centers of cells $\Ups\subset\Lam$ and treated as a subset in $\bbZ^d$. 
The number of lattice sites in $\Lam\subset\bbZ^d$ will coincide with the  Euclidean volume 
$\ups (\Lam)$ of $\Lam\subset\bbR^d$. \def\und{\underline}

We will need some notation for `extended` and `reduced` boxes produced from a given box,
$\Lam$. To this end, for a given unit cell $\Ups$ denote by $\rQ_m(\Ups)$ a cube of integer linear size $m$ concentric with $\Ups$. Then $\Ups = \rQ_1(\Ups)$ and we set:
\beq\begin{array}{c}{^m\Lam} = \bigcup\limits_{\Ups \subset \Lam} \rQ_{2m + 1}(\Ups) \quad {\rm and}\quad
{_m\Lam} = \bigcup\limits_{\Ups: \; \rQ_{2m+1}(\Ups) \subset \Lam} \Ups.\end{array}\label{2.04}\eeq

Let $\ubX\in\cA (\Lam )$ be a particle configuration admissible in $\Lam\subset\bbZ^d$. Then 
each unit cell $\Ups\subset\Lam$ cannot have particles of more than one type in it.
This gives rise to a map 
\beq\ubphi_\Lam\,:\;\Ups\subset\Lam \mapsto \ubphi_\Lam (\Ups ,\ubX^\Lam )\in\{0,1,\ldots,q\}\label{2.05}\eeq
attaching types to unit cells in $\Lam$. (Here $\ubphi_\Lam (\Ups ,\ubX^\Lam) = 0$ means cell $\Ups $ 
has no particles in $\ubX$.) The collection of pairs $(\Ups ,\ubphi_\Lam(\Ups ,\ubX^\Lam))$, where 
$\Ups\subset\Lam$, is called an {\it admissible cell configuration} in $\Lam$
generated by $\ubX$.  Moreover, a map 
\beq\uphi_\Lam:\;\Ups \subset\Lam\mapsto\{0,1,\ldots ,q\}\label{2.06}\eeq 
is called an 
admissible cell configuration
in $\Lam$ with empty boundary conditions if there exists a particle configuration $\ubX\in\cA (\Lam )$ such that $\uphi_{\,\Lam}(\Ups )
=\ubphi_\Lam(\Ups ,\ubX^\Lam)$
$\;\forall\;$ $\Ups \subset\Lam$. We refer to $\ubX$ as a generating particle configuration for $\uphi_{\,\Lam}$. 
Furthermore, 
the value $\uphi_{\,\Lam}(\Ups )$ is called the type of cell $\Ups $ in 
$\uphi_{\,\Lam}$. 

Mnemonically, the difference between $\ubphi_\Lam$ and $\uphi_\Lam$ is that the former is always used
with a reference to a generating particle configuration $\ubX$ whereas the latter can be considered without 
bearing in mind a particular generating particle configuration (which is guaranteed to exist and transfers a number of properties to the generated cell configuration $\uphi_\Lam$).  The same is true about $\uphi$ and $\ubphi$ 
introduced in Eqns \eqref{2.07} and \eqref{2.08} below.

The set  of admissible cell configurations in $\Lam$ with empty boundary conditions  
is denoted by $\cC (\Lam )$. The map $\ubphi_\Lam$ and probability distribution $\mu_\Lam(\,\cdot\,)$ defined in \eqref{1.06} induces a probability distribution 
on $\cC (\Lam )$ which we again denote by $\mu_\Lam(\,\cdot\,)$. 

Similarly, we denote by $\cC(\Lam ||\,i)$ the set of admissible cell 
configurations $\uphi_{^2\Lam} \in \cC ({^2\Lam} )$ generated by 
$\ubX\in\cA (\Lam ||\,i)$. Without loss of generality we denote these configurations $\uphi_{\Lam}$ rather than $\uphi_{^2\Lam}$ as their value is equal to $i$ for each $\Ups \subset {^2\Lam} \setminus \Lam$ . The probability distribution on $\cC(\Lam ||\,i)$ induced by $\mu_\Lam (\,\cdot\,||\,i)$ defined in  \eqref{1.12} is also denoted by $\mu_\Lam (\,\cdot\,||\,i)$.

Now let $\ubX\in\cA$. In a similar manner we define a map 
\beq\ubphi\,:\;\Ups\in\bbZ^d\mapsto\ubphi (\Ups ,\ubX)\in\{0,\ldots ,q\}.\label{2.07}\eeq 
The collection of pairs $(\Ups, \ubphi (\Ups,\ubX)$ is called
an admissible cell configuration in $\bbZ^d$ generated by $\ubX$. Further, a map \beq\uphi :\;\Ups\in\bbZ^d\mapsto\{0,1,\ldots ,q\}\label{2.08}\eeq 
is called an admissible cell configuration in $\bbZ^d$ if there exists
a particle configuration $\ubX\in\cA$ such that $\uphi (\Ups )=\ubphi (\Ups,\ubX)$. Here we again say that $\uphi (\Ups )$ is the type of cell $\Ups$ in $\uphi$. We denote by $\cC$ the set of admissible cell configurations in $\bbZ^d$.  If $\uphi_\Lam \in \cC (\Lam||\,i)$ then its extension $\uphi^*_{\Lam ,i} \in \cC$ is defined as
\beq\label{2.09}\begin{array}{r}
\uphi^*_{\Lam ,i}(\Ups) := \begin{cases} \uphi_\Lam(\Ups) &\mbox{if}\; \Ups \subset \Lam \\
i &\mbox{if}\; \Ups \not\subset \Lam. \end{cases}
\end{array}\eeq

For the rest of the paper we are working mainly in terms of admissible cell configurations (i.e. in terms of discretized lattice model) passing back to particle configurations when necessary. The exact procedures of recalculating statements given in terms of cell configurations to the corresponding statements in terms of particle configurations are self-evident except for few cases detailed later in this section.

\bigskip
{\bf Definition 2.1.}  Given an admissible cell configuration $\uphi\in\cC$, 
we say that a cell $\Ups$ of non-$0$ type $i$ is {\it in phase} $i$ {\it in} $\uphi$ (or in a generating particle configuration $\ubX$) if all of $5^d$ cells in the cube ${^2\Ups}$ have attached type $i$. 

Next, given an admissible cell configuration $\uphi_\Lam\in\cC(\Lam||\,i)$, a cell $\Ups\subset\Lam$ of non-$0$ type $j$ is said to be  {\it in phase $j$ in} $\uphi_\Lam$ (or in a generating particle configuration $\ubX^\Lam$) if each cell of ${^2\Ups} \cap \Lam$ has attached type $j$. 


\bigskip
The property behind Definition 2.1 is that inside a cell in phase $i$ any position of a particle of type $i$ in a generating particle configuration $\ubX$ is allowed, regardless of particle positions in any other cell. In probabilistic terms, restrictions $\bX^{\Ups }_i$ and $\ubX^{\Lam\setminus\Ups}$ are conditionally independent, given that cell $\Ups$ is in phase $i$. Here we refer to any measure $\mu_\Lam\left(\,\cdot\,|\ubY^\complement\right)$, $\mu_\Lam(\,\cdot\,||\,i)$ or $\mu_\Lam(\,\cdot\,)$, with $\Lam$ containing ${^2\Ups}$, provided that the probability of the condition under such a measure is positive. 

Indeed, given  $i,j\in\{1,\ldots ,q\}$ with $i\neq j$, consider the union $\bbU=\operatornamewithlimits{\cup}\limits_{\tx\in\Ups} \bbB(\tx, D(i,j))$. Then, for any choice of points $\ty$ in each of the $5^d-1$ cells in ${^2\Ups} \setminus \Ups$ (at least one point in a cell), the union of Euclidean balls of radius $D(i,j)$ with centers at points $\ty$  contains $\bbU$. (Note that the above statement is not true if ${^2\Ups}$ is replaced with ${^1\Ups}$). This is the {\it screening phenomenon} mentioned in the previous section.

Because of the screening phenomenon, Definition~2.1 does not refer to the interaction radius $\oa$ (or to collection $\Dm$ specifying the potentials $\Phi_{ij}$). In the literature the cube surrounding cell $\Ups$ has traditionally the size larger than the radius of interaction. See, e.g., \cite{BKL, K, Si, Z}. 
\bigskip

{\bf Definition 2.2.} Given an admissible cell configuration $\uphi\in\cC$, consider the union $\rP(\uphi )$ of all cells $\Ups\in\bbZ^d\;$ which are in some (non-$0$) phase in $\uphi$. The complement $\bbZ^d\setminus\rP (\uphi )$ is denoted by $\rC (\uphi )\;$. Similarly for $\uphi_\Lam \in \cC (\Lam||\,i)$ consider  the sets $\rP(\uphi^*_{\Lam ,i})$
and $\rC (\uphi^*_{\Lam ,i}) = \bbZ^d \setminus \rP(\uphi^*_{\Lam ,i})$, where $\uphi^*_{\Lam ,i}$ is defined in \eqref{2.09}.

A {\it \basic contour} $\;\Gam =\Gam (\uphi )$ or $\Gam =\Gam (\uphi_\Lam )$ generated by an admissible cell configuration $\uphi \in \cC$ or $\uphi_\Lam \in \cC (\Lam||\,i)$ is a 
connected component of $\rC(\uphi )$ or a connected component of $\rC (\uphi^*_{\Lam ,i})$ considered with the types of cells that lie in this connected component. The set of unit cells (with discarded types) forming {\BC} $\Gam$ is denoted by $\rB (\Gam )$ and called the {\it base} of {\BC} $\Gam$ (as before, $\rB (\Gam )$ can be identified with a subset of $\bbR^d$). The set $\rB(\Gam )$ contains a subset $\rO (\Gam )$ formed by unit cells with type $0$. $\qquad\blacktriangle$

\bigskip
Note that for a {\BC} $\Gam (\uphi_\Lam )$ the set $\rB (\Gam )$ is always finite and may contain unit cells $\Ups\not\subset\Lam$. However, $\rB(\Gam )\subseteq {^2\Lam}$ and $\rO(\Gam )\subseteq {\Lam}$. The simplest example is where $\uphi_\Lam$ attaches type $0$ to all unit cells in $\Lam$.

It is clear that the Euclidean volumes satisfy:
\beq  0 < 5^{-d}\ups(\rB(\Gam))  \le \ups(\rO(\Gam)) < \ups(\rB(\Gam)),\label{2.10}\eeq 
as any cube of size 5 centered at any cell of $\rB(\Gam )$ contains at least one cell of type 0 (empty cell).
\bigskip

{\bf Definition 2.3.} Given an admissible cell configuration $\uphi\in\cC$ or  $\uphi_\Lam\in\cC (\Lam ||\,i)$ for some $i=1,\ldots ,q$, let $\Gam$ be a {\BC} in $\uphi$ or $\uphi_\Lam$. If $\rB(\Gam)$ is finite then the complement $\rB (\Gam)^\cmp=\bbZ^d\setminus \rB (\Gam )$ has a single connected component which is called the {\it exterior} of $\;\Gam$ and denoted by $\;\rE (\Gam )$. The region $\bbZ^d\setminus\big(\rB (\Gam)\cup\rE (\Gam )\big)$ is denoted by $\rI (\Gam )$ and called the {\it interior} of $\Gam$ (it may be empty); its connected components are denoted by $\rI_s(\Gam )$ and labeled by $s=1,2, \ldots$ in an arbitrary order. $\qquad\blacktriangle$

\bigskip
{\bf Definition 2.4.}  For  $\uphi\in\cC\;$ or $\;\uphi_\Lam\in\cC (\Lam ||\,i)$, let $\Gam$ be a {\BC} in $\uphi\;$ or $\;\uphi_\Lam$ with finite $\rB(\Gam)$.  It is not hard to see that under $\uphi$ all unit cells $\Ups \subset \rB(\Gam)$ adjacent to $\rE(\Gam)$ have the same type $\iota^\tE (\Gam )\in\{1,\ldots ,q\}$ called the {\it external type} for $\Gam$. The same is true under $\uphi^*_{\Lam ,i}$ (cf. \eqref{2.09}); we will employ the same term external type and the same notation $\iota^\tE (\Gam )$. Further, in both situations (with $\uphi$ and $\uphi_\Lam$) the unit cells $\Ups \subset \rB(\Gam)$ adjacent to $\rI_s(\Gam )$ also have the same type $\iota_s(\Gam )= \iota (\rI_s(\Gam ))\in\{1,\ldots ,q\}$. The value $\iota_s(\Gam )$ is called the $s$-th {\it internal type} for $\Gam$. $\qquad\blacktriangle$

\bigskip
{\bf Definition 2.5.} Given $\uphi\in\cC$, the set of \basic contours in $\uphi$ is called the {\it \basic contour collection} in $\bbZ^d$ generated by $\uphi$ and denoted by $\uGam (\uphi )$. (The collection $\uGam (\uphi )$ can be finite or countable.) Likewise, for $\uphi_\Lam\in\cC(\Lam||\,i)$, the collection of {\BCs} in $\uphi^*_{\Lam ,i}$ is denoted by $\uGam (\uphi_\Lam)$ and called the {\BCC} in $\Lam$ generated by $\uphi_\Lam$. (The collection $\uGam(\uphi_\Lam)$ is always finite.) $\blacktriangle$

\bigskip
It is convenient to introduce the maps
\beq\begin{array}{rl}
\ubgam :&\uphi\in\cC\mapsto \uGam\,(\,\uphi\,), \\
\ubgam\circ\ubphi :&\ubX\in\cA\mapsto \uphi\mapsto\uGam (\uphi ),\\ 
\ubgam_\Lam:&\uphi_\Lam\in\cC (\Lam ||i)\mapsto \uGam\,(\uphi_\Lam), \\
\ubgam_\Lam\circ\ubphi_\Lam:&\ubX\in\cA(\Lam ||\,i)\mapsto \uphi_\Lam\mapsto\uGam (\uphi_\Lam).
\end{array} \label{2.11}\eeq

For brevity we will say that $\Gam$ is a {\BC} in/from an admissible cell configuration $\uphi\in\cC$
or  $\uphi\in\cC (\Lam||\,i)$ if $\Gam$ belongs to the {\BCC} $\uGam (\uphi )$ or $\uGam (\uphi _\Lam)$.  

Suppose $\Gam$ is a {\BC} in  $\uphi\in\cC$ with finite $\rB(\Gam)$. Define $\uphi^*_\Gam \in\cC$ as
\beq\label{2.12}\begin{array}{r}
\uphi^*_\Gam(\Ups) := \begin{cases} \uphi(\Ups) &\mbox{if}\; \Ups \subset \rB(\Gam) \\
\iota^\tE (\Gam) &\mbox{if}\; \Ups \subset \rE(\Gam ) \\
\iota_s(\Gam ) &\mbox{if}\; \Ups\subset\rI_s(\Gam ). \end{cases}
\end{array}\eeq
Then, clearly, the {\BCC} $\uGam (\uphi^*_{\Gam})$ consists of a single {\BC} $\Gam$. A similar construction can be performed for any {\BC} $\;\Gam$ from $\uphi_\Lam\in\cC (\Lam ||\,i)$ with $\iota^\tE (\Gam )=i$. The corresponding admissible cell configuration from $\cC (\Lam ||\,i)$ is denoted by $\uphi^*_{\Lam,\Gam}$. Evidently, $\uphi^*_{\Lam,\Gam}$ is just a restriction of $\uphi^*_\Gam$ to $\Lam$. Also, the cell configuration $\uphi^*_{\Lam,\Gam}$ defined in \eqref{2.12} should not be confused with the cell configuration $\uphi^*_{\Lam ,i}$ defined in \eqref{2.09}. 

For any cell configuration $\uphi$ and any box $\Lam$ let $\uphi{\big|}_\Lam$ be the restriction of $\uphi$ to $\Lam$. Note that $\uphi^*_\Gam$ is uniquely determined by its restriction $\uphi^*_\Gam{\big|}_{\rB (\Gam )}$ to $\rB (\Gam )$. 
In fact, it will be convenient to identify a {\BC} $\Gam$ with a pair 
\beq\label{2.13}
\Gam = (\rB,\varphi_\rB ),
\eeq
where $\rB=\rB(\Gam)$ is a connected union of cubes of size $5$ and $\varphi_\rB = \varphi(\Gam)$ is a restriction $\uphi{\big|}_{\rB}$ of a cell configuration $\uphi\in\cC$ having a single {\BC} $\Gam(\uphi)$ with $\rB(\Gam(\uphi)) = \rB$. 

\bigskip

{\bf Definition 2.6.} A {\BC} $\Gam = (\rB, \varphi_\rB)$ is called  {\it non-separating} if $\Gam$ is drawn within a single type, i.e. for each $\Ups \subset \rB$ the value of $\varphi_\rB(\Ups)$ is either $0$ or some $i \in \{1,\ldots,q\}$. Otherwise,  {\BC} $\Gam$ is called {\it separating}. (See Figure 1). $\qquad\blacktriangle$

\begin{center}
\includegraphics[scale=1]{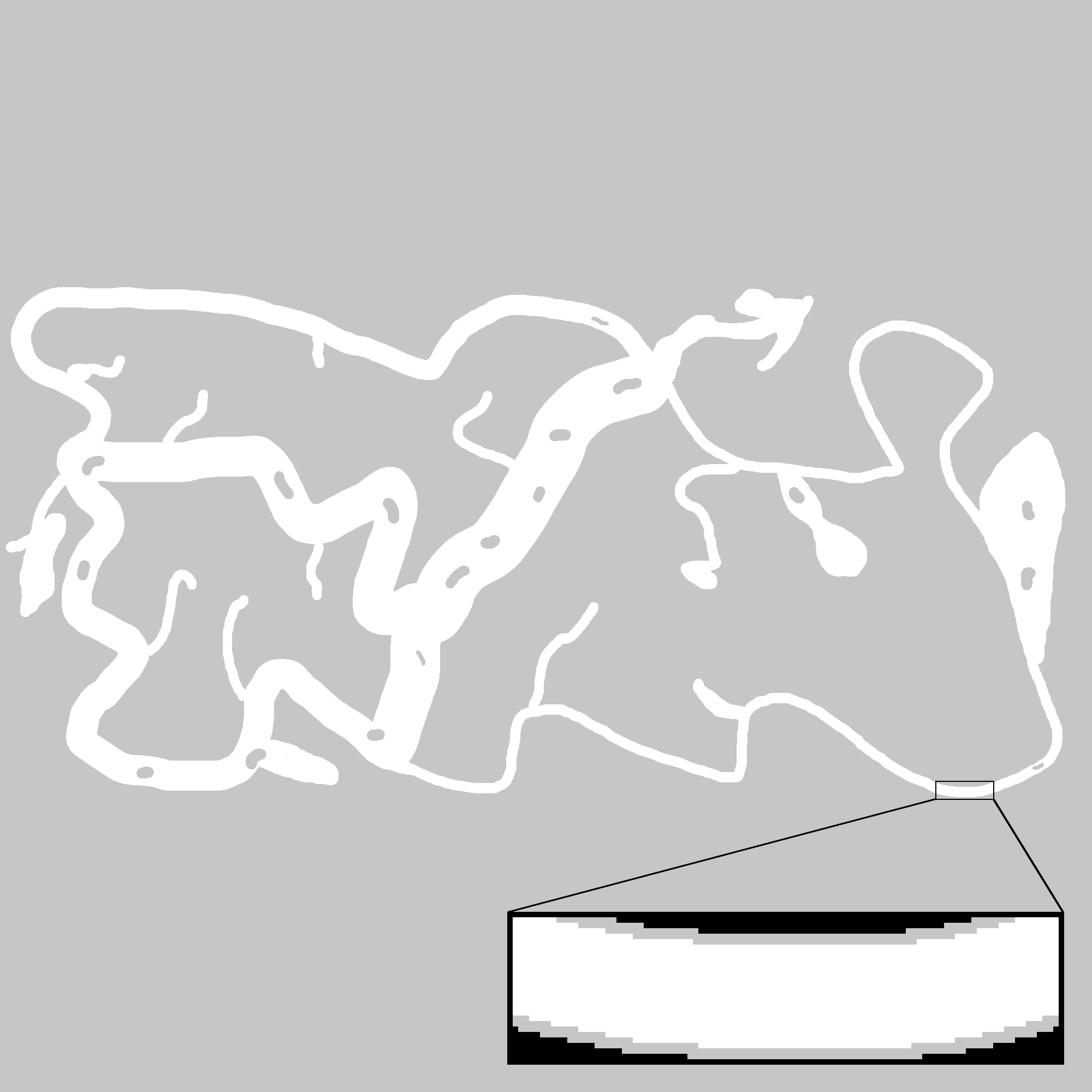}\quad
\includegraphics[scale=1]{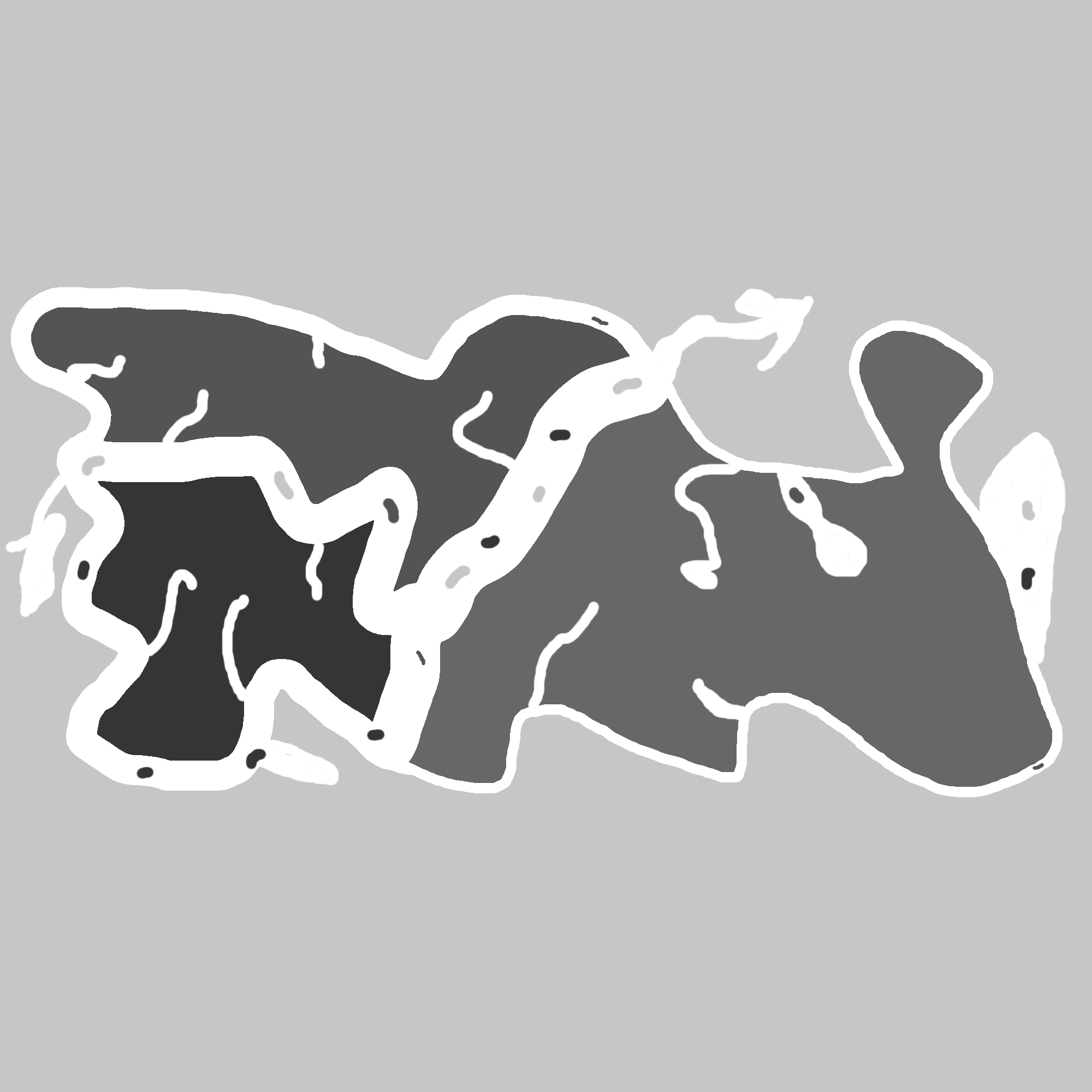}
\cl{{\bf (i)}\qquad\qquad\qquad\qquad\qquad\qquad\qquad {\bf (ii)}}
\cl{}
\cl{\bf Figure 1. Non-separating and separating contours}
\end{center}

In Figure 1 it is assumed that the picture is being drawn on a small size square lattice as is shown in the magnified insert in Figure 1(i). The white color indicates empty cells which form the set $\rO (\Gam)$. The width of $\rO (\Gam )$ respects the hard-core diameters $D(\,\cdot\,,\,\cdot\,)$. Different shades of grey colors represent unit cells containing different non-$0$ particle types. The set $\rB(\Gam)$ is slightly wider than $\rO (\Gam)$ and includes additional strips of width 2 formed by non-$0$ cells, as can be seen in the magnified insert in Figure 1(i). In this insert a small piece of $B(\Gam)$ is drawn in not black colors. Large connected areas colored some shade of grey represent either $\rE(\Gam)$ or connected components of $\rI(\Gam)$ (there are 3 of them). Figure 1(i) shows non-separating \basic contour. The small rectangular area is magnified to show the details. Figure 1(ii) shows separating contour. Small grey colored areas surrounded by white represent unit cells not in any phase. They are drawn out of proportion (thicker than 5 unit cells) to make them visible.

\bigskip
{\bf Definition 2.7.}  Given a cell configuration $\uphi\in\cC$, let $\Gam$ be a {\BC} from $\uphi$. The {\BC} $\Gam$ is called {\it external} (in {\BCC} $\uGam =\uGam (\uphi )$) if there exist a path on $\bbZ^d$ beginning in $\rB (\Gam )$ and going to infinity (i.e., reaching outside any cube with center at the origin) while passing through unit cells in phase $\iota^\tE (\Gam )$. An external {\BC} will be denoted by $\Gam^\tE$. The intersection $\cap\rE (\Gam^\tE)$ taken over all external {\BCs} $\Gam^\tE\in\uGam$ is denoted by $\rE (\uGam )$ and called the exterior of {\BCC} $\uGam$. If $\rE(\uGam )$ is a non-empty connected set then the value $\iota^\tE(\Gam^\tE)$ is the same for all external {\BCs} $\Gam^\tE\in\uGam$, and $\uphi$ assigns this value to all $\Ups\subset\rE (\uGam )$. We denote this value by $\iota^\tE(\uGam)$.

Similarly, for  $\uphi_\Lam\in\cC (\Lam ||\,i)$ a {\BC} $\Gam$ from $\uGam =\uGam(\uphi_\Lam)=\uGam(\uphi^*_{\Lam ,i})$ is called external if $\Gam$ is external in the cell configuration $\uphi^*_{\Lam ,i}\in\cC$ introduced in Eqn \eqref{2.09}. In that case $\iota^\tE (\Gam )=i$. We set $\rE_\Lam (\uGam )=\Lam\cap\rE(\uGam(\uphi^*_{\Lam ,i}))$. Clearly, $\iota^\tE(\uGam)=i$. $\qquad\blacktriangle$ 

\bigskip
Denote
\beq\label{2.14}\begin{array}{rl}
\cCi:=\Big\{\uphi\in\cC: &\hbox{$\rE (\uGam (\uphi )) \not = \varnothing$ is connected,}\\ &\hbox{and}\;\uphi (\Ups )=i\; \hbox{for all}\;\Ups\subset\rE (\uGam (\uphi ))\Big\}.\end{array}\eeq
Pictorially, a cell configuration $\uphi\in\cCi$ consists of an `external sea` of cells $\Ups$ in phase $i$ and finitely or countably many `islands` of a finite size represented by external {\BCs} $\Gam^\tE$, all of them having $\iota^\tE(\Gam^\tE )=i$.

Next, 
\beq\label{2.15}
\cDi := \Big\{\uGam:\; \uGam = \uGam(\uphi),\quad \uphi\in \cCi \Big\}
\eeq
Equivalently, $\cDi$ consists of finite or countable  {\BCCs} $\uGam$ such that any unit cell $\Ups\in\bbZ^d$ is either in phase $i$ or belongs to the union $\rI (\Gam^\tE )\cup\rB (\Gam^\tE )$ where $\Gam^\tE$  is an external contour in $\uGam$ with $\iota^\tE (\Gam^\tE )=i$.

Figure 2 below shows a {\BCC} including 13 \BCs, of which 7 are external. We should 
stress that all figures in the paper should be viewed as drawn on a small-size square lattice.    

\begin{center}
\includegraphics[scale=1]{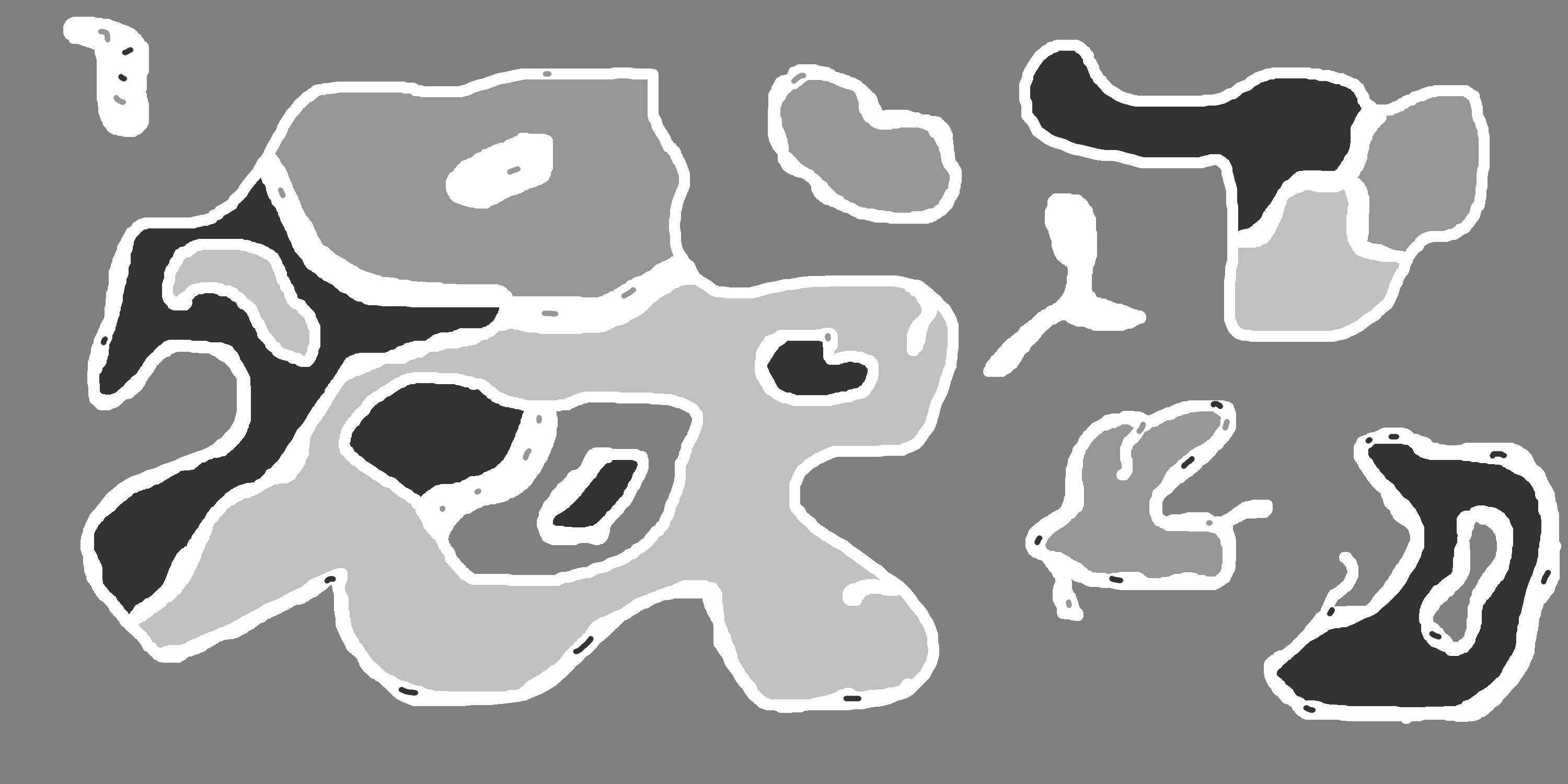}
\cl{\bf Figure 2. A collection of \basic contours}
\end{center}
\bigskip

{\bf 2.3. Compatible contour collections.} In this section we complete the discretization of the model. We define compatible collections of \basic contours and statistical weights of \basic contours. Then we rewrite the partition functions originally defined in terms of particle configurations as partition functions over compatible collections of \basic contours (or equivalently cell configurations). Finally, we define Gibbs/DLR measures on compatible collections of \basic contours (or equivalently cell configurations) and explain how the original measures on particle configurations can be reconstructed from them.

\bigskip
{\bf Definition 2.8.} Let $\uGam$ be a {\BCC} generated by $\uphi\in\cC$ or $\uphi^*_{\Lam ,i}$
for $\uphi_\Lam\in\cC (\Lam ||\,i)$. Let $\Gam_1,\Gam_2\in\uGam$ be  two  
different contours. We say that $\Gam_1$ and $\Gam_2$ are {\it separated} (in $\uGam)$) 
if there is a {\BC} $\Gam_3\in\uGam$ different from $\Gam_1$ and $\Gam_2$ such that
there is no path on $\bbZ^d$ joining $\rB (\Gam_1)$ and $\rB (\Gam_2)$ and avoiding
$\rB (\Gam_3)$.  In this case we say that $\Gam_3$ {\it separates} $\Gam_1$ and $\Gam_2$.

In the opposite case (when $\Gam_1$ and $\Gam_2$ are not separated) there are two 
disjoint possibilities: 
\begin{description} \setlength{\itemsep}{0pt} \setlength{\parskip}{0pt} 
\item[(\phantom{i}i)] $\Gam_1$ and $\Gam_2$ are {\it nested}, i.e. either $\rB (\Gam_1)\subset\rI (\Gam_2)$ or $\rB (\Gam_2)\subset\rI (\Gam_1)$.
\item[(ii)] $\Gam_1$ and $\Gam_2$ are {\it mutually external}, i.e. $\rB (\Gam_1)\cap\rI (\Gam_2)= \rI (\Gam_1)\cap\rB (\Gam_2)=\varnothing. \blacktriangle$
\end{description}

It is not hard to see that in a {\BCC} $\uGam$ generated by $\uphi$ or $\uphi^*_{\Lam ,i}$ 
the following {\it compatibility conditions} are always fulfilled:
\begin{description} \setlength{\itemsep}{0pt} \setlength{\parskip}{0pt} 
\item[(\phantom{ii}i)] For any two different {\BCs} $\Gam_1, \Gam_2 \in \uGam$,
\beq \rB (\Gam_1)\cap\rB (\Gam_2)=\varnothing.\label{2.16}\eeq
\item[(\phantom{i}ii)] For any two mutually external  {\BCs} $\Gam_1, \Gam_2 \in \uGam$,
\beq\iota^{\tE}(\Gam_1) = \iota^{\tE}(\Gam_2).  \label{2.17}\eeq
\item[(iii)] For any two nested  {\BCs} $\Gam_1, \Gam_2 \in \uGam$ such that
$\rB (\Gam_1)\subset\rI_s(\Gam_2)$,
\beq\iota^{\tE}(\Gam_1) = \iota_s(\Gam_2).
\label{2.18}\eeq
\end{description}

Observe that conditions \eqref{2.16}--\eqref{2.18} are expressed in terms of bases $\rB (\Gam)$ and types $\varphi (\Gam)$ only. Therefore, one can think of compatible collections $\uGam$ of pairs $\Gam = (\rB,\varphi_{\rB})$. In fact, for any such finite or countable collection of pairs $\uGam$ there exists a unique admissible cell configuration 
\beq
\uphi^*_{\uGam} = \ubgam^{-1} (\uGam).
\label{2.19}\eeq
Cf. Eqn \eqref{2.12}. Thus, we can speak about contours and compatible collection of contours with or without specifying explicitly the generating cell configuration $\uphi$. If this configuration is not specified then we always assume that it is equal to $\uphi^*_{\uGam}$.

Now, given a box $\Lam$, we say that a (finite) compatible collection $\uGam$ is associated with $\cC (\Lam ||\,i)$ if $\uphi^*_{\uGam}$ attaches type $i$ to each cell $\Ups\not \subset\Lam$, i.e. $\iota^\tE(\uGam)=i$. The set of such collections is denoted by $\cD (\Lam\|\,i)$. It is not hard to see that 
\beq 
\cD (\Lam\|\,i)=\ubgam_\Lam\circ\ubphi_\Lam (\cA (\Lam\|\,i)),
\;1\leq i\leq q. \label{2.20}\eeq
Observe that for $\uGam\in\cD (\Lam ||\,i)$ and $\Gam\in\uGam$ we have $\rO (\Gam )\subseteq\Lam$. Moreover, for $\uGam\in\cD (\Lam ||\,i)$, and $\Gam\in\uGam$, base $\rB (\Gam)\subseteq {^2\Lam}$. 

We use Eqn \eqref{2.20} for making a passage from measures $\mu_\Lam (\,\cdot\,||\,i)$ to measures on compatible collections $\uGam \in \cD (\Lam\|\,i)$. Referring to \eqref{1.13}, consider the event 
\beq\label{2.21}\begin{array}{l}
\cA (\Lam, =\, i)=\Big\{\ubX:\;\ubX^\Lam\in\cA (\Lam \|\,i),\; \ubphi_\Lam (\ubX^\Lam) \equiv i\Big\}.\end{array}\eeq
(The condition $\ubphi_\Lam (\ubX^\Lam) \equiv i$ is equivalent to $\ubgam_\Lam\circ\ubphi_\Lam (\ubX^\Lam)=\varnothing$.) Then
\beq  {\bbP}\big(\cA (\Lam ,\,=\,i)\big) = e^{-z(q-1)\ups (\Lam)}(1 - e^{-z})^{\ups (\Lam)}. \label{2.22}\eeq
Note that the RHS in \eqref{2.22} does not depend on $i$. Throughout the paper (beginning
with Eqn \eqref{2.31}), the value
\beq\label{2.23}\rv=e^{-z}\eeq
plays the role of a small parameter. With this notation the probability  ${\bbP}\big(\cA (\Lam ,\,=\,i)\big) = \big( \rv^{q-1}(1 -\rv)\big)^{\ups (\Lam)}$.

Further, for $\uGam\in\cD (\Lam ||\,i)$ we set: 
\beq
\cA (\uGam,\Lam\|\,i) =
\Big\{\ubX\in\cA (\Lam\|\,i):\; \ubgam_\Lam\circ\ubphi_\Lam (\ubX^\Lam)=\uGam \Big\},
\label{2.24}\eeq
and for any $\Gam$ with $\rO(\Gam) \subseteq \Lam$ 
\beq
\cA (\Gam,\Lam) = \Big\{\ubX\in\cA (\Lam\|\,\iota^{\tE}(\Gam)):\; \ubgam_\Lam\circ\ubphi_\Lam(\ubX^\Lam)=\Gam \Big\}. \label{2.25}\eeq
In other words,  \eqref{2.25} means that the whole contour configuration $\ubgam_\Lam\circ\,\uphi (\ubX^\Lam)$ is reduced to a single {\BC} $\Gam$.

Probability distributions on admissible particle configurations and probability distributions on compatible {\BCCs} are closely related. The measure $\mu_\Lam (\cB\|\,i)$ defined on events $\cB \in \cA (\Lam\|\,i)$ (see \eqref{1.12}) induces the measure $\mu_\Lam (\uGam\|\,i)$ on compatible {\BCCs} $\uGam\in\cD(\Lam\|\,i)$:
\beq
\mu_\Lam (\uGam\|\,i):=\mu_\Lam\Big(\cA (\uGam ,\Lam ||\,i)\Big\|\,i\Big).
\label{2.26}\eeq
Recall that according to \eqref{1.12})
\beq
\mu_\Lam\Big(\cA (\uGam ,\Lam ||\,i)\Big\|\,i\Big)=\frac{{\bbP}
\big(\cA (\uGam,\Lam\|\,i)\big)}{{\bbP}\big(\cA (\Lam\|\,i)) \big)}.
\label{2.27}\eeq

Going back from contours to particles, given {\BCC} $\uGam\in\cD (\Lam||\,i)$, the conditional distribution of the generating particle configuration $\ubX:\; \ubgam_\Lam\circ\ubphi_\Lam (\ubX^\Lam)=\uGam$ under measure $\mu_\Lam (\,\cdot\,\|\,i)$ is determined by 
\beq
\mu_\Lam\Big(\,\cdot\,\|\,i; \uGam\Big)=\frac{\bbP\left(\,\cdot\,\cap\cA (\uGam, \Lam ||\,i) \right)}{\bbP (\cA (\uGam, \Lam\|\,i))}. 
\label{2.28}\eeq
For an events $\cB$ localized inside $\rE_{\Lam}(\uGam)$ we have that 
\beq
\mu_\Lam\Big(\cB\|\,i;\uGam\Big)=\frac{\bbP\left(\cB\cap\cA (\Lam, =\, i) \right)}{\bbP (\cA (\Lam, =\, i))}, 
\label{2.29}\eeq
which is a reformulation of the screening phenomenon: conditional on cell $\Ups \subset\Lam$ being in phase $i\neq 0$, the positions of (type $i$) particles inside $\Ups$  are  independent of particle positions inside any other cell.

A more convenient expression for $\mu_\Lam (\uGam\|\,i)$ can be derived in terms of statistical weights of contours. A {\it statistical weight of a \BC} $\Gam$ is defined as
\beq
\rw (  \Gam):=
\frac{{\bbP}\big(\cA (\Gam,\Lam)\big)}{{\bbP}\big(\cA (\Lam,\,=\,\iota^{\tE}(\Gam))\big)}.
\label{2.30}\eeq
Referring to Eqn \eqref{2.22} and the screening phenomenon, we conclude that the ratio in th RHS of \eqref{2.30} does not depend on $\Lam\supseteq\rO (\Gam )$  (or equivalently ${}^2\Lam\supseteq\rB (\Gam )$). That is why $\Lam$ is absent in the LHS of \eqref{2.30}.

For $z>0$ large enough, $\rw ( \Gam )$ is exponentially small in volume of $\rB(\Gam)$ as it satisfies straightforward bounds
\beq   
\left(\frac{\rv}{1-\rv} \right)^{\ups (\rB(\Gam))} < \rw (  \Gam) <
\left(\frac{\rv}{1-\rv} \right)^{\ups (\rO(\Gam))} ,\label{2.31}\eeq
cf. \eqref{2.10}.

A {\it statistical weight of a compatible \BCC} $\uGam \in\cD (\Lam\|\,i)$ is defined as
\beq
\rw (  \uGam)=\frac{{\bbP}\big(\cA (\uGam,\Lam \|i)\big)}{{\bbP}\big(\cA (\Lam ,\,=\,i) \big)}.
\label{2.32}\eeq
The ratio in the RHS of \eqref{2.32} also does not depend on $\Lam$ if $\Lam\supseteq\rO(\Gam )$ for all $\Gam\in\uGam$. 

The normalizing factor ${\bbP}\big(\cA (\Lam ,\,=\,i) \big)$ used in \eqref{2.32}, \eqref{2.30} represents the statistical weight of a `ground state` cell configuration $\uphi_{\Lam} \equiv i$ such that the statistical weight \eqref{2.32} of empty {\BCC} is $1$. 

The following factorization property is crucial for the forthcoming analysis 
\beq
\rw (  \uGam)= \prod_{\Gam \in \uGam} \rw (  \Gam).  
\label{2.33}\eeq
This factorization holds because of the triangle property \eqref{1.03} and the screening phenomenon. In a sense, the `contour language' owes its convenience to \eqref{2.33}.

The set $\cD (\Lam\|\,i)$ of {\BCCs} with the statistical weights $\rw (\uGam )$ yields the {\it {\BC} ensemble} in $\Lam$ with the external type $i$. For this ensemble, following \cite{Si} (see also \cite{BKL}), we define the `rarefied` partition function in $\Lam$ with the external type $i\in\{1,\ldots ,q\}$:
\beq
\Xi (\Lam\|\,i):=\frac{{\bbP}\big(\cA (\Lam\|\,i)) \big)}{{\bbP}\big(\cA (\Lam ,\,=\,i)\big)}\,
=\sum_{\uGam \in \cD (\Lam\|\,i)}\;\; \prod_{\Gam \in \uGam} \rw (  \Gam).
\label{2.34}\eeq
An empty collection $\uGam$ enters the RHS of \eqref{2.34} with the statistical weight $1$. In case $\Lam = \varnothing$, we set $\Xi (\Lam\|\,i)=1$.  Representation \eqref{2.34} (and similar representation \eqref{2.40} below) makes possible the use of established  methods of cluster or polymer expansions. 

A specific feature of the model under consideration (with $q\leq 4$) is that $\Xi (\Lam\|\,i)$ gives the same value $\forall$ $i\in\St$ (for all values of $z$). This fact is established in Lemma 3.3 and repeatedly used in Sections~4~and~5. It would be possible to expect that $\Xi (\Lam\|\,i)$ is maximized when $i\in\St$. This property would have simplified our proofs,  but we couldn`t verify it. (It turns out to be false for $z$ small and some $\tt D$.)

It is also convenient to use the {\it external {\BC} ensemble} in box $\Lam$ with the external type $i$. To this end, for any contour $\Gam =(\rB,\varphi_{\rB})$ with finite $\rB$ we introduce the {\it external statistical weight}, otherwise known as the `crystalline` partition function
\beq
w(\Gam ):=\diy\frac{{\bbP}\Big\{\cA \big(\Gam ,\rB (\Gam )
\cup \rI (\Gam)\|\,\iota^\tE(\Gam )\big) \Big\}}
{{\bbP}\Big\{\cA\big(\rB (\Gam )
\cup\rI (\Gam ),\,=\,\iota^\tE(\Gam )\big)\Big\}}\,.    
\label{2.35}\eeq 
A useful consequence of \eqref{2.34} is that
\beq w(\Gam )=\rw (\Gam )\prod_s \Xi\left({_3\rI_s(\Gam )}\,\big\|\,\iota_s(\Gam )\right)
\label{2.36}\eeq
where ${_3\rI_s(\Gam )}$ is defined in accordance with \eqref{2.04}.
\vskip .5 truecm

{\bf Definition 2.9.} We say that a  {\BCC} from $\cD (\Lam ||\,i)$ is {\it external}\; if 
it consists of mutually external  contours. In this case we  use  the notation $\uGam^\tE$.
An equivalent definition is that the following properties hold:
\beq  
\big(\rB(\Gam_1^\tE) \cup \rI(\Gam_1^\tE)\big) \cap 
\big(\rB(\Gam_2^\tE) \cup \rI(\Gam_2^\tE)\big) = \vnth , 
\label{2.37}\eeq
and
\beq \iota^{\tE}(\Gam_1^{\tE}) = \iota^{\tE}(\Gam_2^{\tE})=i.\;\,
\label{2.38}\eeq
The set of  external {\BCCs} from $\cD (\Lam ||\,i)$ is denoted by $\cD(\Lam\|\,i; \tE)$. 
$\blacktriangle$
\vskip .5 truecm

Given an external {\BCC} $\uGam^\tE\in\cD(\Lam\|\,i; \tE)$, we set:   
\beq  
w(\uGam^\tE):= \prod_{\Gam^\tE \in \uGam^\tE} \; w(\Gam^\tE).  
\label{2.39}\eeq
Together with statistical weight \eqref{2.39}, set $\cD(\Lam\|\,i; \tE)$ forms an {\it external {\BC} ensemble} in 
$\Lam$ with the external type $i$.

With this definitions at hand, we obtain the following representation for the partition function $\Xi  (\Lam\|\,i)$ in \eqref{2.34}:
\beq
\Xi  (\Lam\|\,i)
=\sum_{\uGam^\tE \in \cD(\Lam\|\,i; \tE)}\;w(\uGam^\tE) = 
\sum_{\uGam^\tE \in \cD(\Lam\|\,i; \tE)}\;\prod_{\Gam^\tE \in \uGam^\tE} w(\Gam^\tE).  
\label{2.40}\eeq
Substituting of \eqref{2.36} into the RHS of Eqn \eqref{2.40} reveals the following recursive property of rarefied partition functions:
\beq\label{2.41}
\Xi  (\Lam\|\,i)
=\sum_{\uGam^\tE \in \cD(\Lam\|\,i; \tE)}\;\prod_{\Gam^\tE \in \uGam^\tE} \rw(\Gam^\tE)
\prod_s \Xi\left({_3\rI_s(\Gam^\tE)}\,\big\|\,\iota_s(\Gam^\tE)\right).
\eeq
As in \eqref{2.36}, the reduced box ${_3\rI_s(\Gam^\tE)}$ is obtained from the interior component $\rI_s(\Gam^\tE)$ in accordance with  \eqref{2.04}. The probability $\mu_\Lam (\uGam^\tE\|\,i)$ of having $\uGam^\tE$ as a collection of external {\BCs} has the form
\beq
\mu_\Lam (\uGam^\tE\|\,i)=\frac{w(\uGam^\tE)}{\Xi  (\Lam\|\,i)}. 
\label{2.42}\eeq
Denote by 
\beq
\diy\cD(\uGam^\tE) = \big\{ \uGam^{',\tE} \in \cD(\Lam\|\,i; \tE):\; \Gam^{\tE} \in \uGam^{',\tE}, \quad \forall \Gam^{\tE} \in \uGam^{\tE} \big\}
\label{2.43}\eeq
the subset of $\cD(\Lam\|\,i; \tE)$ consisting of external {\BCCs} having $\uGam^\tE$ as a subcollection. Then it follows from \eqref{2.41} and \eqref{2.42} that
\beq
\mu_\Lam (\cD(\uGam^\tE)\|\,i) \le \prod_{\Gam^\tE \in \uGam^\tE} \rw(\Gam^\tE)
\frac{\prod_s \Xi\left({_3\rI_s(\Gam^\tE)}\,\big\|\,\iota_s(\Gam^\tE)\right)}
{\prod_s \Xi\left({_3\rI_s(\Gam^\tE)}\,\big\|\,\iota^{\tE}(\Gam^\tE)\right)}. 
\label{2.44}\eeq
This inequality is known as the Peierls bound. We will reformulate it in Section 4 (cf. \eqref{4.23}) and use repeatedly  in Section~5; see, e.g., Eqn \eqref{5.24}.

Concluding this section, we make the following remark.  The machinery developed in Sections~3~and~4 will allow us to construct, for each $i\in\St$, limiting DLR measures concentrated on the space of compatible {\BCCs} $\cDi$ (see Eqn \eqref{2.15}), or equivalently, on the space of cell configurations $\cCi$ (see Eqn \eqref{2.14}). We can denote this measures by $\mu (\rd\uGam \|\,i)$ and $\mu (\rd\uphi \|\,i)$, respectively. In turn, measures $\mu (\rd\uGam \|\,i)$ and $\mu (\rd\uphi \|\,i)$ will generate according to \eqref{2.28} and \eqref{2.29} the measure $\mu (\rd\ubX \|\,i)$ sitting on the set 
\beq\label{2.45}
\cAi=\big\{\ubX \in \cA:\; \ubgam\circ\ubphi(\ubX) \in  \cDi\big\}.
\eeq
(Recall, map $\ubphi$ has been introduced in \eqref{2.07}.)
Measure $\mu (\rd\ubX \|\,i)$ is given by the limit \eqref{1.20} and satisfies the DLR equation \eqref{1.14}. The key fact here is that, with $\mu (\rd\uphi \|\,i)$-probability 1, for any bounded box $\Lam$ there exists a random bounded box $\Lam^*(\uphi,\Lam )$ $\supset\Lam$ such that in cell configuration $\uphi$ all unit cells of annulus ${}^2\Lam^*(\uphi,\Lam ) \setminus \Lam^*(\uphi,\Lam )$ are in phase $i$. Therefore, given function $f:\ubX \to\bbR$ depending on $\ubX^\Lam$, 
\beq\label{2.46}
\int\limits_{\cAi} f(\ubX)\mu (\rd\ubX \|\,i)=\int\limits_{\cCi}\mu (\rd\uphi \|\,i)\int\limits_{\cA(\Lam^*(\uphi ,\Lam)\|i)} f(\ubX )\;\mu_{{\Lam^*(\uphi ,\Lam)}} \big(\rd\ubX \|\,i; \uGam(\uphi)\big),
\eeq
which implies DLR property for $\mu (\rd\ubX \|\,i)$. 

\bigskip
\section{The ensemble of \SBCs}

In Section~3 we use concepts and terminology defined in Sections~6. More precisely, in
Section 6 we collect results about 
abstract polymer models. (Section 6 is a shortened version of Section~3  from \cite{MS2}.) The 
point is that if a specific ensemble of contours satisfies the generic conditions of Theorem~6.1 (mainly 
\eqref{A.03}) then it provides a `full control' over the corresponding finite-volume Gibbs distributions and 
their infinite-volume limits. We note that in this paper the polymer expansion analysis follows  \cite{KP}, 
with (technical) modifications proposed in \cite{MS2}. There exist several improvements of this approach; 
see \cite{FP} and the references therein; these improvements may help in a more accurate assessment of 
the threshold value $z_0$ in Theorems~1.1~and~1.2. 

\bigskip
{\bf 3.1. The definition of a small \BC\!\!.} In this subsection we define an ensemble of `small' contours. Then in section
3.2  we compare free energies of these ensembles for different external particle types. It appears that the free energy is maximal for the stable external types. An unstable particle type has a deficiency in the free energy, which gives a quantitative characterization of the stable type dominance. See Lemma~3.4 for more details.

\bigskip
{\bf Definition 3.1} A \basic contour $\Gam$ is called {\it small}\; if\; ${\rm{diam}}\,\rI (\Gam )$, the diameter of its interior $\rI(\Gam)$, is less than $\oa^{\,2d}$ (see \eqref{1.17} for the definition of $\oa$). We employ the notation $\Gam^\tS$ for small contours; the index or argument $\tS$ is used for the same purpose in a number of places below. In particular, $\uGam^\tS$ stands for a collection  containing only (mutually external) small contours. Contours that are not small are referred to as {\it large}.  $\qquad\blacktriangle$

\bigskip
The threshold $\oa^{\,2d}$ is selected (rather arbitrarily) to be  larger than $\ups(\bbB (\oa))$;
cf. Eqn \eqref{3.11}. This is important for the analysis in Sections~4~and~5. Viz., see the proof of Lemma 4.1
(specifically, Eqn \eqref{4.14}. Any independent on $z$ number will work for the analysis in the current section. 

\bigskip
Let $\cD(\Lam\|\,i; \tS)\subset\cD(\Lam\|\,i;\tE)$ be the set of {\BCCs} \;$\uGam^{\tS}$ in $\Lam$ with the external type $i$. (Recall, the set $\cD( \Lam \|\,i)$ was defined in Eqn \eqref{2.20} and its subset $\cD(\Lam\|\,i;\tE)$ was introduced in Definition~2.9.) Then, for any {\SBC} $\Gam^\tS\in\uGam^{\tS}\in\cD(\Lam\|\,i; \tS)$, the external type 
$\iota^\tE (\Gam^\tS)=i$. Accordingly, with $\ubphi$ as in \eqref{2.07}, set:
\beq\label{3.01}
\cA (\Lam\|\,i; \tS)=\big\{\ubX \in \cA(\Lam\|\,i):\; \ubgam\circ\ubphi(\ubX) \in  \cD(\Lam\|\,i; \tS)\big\}.
\eeq

The compatibility condition for {\SBCs} is given by \eqref{2.37}--\eqref{2.38} and the corresponding partition function is defined by
\beq\Xi (\Lam\|\,i;\tS):=\sum_{\uGam^{\tS} \in \cD(\Lam\|\,i; \tS)}\;\prod_{\Gam^\tS\in\uGam^{\tS}} w(\Gam^\tS).
\label{3.02}\eeq
Here $w(\Gam )$ is the external statistical weight determined by \eqref{2.35} or equivalently \eqref{2.36}. We call $\Xi (\Lam\|\,i;\tS)$ the partition function in the {\it {\SBC} ensemble} in $\Lam$.

\bigskip
{\bf Lemma 3.1.} {\sl Suppose $z$ is large enough. Then, for any finite box $\Lam$ and $\rv$ as in \eqref{2.23}, 
\beq \diy\Xi (\Lam\|\,i;\tS) < \Xi (\Lam\|\,i) 
\le\left(\frac{1 + \rv^{\,\uA}}{1-\rv}\right)^{\ups(\Lam)}.\label{3.03}\eeq
Here and below, 
\beq \uA=\ups (\bbB (\ua -2)), \label{3.04}\eeq
where $\ua$ is defined in \eqref{2.02}.}

\bigskip
{\bf Proof of Lemma 3.1.} To estimate the partition function $\Xi (\Lam\|\,i)$ from above we use the representation \eqref{2.34} and the upper estimate in \eqref{2.31}: 
\beq  0 < \rw (  \Gam) < \left(\frac{\rv}{1-\rv} \right)^{\ups(\rO(\Gam))}.  \label{3.05}\eeq
Relaxing a compatibility condition can only increase the partition function. To that end, we introduce a partition function ${\wt\Xi} (\Lam\|\,i)$ with relaxed compatibility condition for contributing contours. Namely, in ${\wt\Xi}(\Lam\|\,i)$ we allow the separating {\BCs} to overlap with each other and with non-separating contours. Also we allow inside $\Lam$ separating {\BCs} $\Gam$ with $\iota^\tE (\Gam)) \not = i$, i.e. we remove both restrictions \eqref{2.37} and \eqref{2.38} for separating contours, making those {\BCs} completely independent from each other and from 
non-separating contours. The relaxed partition function ${\wt\Xi}(\Lam\|\,i)$ can be calculated exactly:
\beq  {\wt\Xi} (\Lam\|\,i) =(1-\rv)^{-\ups(\Lam)} 
\prod\limits_{\Gam:\;\rB(\Gam)\subseteq \Lam} \left(1+ \rw ( \Gam)\right),  \label{3.06}\eeq
where the product is taken over all separating {\BCs} $\Gam$ inside $\Lam$.

The trademark of the contour techniques is a summation over weighted connected lattice subsets. Typically, 
the weights are exponentially small compared to the number of lattice sites in the subset. On the other hand, the
number of different connected lattice subsets of cardinality $N$ containing the origin is at most $c^N$, where 
$c$ depends 
on dimension $d$ and the radius of connectivity. Indeed, each connected subset has a spanning tree rooted at 
the origin. The spanning tree can be traversed by a path starting at the root and passing through every link of the
tree only twice. Therefore, the length of the path is at most $2N$. Obviously, the number of $r$-connected lattice
paths of length $2N$ is smaller than $c(d,r)^{2N}$, where $c(d,r)$ is a number of $r$-neighbors for a given lattice
 site.

We apply these considerations to the sum $\sum\limits_{\Gam :\;\rO (\Gam )\supseteq\Ups (0)}\rw (\Gam )$, where
all $\Gam$ are separating. Each  {\BC}  $\Gam$  contributing to this sum has $\ups(\rO(\Gam)) >\uA +1$. In 
addition, $\rO(\Gam )$ is 5-connected. Finally, $\rB(\Gam) \subseteq {}^2\rO(\Gam)$ and therefore there is at most
$q^{5^d \ups(\rO(\Gam))}$ possibilities to reconstruct $\rB(\Gam)$ from $\rO(\Gam)$ and specify a configuration 
$\varphi_{\rB(\Gam)}$. Thus,
\beq \sum_{\Gam :\;\rO (\Gam )\supseteq\Ups (0)}\rw (\Gam ) \le \sum_{n=\uA +1}^{\infty}\left(q^{5^d}c\,
\diy\frac{\rv}{1-\rv}\right)^n \le \rv^{\uA +1/2}, \label{3.07}\eeq
where the last inequality is true for $z$ large enough. Moreover, 
\beq\begin{array}{l}
\diy\prod\limits_{\Gam :\;\rB(\Gam ) \subseteq \Lam}\left(1 + \rw (\Gam )\right)
<\left(1+2\sum_{\Gam :\;\rO (\Gam )\supseteq\Ups (0)}\rw (\Gam )\right)^{\ups (\Lam )}
<\left(1 + \rv^{\,\uA}\right)^{\ups(\Lam)},\end{array} \label{3.08}\eeq
where both inequalities are true for $z$ large enough. (Variations of \eqref{3.07} and \eqref{3.08} could be seen in a 
number of forthcoming arguments.)
$\qquad\blacksquare$

\bigskip
Lemma~3.1 is instrumental in estimating statistical weights $w(\Gam^\tS)$ as it provides the control over 
$\prod_s \Xi\left({_3\rI_s(\Gam )}\,\big\|\,\iota_s(\Gam )\right)$ in \eqref{2.36}.

\bigskip
{\bf Lemma 3.2.} {\sl For any {\SBC} $\Gam^\tS$, the external statistical weight $w(\Gam^\tS)$ (see 
\eqref{2.35}) obeys
\beq  w(\Gam^\tS) > \left(\frac{\rv}{1-\rv} \right)^{\ups (\rB(\Gam^\tS))} 
(1-\rv)^{- \ups (\rI(\Gam^\tS))}  \label{3.09}\eeq
and
\beq  w(\Gam^\tS) < \left(\frac{\rv}{1-\rv}\right)^{\ups (\rO(\Gam^\tS))} 
\left(\frac{1+\rv^{\uA}}{1-\rv}\right)^{\ups (\rI(\Gam^\tS))} \label{3.10}\eeq
where $\rv=e^{-z}$ and $\uA$ is defined in Eqn {\rm{\eqref{3.04}}}.}

\bigskip
{\bf Proof of Lemma 3.2.} The lemma is a direct consequence of \eqref{2.31} and \eqref{3.03}.  
$\qquad\blacksquare$

\bigskip
The first factor in both \eqref{3.09} and \eqref{3.10} is the main part of the estimate as the second factor is 
close to $1$. Indeed, for $z$ large enough, 
\beq  1 < \left(\frac{1 + \rv^{\uA}}{1-\rv}\right)^{\ups (\rI(\Gam^\tS))}
< \left(\frac{1 + \rv^{\uA}}{1-\rv} \right)^{\oa^{\,2d^2}} < 1 + \rv^{0.9} \label{3.11}\eeq
as $\ups (\rI(\Gam^\tS)) < \oa^{\,2d^2}$.  (Among all positive numbers less than 1 the value ${0.9}$ 
has been selected only for definiteness.) 

\bigskip
{\bf 3.2. Free energies of {\SBC} ensembles.} In this subsection we apply polymer expansion technique (cf. Section~6) to calculate free energies of {\SBC} ensembles. This analysis culminates in Lemma~3.4 and its Corollary~3.5 which demonstrate that the the stable, in a sense of \eqref{1.19}, particle types have maximal free energies of {\SBC} ensembles.

Representation \eqref{3.02} and upper bound \eqref{3.10} imply that $\log\;\Xi (\Lam\|\,i;\tS)$ can be calculated using Theorem~6.1. Indeed, the symmetric compatibility relation $\Gam^\tS_1\sim\Gam^\tS_2$ for {\SBCs}
$\Gam^\tS_1$ and $\Gam^\tS_2$ is given by \eqref{2.37} and \eqref{2.38}. The function $a(\,\cdot\,)$ in Theorem 6.1 can be selected as
\beq  a(\Gam^\tS) := \ups (\rB(\Gam^\tS)) \log\,(1 +\rv^{0.9}). \label{3.12}\eeq
The bound \eqref{A.03} follows from the inequality
\beq \sum_{\Gam^\tS:\; \rB(\Gam^\tS)\cap\bbB\left(\oa^{2d}\right)\neq\varnothing}\;\; 
w(\Gam^\tS)(1+\rv^{0.8})^{ \ups(\rB(\Gam^\tS))} \le \log\,(1+\rv^{0.9}), \label{3.13} \eeq 
which can be verified for $z$ large enough (with respect to $\oa$) using the same enumerating arguments as in the proof of Lemma~3.1, together with estimates \eqref{3.10} and \eqref{3.11}. 

For a polymer $\Pi^\tS=[[\,\Gam^\tS\,]]$ formed by {\SBCs} $\Gam^\tS$ (see Section 6) we define the {\it base} $\rB(\Pi^\tS) =\operatornamewithlimits{\cup}\limits_{\Gam^\tS\in\Pi^\tS} \rB(\Gam^\tS)$ and the {\it statistical weight} $w(\Pi^\tS)$ as in \eqref{A.05}. Then, due to shift-invariance of the statistical weights $w(\Gam )$, the representation \eqref{A.04} can be rewritten as
\beq\log\,\Xi (\Lam\|\,i;\tS) = \ups (\Lam ) f(i;\tS) + r(\Lam\|\,i;\tS).  \label{3.14}\eeq
Here the principal term $f(i;\tS)$ represents the free energy of the (infinite-volume) {\SBC} ensemble with the 
external type $i$:
\beq  f(i;\tS) = \sum\limits_{\Pi^\tS:\; \rB(\Pi^\tS) \supseteq\Ups (0)} \;\frac{w(\Pi^\tS)}{\ups (\rB(\Pi^\tS))}  
\label{3.15}\eeq
and $r(\Lam\|\,i;\tS)$ is the remainder:
\beq r(\Lam\|\,i;\tS) = -\sum\limits_{\substack{\Pi^\tS:\;\rB(\Pi^\tS) \cap \Lam \not = \vnth ,\\ 
\rB(\Pi^\tS) \cap \Lam^\cmp \not = \vnth}}\,w(\Pi^\tS)\,
\frac{\ups (\rB(\Pi^\tS) \cap \Lam )}{\ups (\rB(\Pi^\tS))}\,.  \label{3.16}\eeq
To assess $r(\Lam\|\,i;\tS)$, we apply the bound from Eqn \eqref{A.06} and plug into it the definition \eqref{3.12} and bound \eqref{3.10}. Then one can see that for $z$ large enough
\beq  |r(\Lam\|\,i;\tS)| < \ups (\partial \Lam )\, \rv^{0.9}  \label{3.17}\eeq
where $\partial \Lam =\Lam\setminus {_1\Lam}$.

With representation \eqref{3.15} at hand, we are ready to compare the free energies $f(i; \tS)$ 
for different types $i$. Recall, that collection $\St$ is defined in \eqref{1.19} using vectors $\unn (i)$ specified in \eqref{1.18}. For $q=4$, a given type $i$ is also characterized by an ordered `incident'  collection of $3$ hard-core exclusion diameters. We assume that it is an increasing order. 

\bigskip
{\bf Lemma 3.3.} {\it Suppose that  for two types, $1$ and $2$, we have coinciding
collections of ordered incident hard-core exclusion diameters:  
\beq  
D(1, j_1) = D(2, k_1) \le D(1, j_2) = D(2, k_2) \le D(1, j_3) = D(2, k_3),
\label{3.18}\eeq  
where
\beq
\{j_1,j_2,j_3\}=\{2,3,4\}\;{\rm and}\;\{k_1,k_2,k_3\}=\{1,3,4\}. 
\label{3.19} \eeq 
Then, for any box $\Lam$, the permutation $\{1,2,3,4\}\to\{1,2,3,4\}$ with
\beq
1\mapsto 2,\;j_1\mapsto k_1,\;j_2\mapsto k_2,\;j_3\mapsto k_3 
\label{3.20} \eeq 
defines a {\rm{1-1}} map between events $\cA (\Lam\|\,1; \tS)$ and $\cA (\Lam\|\,2; \tS)$ as well as a {\rm{1-1}} map between events $\cA (\Lam\|\,1)$ and $\cA (\Lam\|\,2)$ (modulo subsets of ${\bbP}$-probability zero). These maps preserve the probability distribution ${\bbP}$, which implies that
\beq \Xi (\Lam\|\,1;\tS) = \Xi (\Lam\|\,2;\tS)\label{3.21} \eeq
and
\beq \Xi (\Lam\|\,1) = \Xi (\Lam\|\,2). \label{3.22} \eeq}

{\bf Proof of Lemma 3.3.} It is enough to consider only two possibilities for the map \eqref{3.20}:
\beq
1\leftrightarrow 2,\;3\leftrightarrow 4 \quad {\rm or} \quad 1\leftrightarrow 2,\;3\leftrightarrow 3,\;4\leftrightarrow 4. 
\label{3.23} \eeq 
After that observation the claim is verified directly, by inspecting all possible relations between values $D(i,j)$ listed in \eqref{3.18}.  $\qquad\blacksquare$

\bigskip
{\bf Remark 3.1.} A direct analog of Lemma 3.3 for $q=5$ fails: the fact that types $1$ and $2$ 
have $\unn (1)=\unn (2)$ does not imply that there exists a permutation with the above properties.

\bigskip
For $1\leq i<j\leq 4$ introduce the quantities
\beq  A(i,j)= \ups(\bbB (D(i,j)-2)),\; B(i,j)=\ups(\bbB (D(i,j)+2)).\label{3.24}\eeq

\bigskip
{\bf Lemma 3.4.} {\sl Suppose that  for two types, $1$ and $2$, we have different ordered
collections of incident hard-core exclusion diameters: 

\noindent
\centerline{$D(1, j_1) \le D(1, j_2) \le D(1, j_3)$ and $D(2, k_1) \le D(2, k_2) \le D(2, k_3)$.  }

\medskip\noindent
If $D(1, j_1) < D(2, k_1)$ then
\beq  f(1; \tS) - f(2; \tS) > 0.5 \rv^{B(1, j_1) }.  \label{3.25}\eeq
If $\;D(1, j_1) = D(2, k_1)$ but $D(1, j_2) < D(2, k_2)$ then
\beq  f(1; \tS) - f(2; \tS) > 0.5 \rv^{B(1, j_2)}.  \label{3.26}\eeq
Finally, if $D(1, j_1) = D(2, k_1)$, $D(1, j_2) = D(2, k_2)$ but $D(1, j_3) < D(2, k_3)$ then
\beq  f(1; \tS) - f(2; \tS) > 0.5 \rv^{B(1, j_3)}.  \label{3.27}\eeq
}

{\bf Proof of Lemma 3.4.} The statement of the lemma is a straightforward (although tedious) 
consequence of representation \eqref{3.15} and bounds \eqref{3.09}--\eqref{3.10}.  

In fact, suppose that  $D(1, j_1) < D(2, k_1)$. Then the summands in the RHS of \eqref{3.15} for  $f(1; \tS)$ and
$f(2; \tS)$ generated exclusively by non-separating {\SBCs} (i.e. the polymers  consisting only of non-separating
small contours) are identical. Hence, these summands denoted by $f_{\rm ns}(1; \tS)$ and $f_{\rm ns}(2; \tS)$
respectively, cancel each other  when one takes the difference $f(1; \tS) - f(2; \tS)$. 

Further,  according to \eqref{A.06}, for $z$ large enough, the contribution $f_{\rm s}(2; \tS)$ to 
$f(2; \tS)$ provided by polymers containing at least one separating {\SBC} does not exceed
\beq\begin{array}{l}
\diy\sum_{\substack{\Gam^{\tS}:\; \rB(\Gam^{\tS}) \supseteq \Ups(0),\\ 
\Gam^\tS\;\hbox{\small{is separating}}}} w(\Gam^\tS)\; e^{a(\Gam^\tS)} \le \\
\diy\sum_{\substack{\Gam^{\tS}:\; \rB(\Gam^{\tS}) \supseteq \Ups(0),\\ 
\Gam^\tS\;\hbox{\small{is separating}}}} 
\left(\frac{\rv }{1-\rv}\right)^{\ups (\rO(\Gam^\tS))} 
\left(\frac{1+\rv^{\,\uA}}{1-\rv } \right)^{\ups (\rI(\Gam^\tS))}\left(1+ \rv^{0.9}\right)^{\ups (\rB(\Gam^\tS))},
\end{array} \label{3.28}\eeq
where the inequality is due to \eqref{3.12} and \eqref{3.10}, and $\uA$ is as in \eqref{3.04}. According to 
\eqref{3.11}, for $z$ large enough 
\beq
\left(\frac{1 + \rv^{\,\uA}}{1-\rv\;}\right)^{\ups(\rI(\Gam^\tS))} < 
(1 + \rv^{0.9}) < (1 + \rv^{0.8})^{0.5}. 
\label{3.29}\eeq
Next, $A(2, k_1)$ yields a lower bound for the amount of empty cells in a separating 
{\SBC} $\Gam^\tS$ contributing to $f(i_2; \tS)$ and, similarly to \eqref{3.07}, for $z$ large enough
\beq\begin{array}{l}
\diy\sum_{\substack{\Gam^{\tS}:\; \rB(\Gam^{\tS}) \supseteq \Ups(0),\\ 
\Gam^\tS\;\hbox{\small{is separating}}}} 
\left(\frac{\rv}{1-\rv}\right)^{\ups(\rO (\Gam^\tS))}\left(1+\rv^{0.9}\right)^{\ups(\rB (\Gam^\tS))} \\
\diy < \sum_{n = A(2,k_1) + 1}^{\infty}\left(q^{5^d} c\, ( 1+\rv^{0.9})^{5^d}\frac{\rv}{1-\rv}\right)^n 
\diy < \rv^{A(2, k_1)} (1 + \rv^{0.8})^{0.5}.\end{array} \label{3.30}\eeq
Combining \eqref{3.29} and \eqref{3.30}, we conclude
\beq
\diy f_{\rm s}(2; \tS) < \sum_{\substack{\Gam^{\tS}:\; \rB(\Gam^{\tS}) \supseteq \Ups(0),\\ 
\Gam^\tS\;\hbox{\small{is separating}}}} w(\Gam^\tS)\; e^{a(\Gam^\tS)}
< \rv^{A(2, k_1)} (1 + \rv^{0.8}).\label{3.31}\eeq

On the other hand, $f(1; \tS)$ incorporates the contribution from {\SBCs} which are contained in a ball of 
radius $D(1, j_1) + 2 $ with a single cell in the middle occupied by type $j_1$. We call them {\it dominating} small 
contours. To estimate $f(1; \tS)$ from below we can consider an ensemble containing only dominating {\SBCs} 
and all non-separating small contours. Denote the corresponding free energy by ${\wt f}(1; \tS)$. Clearly, 
${\wt f}(1; \tS) < f(1; \tS)$. The contribution $f_{\rm ns}(1; \tS)$ provided by polymers consisting only from 
non-separating {\SBCs} is identical for both $f(1; \tS)$ and ${\wt f}(1; \tS)$. The contribution 
${\wt f}_{\rm sd}(1; \tS)$ to ${\wt f}(1; \tS)$ from the polymers consisting of a single dominating {\SBC} 
$\Gam^{\tS}$ satisfies the lower bound
\beq
{\wt f}_{\rm sd}(1; \tS) = w(\Gam^{\tS}) > \rv^{B(1,j_1)} 
\label{3.32}\eeq
because of Eqn \eqref{3.09}.

The remaining part ${\wt f}_{\rm nsd}(1; \tS)$ of ${\wt f}(1; \tS)$ collects the contribution of polymers containing 
at least two {\SBCs} with at least one of them being a dominating small contour. By virtue of Eqn \eqref{A.06}, 
the sum of absolute values of statistical weights of all polymers containing a dominating {\SBC} $\Gam^{\tS}$ 
and at least one other {\SBC} does not exceed
\beq\begin{array}{rll}
{\wt f}_{\rm nsd}(1; \tS) &< w(\Gam^{\tS})(e^{a(\Gam^{\tS})} - 1) &< 2w(\Gam^{\tS}) a(\Gam^{\tS})\\
\diy &< \; 2w(\Gam^{\tS})\, B (1, j_1) \,\rv^{0.9}\; &<\; 0.1\,w(\Gam^{\tS}),
\end{array}  \label{3.33}\eeq
where $a(\Gam^{\tS})$ is defined in \eqref{3.12} ans $z$ is so large that  $2B(1, j_1) \rv^{0.9} < 0.1$. 

Thus, 
\beq\begin{array}{rl}  f(1; \tS) &>{\wt f}(1; \tS) \\
&> f_{\rm ns}(1; \tS) +{\wt f}_{\rm sd}(1; \tS) - {\wt f}_{\rm nsd}(1; \tS)\\
&> f_{\rm ns}(1; \tS) + 0.9\,\rv^{B(1, j_1)}. 
\end{array}\label{3.34}\eeq
Comparing \eqref{3.34} to \eqref{3.31} yields \eqref{3.25}, provided that $z$ is large enough and the difference
$D(1, j_1) - D(2, k_1) > 4$ (which is guaranteed by \eqref{2.01}). 

The proof of \eqref{3.26} and \eqref{3.27} is similar: we first identify common parts in partition functions 
$f(1; \tS)$ and $f(2; \tS)$ and then compare remaining parts. For instance, in the case of \eqref{3.26} the
contribution to $f(1; \tS)$ from all polymers containing only non-separating {\SBCs} and {\SBCs} separating 
$1$ and $j_1$ is the same as the contribution to $f(2; \tS)$ from all polymers containing only  non-separating
{\SBCs} and {\SBCs} separating $2$ and $k_1$. On the other hand, in this case $f(1; \tS)$ incorporates the 
contribution from dominating {\SBCs} which are contained in a ball of radius $D(1, j_2) + 2 $ with a single cell 
in the middle occupied by type $j_2$. As in \eqref{3.34}, this contribution is bounded from below by 
$0.9\,\rv^{B(1, j_2)}$. As to the remaining  part in $f(2; \tS)$, it incorporates polymers containing at least one 
{\SBC} separating $2$ and $k_2$ or at least one {\SBC} separating $2$ and $k_3$. This contribution is 
upper-bounded by an expression similar to \eqref{3.31}, with $A(2,k_1)$ replaced by $A(2,k_2)$. 
$\qquad\blacksquare$

\bigskip
{\bf Corollary 3.5.} {\sl The types $i\in\{1,2,3,4\}$ generating the maximal value of 
$f(\,\cdot\,; \tS)$ are precisely those forming set $\St$ defined in \eqref{1.19}.}

\bigskip
{\bf Proof of Corollary 3.5.} Comparing assumption on collection $\Dm$ used in Lemma~3.4 with the definition of $\St$
implies the corollary. $\qquad\blacksquare$

\bigskip
\section{Unrestricted partition functions $ \Xi (\Lam\|\,i)$}

The purpose of this section is to analyze complete ensembles of {\BCs}. In particular, we derive a polymer 
expansion for $\log \Xi (\Lam\|\,i)$, $i\in\St$, similar to that in Eqns \eqref{3.14}--\eqref{3.17} for 
$\log \Xi (\Lam\|\,i;\tS)$. Recall, the rarefied partition function $\Xi (\Lam\|\,i)$ was defined in Eqns \eqref{2.34}, 
\eqref{2.40} and 
its restricted version $\Xi (\Lam\|\,i;\tS)$ in \eqref{3.02}. We have to distinguish between stable partition functions 
$\Xi (\Lam\|\,i)$, where 
$i \in \St$, and unstable partition functions $\Xi (\Lam\|\,i)$, with $i \not \in \St$. 
For unstable partition function we need upper bounds. In these bounds we employ new objects which we call
`boundary layers': these are groups of {\LBCs} (see Definition~3.1) organized in a certain manner; cf. Definition 
4.1. For stable partition functions we construct the aforementioned expansion in terms of {\BCs} with renormalized
statistical weights  \eqref{4.20}. Upper bounds for unstable partition functions are needed to control these 
renormalized statistical weights. The desired representation of \;$\Xi (\Lam\|\,i)$, $ i \in \St$, suitable for applying 
polymer expansion Theorem~6.1 is achieved in Eqn \eqref{4.22}.

\bigskip
{\bf 4.1. Upper bounds for $\Xi (\Lam\|\,i)$, $i\not \in \St$. Boundary layers.} We start with the definition of a 
boundary layer. Let $\cD( \Lam \|\,i)$ be as in Eqn \eqref{2.20}. Given a compatible contour collection 
$\uGam\in\cD( \Lam \|\,i)$, we use a procedure of  `erasing` the small \basic contours $\Gam^{\tS}\in\uGam$.
Namely, {\it erasing} an {\SBC} $\Gam^{\tS}$ means that $\rB(\Gam^{\tS}) \cup \rI(\Gam^{\tS})$ is filled with the type
$\iota^\tE (\Gam^\tS)$ in the corresponding cell configuration. Equivalently erasing means removing 
$\Gam^{\tS}$ from the collection $\uGam$. 

\bigskip
{\bf Definition 4.1.} Consider a compatible {\BCC} $\uGam\in\cD( \Lam \|\,i), i \not \in \St$ and  erase all {\SBCs} $\Gam^{\tS}\in\uGam$. The remaining {\LBCs} from $\uGam$ form a collection $\uGam^{\rL}$ which determines a cell configuration, $\uphi^*_{\uGam^{\rL}} \in \cCi$, cf.  \eqref{2.19}. By construction, $\uphi^*_{\uGam^{\rL}}\big|_{\Lam^\cmp} \equiv i$. We take the union ${\rm U}={\rm U}\left(\uphi^*_{\uGam^{\rL}}\right)$ of all cells $\Ups\subset\Lam$ that  that are not in a phase $l\in\St$ in $\uphi^*_{\uGam^{\rL}}$. A connected component of set ${\rm U}$ which is adjacent to $\Lam^\cmp$ is called the {\it base} of the boundary layer in $\uGam$. (It is not hard to see that among connected components of ${\rm U}$ there is only one adjacent to  $\Lam^\cmp$). This connected component, considered together with the types of unit cells belonging to it, is called a {\it boundary layer} in $\uGam$. A boundary layer is denoted by $\Gam^\rU(\uGam )$ and its base by $\rB\left(\Gam^\rU(\uGam )\right)$;  index $\rU$ means unstable. We use a shorter notation $\Gam^\rU$ when it does not create a confusion. 

The boundary layer can also be considered for a particle configuration $\ubX\in\cA (\Lam\|\,i)$
(by referring to maps $\ubgam_\Lam$ and  $\ubphi_\Lam$ from \eqref{2.05} and \eqref{2.11}); this point of view 
will be particularly convenient in Section 5 (see Definition 5.2).
$\qquad\blacktriangle$

\bigskip 
An alternative understanding of a boundary layer $\Gam^\rU$ is that it is a compatible 
collection of {\LBCs} $\Gam_m^{\rL}\left(\Gam^\rU\right)$ such that: 
\begin{description} \setlength{\itemsep}{0pt} \setlength{\parskip}{0pt}
\item[(\phantom{ii}i)] $\Gam^\rU$ contains 0 or more {\LBCs} $\Gam^{\rL,\tE}_k\left(\Gam^\rU\right)$ such that at least one of internal types $\iota_s\left(\Gam^{\rL,\tE}_k\left(\Gam^\rU\right)\right)\in\St$, i.e. is stable. We denote the corresponding interiors 
$\rI_s\left(\Gam^{\rL,\tE}_k\left(\Gam^\rU\right)\right)$ of $\Gam_m^{\rL}\left(\Gam^\rU\right)$ by $\rI_{k,s}(\Gam^\rU)$ and the types $\iota_s\left(\Gam^{\rL,\tE}_k\left(\Gam^\rU\right)\right)$ by 
$\iota_{k,s}\left(\Gam^\rU\right)$.

\item[(\phantom{i}ii)] $\Gam^\rU$ contains 0 or more {\LBCs} $\Gam^{\rL,\tI}_n\left(\Gam^\rU\right)$ which are different 
from {\LBCs} $\Gam^{\rL,\tE}_k\left(\Gam^\rU\right)$ described in item {\bf (i) } above. 

\item[(iii)] Each connected component $\rV_j\left(\Gam^\rU\right)$ of the set 
$\rB\left(\Gam^\rU\right) \setminus\operatornamewithlimits{\cup}\limits_m \rB\left(\Gam_m^{\rL} 
\left(\Gam^\rU\right)\right)$ has the same type $\iota\left(\rV_j\left(\Gam^\rU\right)\right)$ for all constituting unit cells, and this type is unstable.
\end{description}

\bigskip\noindent
Here the index $\rL$ means large, $\tE$ external and $\tI$ internal. Item {\bf (iii)} is actually a consequence of items {\bf (i)}, {\bf (ii)} and the compatibility  of contours $\Gam^{\rL}_m\left(\Gam^\rU\right)$ with each other. The meaning of notations
\beq\begin{array}{c}
\diy \rB^{\rL}(\Gam^\rU) = \operatornamewithlimits{\cup}\limits_m \rB\left(\Gam_m^{\rL} \left(\Gam^\rU\right)\right), 
\quad \rO^{\rL}(\Gam^\rU) = \operatornamewithlimits{\cup}\limits_m \rO\left(\Gam_m^{\rL} \left(\Gam^\rU\right)\right), \\
\diy V(\Gam^\rU) = \operatornamewithlimits{\cup}\limits_j V_j(\Gam^\rU) = \rB\left(\Gam^\rU\right) \setminus \rB^{\rL}(\Gam^\rU)\end{array}\label{4.01}\eeq
is straightforward.

Figure~3 below illustrates (for $d=2$) properties of a boundary layer $\Gam^\rU$ obtained after erasing the small contours. 

\begin{center}
\includegraphics[scale=1.4]{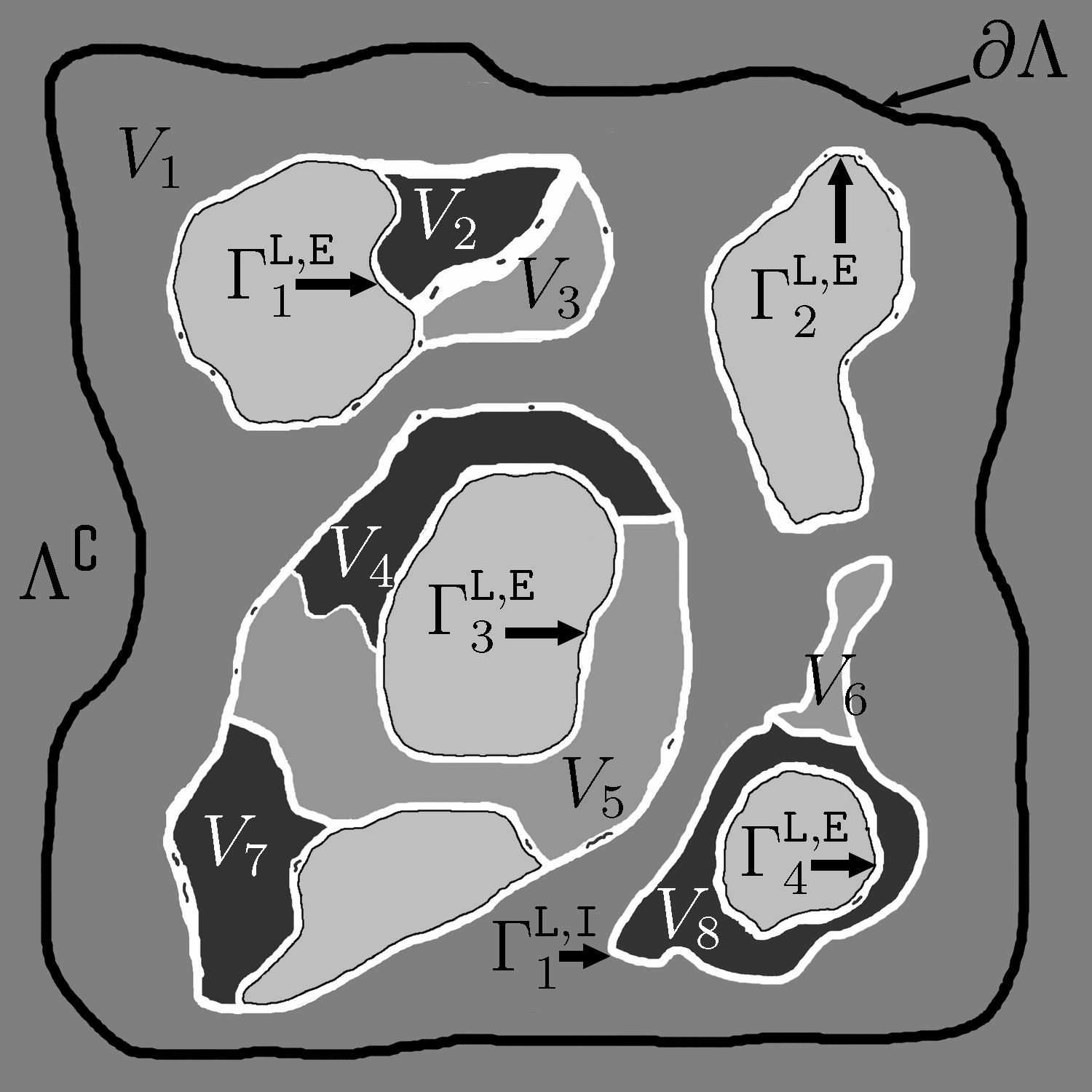}\;
\cl{\bf Figure 3. {A boundary layer}}
\end{center}

There are four {\LBCs} $\Gam^{\rL,\tE}_k=\Gam^{\rL,\tE}_k\left(\Gam^\rU \right)$: each of
them has at least one stable internal type. 
One {\LBC} does not have a stable internal type: this is $\Gam^{\rL,\tI}_1=\Gam^{\rL,\tI}_1\left(\Gam^\rU \right)$.
Together, the {\LBCs} $\Gam^{\rL,\tE}_k$, $1\leq k\leq 4$, and $\Gam^{\rL,\tI}_1$ form the {\BCC} 
$\;\Big\{\Gam^{\rL}_m\left(\Gam^\rU \right)$, $1\leq m\leq 5\Big\}$.
Here set $\St$ consists of a single type represented by a light gray color. 
The connected areas $V_j=V_j\left(\Gam^\rU \right)$ are filled with unstable types which are 
represented by three different shades of darker gray color. 
The unstable type of $V_1$ is the one specified in the boundary condition.
Anything drawn or written in black represents a notation and is not a part of the boundary layer. 
Specifically the thick black line is the boundary of $\Lam$ which is also the external boundary of $\rB(\Gam^\rU)$. 
The thin black lines point to the internal boundary of $\rB(\Gam^\rU)$.
The order of numeration for the {\LBCs} $\Gam^{\rL,\tE}_k$, $\Gam^{\rL,\tI}_n$ or $\Gam^{\rL}_m$ plays no role; 
the same is true about the numeration for areas $V_j$. 

\bigskip
Upon returning the erased {\SBCs} we define
\beq
\rw\left(\Gam^{\rU}\right):= \prod\limits_m \rw \left(\Gam_m^{\rL} \left(\Gam^{\rU}\right)\right)  
\prod\limits_j\diy \Xi \left(\rV_j\left(\Gam^{\rU}\right)\|\,\iota\left(\rV_j\left(\Gam^{\rU}\right)\right);\tS\right).
\label{4.02}\eeq
Here the product over $m$ collects the contribution of {\LBCs} constituting $\Gam^{\rU}$ 
while the product over $j$ collects the contribution of  {\SBCs} that we put back in their places. Then the definition of $\Gam^{\rU}$ implies that
for $i \not \in \St$
\beq
\Xi  (\Lam\|\,i)=\sum_{\Gam^{\rU} \in \cD(\Lam\|\,i)} \rw\left(\Gam^{\rU}\right)
\prod_{k,s} \Xi\left({_3\rI_{k,s}(\Gam^{\rU} )}\,\Big\|\,\iota_{k,s}\left(\Gam^\rU\right)\right),
\label{4.03}\eeq
cf. \eqref{2.41}. Define the {\it renormalized statistical weight} of boundary layer $\Gam^{\rU}$:
\beq
\rW\left(\Gam^{\rU}\right):= \prod\limits_m \rw \left(\Gam_m^{\rL}
\left(\Gam^{\rU}\right)\right)  \prod\limits_j\diy
\frac{\Xi \left(\rV_j\left(\Gam^{\rU}\right)\|\,\iota\left(\rV_j\left(\Gam^{\rU}\right)\right);\tS\right)}
{\Xi \left(\rV_j\left(\Gam^{\rU}\right)\|\,l;\tS\right)},
\label{4.04}\eeq
where type $l \in \St$. Note that, according to Lemma~3.3, partition functions\\ $\Xi \left(\rV_j\left(\Gam^{\rU}\right)\|\,l;\tS\right)$ 
and $\Xi \left(\rV_j\left(\Gam^{\rU}\right)\|\,l\right)$ do not depend on $l \in \St$ as long as $q\leq 4$.
On the other hand, for any $\Gam^{\rU} \in \cD(\Lam\|\,i)$,
\beq
\Xi  (\Lam\|\,l) \ge \prod_j \Xi \left(\rV_j\left(\Gam^{\rU}\right)\|\,l;\tS\right)
\prod_{k,s} \Xi\left({_3\rI_{k,s}(\Gam^{\rU} )}\,\Big\|\,l\right).\label{4.05}\eeq
Dividing \eqref{4.03} by \eqref{4.05} and using definition \eqref{4.04}, we estimate
\beq
\Xi(\Lam\|\,l)^{-1}\;\rw\left(\Gam^{\rU}\right)
\prod_{k,s} \Xi\left({_3\rI_{k,s}(\Gam^{\rU} )}\,\Big\|\,\iota_{k,s}\left(\Gam^\rU\right)\right) \le\rW\left(\Gam^{\rU}\right)
\label{4.06}\eeq
and
\beq
\frac{\Xi(\Lam\|\,i)}{\Xi(\Lam\|\,l)} \le \sum_{\Gam^{\rU} \in \cD(\Lam\|\,i)} \rW\left(\Gam^{\rU}\right),
\label{4.07}\eeq
where $i \not \in \St$ and $l \in \St$.

\bigskip
{\bf Lemma 4.1.} {\sl Let
\beq\begin{array}{l}\diy a\left(\Gam^{\,\rU}\right)  
= \ups\left(\rB^{\rL}\left(\Gam^{\,\rU} \right)\right) \log\,(1 +\rv^{0.8}) 
+ \ups\left(V\left(\Gam^{\,\rU} \right)\right) \log\,(1 + \rv^{\,\oa^{\,2d}})\,.
\end{array}\label{4.08}\eeq
Then, assuming  $z$ large enough, we have that for any box $\Lam$, } 
\beq
\sum_{\Gam^{\rU} \in \cD(\Lam\|\,i)} \rW\left(\Gam^{\rU}\right) e^{a\left(\Gam^{\,\rU}\right)}
< \Big( 1 + \rv^{\,0.5\oa^{\,2d}}  \Big)^{ \ups(\partial\Lam)}.\label{4.09}\eeq

\bigskip
{\bf Proof of Lemma 4.1.} We prove Lemma 4.1 by organizing the {\LBCs} $\Gam_m^{\rL}\left(\Gam^\rU\right)$ of a given boundary layer $\Gam^\rU$ in a tree-like structure $\cT=\cT\left(\Gam^\rU\right)$. 

\begin{center}
\includegraphics[scale=1.1]{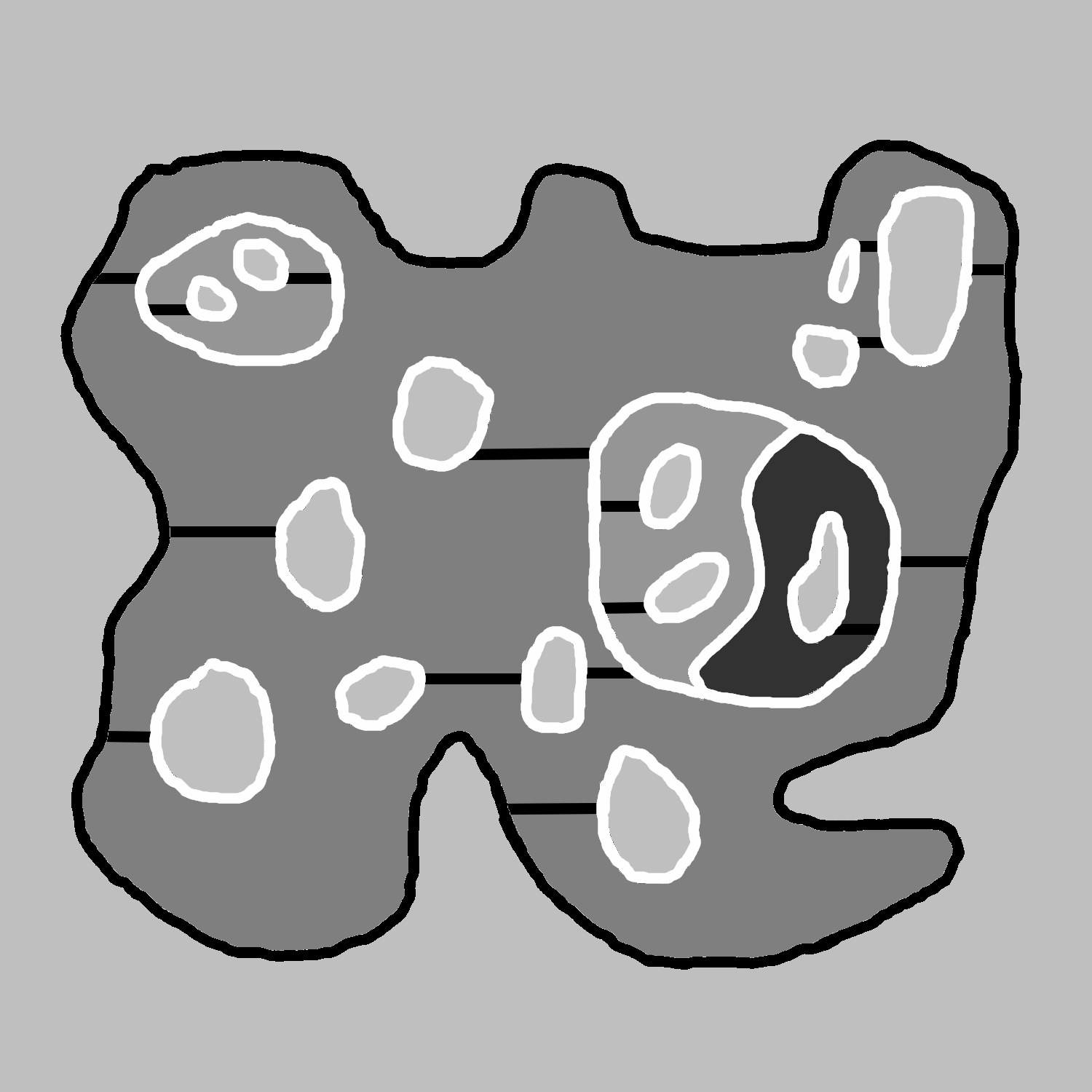}\;
\bigskip
\cl{\bf Figure 4. The tree $\cT\left(\Gam^\rU\right)$}
\end{center}

\noindent The root of the tree is represented by the boundary $\partial \Lam$. Next, the first-level vertices are identified with those {\LBCs} $\Gam_m^\rL=\Gam_m^{\rL}\left(\Gam^\rU\right)$ whose base $\rB\left(\Gam_r^\rL\right)$ can be connected to $\partial \Lam$ by a sequence  of unit cells from $V\left(\Gam^{\,\rU} \right)$ which is parallel to the first coordinate axis and not intersecting $\rB^{\rL}\left(\Gam^{\,\rU} \right)$, cf \eqref{4.01}. For every first-level {\LBC} $\Gam_m^\rL$ we choose a particular sequence of unit cells satisfying the above non-intersection condition and treat it as an edge $\bfe\left(\partial \Lam, \Gam_m^\rL\right)$ joining the root $\partial \Lam$ and the vertex $\Gam_m^\rL$. 

The second-level vertices are those among the  remaining {\LBCs} $\Gam_m^\rL$ whose base can be connected to the base of a first-level {\LBC} with a horizontal sequence of unit cells from $V\left(\Gam^{\,\rU} \right)$ still not intersecting $\rB^{\rL}\left(\Gam^{\,\rU} \right)$. Again,  for every such {\LBC} we choose a sequence of unit cells with these properties and treat it as an edge. This process can be continued further, until all {\LBCs} of $\Gam^\rU$ are included in $\cT$. By construction, the sequences of units cells representing edges of the tree are pair-wise disjoint. See Figure 4.

Next, we take the definition \eqref{4.04} of renormalized statistical weight $\rW\left(\Gam^{\rU}\right) $ and plug in it bound 
\eqref{2.31}, representation \eqref{3.14}, bound \eqref{3.17}, and bounds \eqref{3.25}--\eqref{3.27}. The result is a series of inequalities in Eqns \eqref{4.10}--\eqref{4.18} involving $\rW\left(\Gam^{\rU}\right)$, in one form or another. First,
\beq\begin{array}{rl}\diy \rW\left(\Gam^\rU\right) 
&< \diy\left(\frac{\rv}{1-\rv}\right)^{\ups(\rO^{\rL}(\Gam^{\,\rU} ))} 
e^{\rv^{0.9} \ups (\rB^{\rL} (\Gam^{\,\rU} )) 
-\rv^{\,(2\oa)^{\,d}}  \ups(V(\Gam^{\,\rU} ))} \\
&\diy\le \left(\rv(1+ 5^d 2 \rv^{0.9})\right)^{\ups(\rO^{\rL}(\Gam^{\,\rU} ))} 
e^{- \rv^{\,(2\oa )^{\,d}} \ups(V(\Gam^{\,\rU} ))},   \label{4.10}\end{array}\eeq
because $B(1,j_l)$ in \eqref{3.25}--\eqref{3.27} is smaller than $(2\oa)^{\,d}$.
Consequently, from \eqref{4.10}: 
\beq\begin{array}{l}
\diy\rW\left(\Gam^\rU\right)e^{a(\Gam^\rU)}  
\diy\le \left(\rv(1+ 5^d 2 \rv^{0.8})\right)^{\ups(\rO^{\rL}(\Gam^{\,\rU} ))}\; e^{-0.5 \rv^{\,(2\oa )^{\,d}} \ups(V(\Gam^{\,\rU}))}.\end{array} \label{4.11}\eeq
By construction, if $\cE\left(\Gam^\rU\right)$ denotes the collection of all edges of $\cT$ and
$\ups\left(\cE\left(\Gam^\rU\right)\right)$ stands for the volume of $\cE\left(\Gam^\rU\right)$ then
\beq\ups(\cE(\Gam^\rU)) \le \ups(V(\Gam^{\,\rU})).  \label{4.12}\eeq
Bounds \eqref {4.11} and \eqref{4.12} imply that  
\beq\begin{array}{r}
\diy\rW\left(\Gam^\rU\right)e^{a\left(\Gam^\rU\right)} \le 
\left(\rv(1+ 5^d 2 \rv^{0.8})\right)^{\ups(\rO^{\rL}(\Gam^{\,\rU}))}
e^{-0.5 \rv^{\,(2\oa )^{\,d}} \ups(\cE(\Gam^\rU))}
\end{array}\label{4.13}\eeq

We finish the proof of Lemma 4.1 by induction in the number of levels in $\cT$. 
The root node of a tree is different from the remaining nodes as it is given by $\partial\Lam$ and does 
not have an associated large contour. 
A subtree rooted at a first-level node has only nodes which all are large contours. From now on and until it is said otherwise we consider such `homogeneous` trees.
Later we perform the final estimate including the root $\partial\Lam$. 

A single-level homogeneous tree consists of a single {\LBC} $\Gam^{\rL}$. Any {\LBC} $\Gam^{\rL}$ has
\beq \ups(\rO(\Gam^{\rL})) > \oa^{\,2d} + 1 \, .\label{4.14}\eeq 
Therefore for $z$ large enough, the following bound can be verified similarly to \eqref{3.30}
\beq
\sum\limits_{\Gam^{\rL}:\; \rE(\Gam^{\rL})^\cmp \supseteq \Ups(0)} 
\left(\rv(1+ 5^d 2\rv^{0.8})\right)^{ \ups(\rO (\Gam^{\rL}))} \le \rv^{\oa^{\,2d} + 1}.  
\label{4.15}\eeq

Denote by $\Gam_*^{\rL}(\cT)$ the root {\LBC} of a homogeneous  tree and suppose that the estimate 
\beq
\sum\limits_{\cT:\; \rE(\Gam^{\rL}_*)^\cmp \supseteq \Ups(0)} 
\left(\rv(1+ 5^d 2\rv^{0.8})\right)^{ \ups(\rO (\Gam^{\rL}))} \le \rv^{\oa^{\,2d}}.  
\label{4.16}\eeq
has been verified for homogeneous trees with at most $N$ levels.  A tree with at most $N+1$ levels can be decomposed into:
\begin{description} \setlength{\itemsep}{0pt} \setlength{\parskip}{0pt}
\item[(\phantom{ii}i)] the root {\LBC} $\Gam_*^{\rL}\,$,
\item[(\phantom{i}ii)] at most $\ups(\rB(\Gam_*^{\rL}))$ edges issued from the root {\LBC} $\Gam_*^{\rL}\,$ and leading to the first-level {\LBCs} $\Gam_r^{\rL}$, and
\item[(iii)] a collection of sub-trees with maximum $N$ levels rooted at each of $\Gam_r^{\rL}$.
\end{description}
The volume  
$\ups(\bfe (\Gam_*^{\rL}, \Gam_r^{\rL}))$ occupied by the edge  $\bfe (\Gam_*^{\rL}, \Gam_r^{\rL} )$ joining 
LBCs $\Gam_* ^{\rL}$ and $\Gam_r^{\rL}$ can be any 
positive integer $t$.  Thus, the sum of the weights taken from the RHS of \eqref{4.13}, over all trees with at most $N+1$ levels, does not exceed
\beq\begin{array}{l}
\phantom{\le}\sum\limits_{\Gam^{\rL}_*:\; \rE(\Gam^{\rL}_*)^\cmp \supset \Ups(0)}
\left(\rv(1+ 5^d 2 \rv^{0.8})\right)^{ \ups(\rO (\Gam^{\rL}_*))} 
\Big( 1 + \rv^{\oa^{\,2d}} \sum\limits_{t= 1}^{\infty} e^{-0.5\,t\,\rv^{\,(2\oa)^{\,d}}} \Big)^{ \ups(\rB(\Gam^{\rL}_*))} \\
\le \sum\limits_{\Gam^{\rL}_*:\; \rE(\Gam^{\rL}_*)^\cmp \supset \Ups(0)}
\left(\rv(1+ 5^d 2 \rv^{0.8})\right)^{ \ups(\rO (\Gam^{\rL}_*))} \Big( 1 + \rv^{\,0.5\,\oa^{\,2d}}  \Big)^{ \ups(\rB(\Gam^{\rL}_*))}\;
\le\; \rv^{\oa^{\,2d}}, \end{array} \label{4.17}\eeq
which reproduces \eqref{4.16}. Thus, the estimate \eqref{4.16} is true for any homogeneous tree. 

Finally, we return to our inhomogeneous tree (where the root is represented by the boundary $\partial\Lam$). By using \eqref{4.13}, we estimate the contribution of all uniform trees growing from the root $\partial\Lam$:
\beq\begin{array}{rl}
\diy\sum_{\Gam^{\rU} \in \cD(\Lam\|\,i)} \rW\left(\Gam^{\rU}\right) e^{a\left(\Gam^{\,\rU}\right)} &\le \Big( 1 + \rv^{\oa^{\,2d}} \sum\limits_{r= 1}^{\infty} e^{-0.5\,r\,\rv^{\,(2\oa)^{\,d}}} 
\Big)^{ \ups(\partial\Lam)} \\
&\le \Big( 1 + \rv^{\,0.5\,\oa^{\,2d}}  \Big)^{ \ups(\partial\Lam)}. 
\end{array} \label{4.18}\eeq
This finishes the proof of Lemma~4.1.\quad\quad $\blacksquare$

\bigskip
{\bf Corollary 4.2.} {\sl For any $i \not \in \St$ and any $l \in \St$
\beq\frac{\Xi(\Lam\|\,i)}{\Xi(\Lam\|\,l)}
< \Big( 1 + \rv^{\,0.5\,\oa^{\,2d}}  \Big)^{ \ups(\partial\Lam)}\label{4.19}\eeq
for $z$ large enough.}

\bigskip
{\bf Proof of Corollary 4.2.} Substitute \eqref{4.09} into RHS of \eqref{4.07}. \quad\quad $\blacksquare$

\bigskip
{\bf 4.2. Polymer expansions for $\Xi(\Lam\|\,i)$, $i \in \St$,  and Theorem 1.1(I).} The purpose of this section is to prove assertion(I) of Theorem~1.1 by deriving for $\Xi(\Lam\|\,i), i \in \St$, the representation \eqref{4.22} of the type \eqref{A.01} and then applying Theorem~6.1. 

In analogy with \eqref{4.03}--\eqref{4.05}, define a {\it renormalized statistical weight} $\rW(\Gam)$ of {\BC} $\Gam$:
\beq
\rW(\Gam) = \rw(\Gam) \prod_s \frac{\Xi\left({_3\rI_s(\Gam)}\,\big\|\,\iota_s(\Gam)\right)}
{\Xi\left({_3\rI_s(\Gam)}\,\big\|\,\iota^{\tE}(\Gam)\right)}\label{4.20}\eeq
and rewrite  \eqref{2.40} as
\beq
\Xi  (\Lam\|\,i) = \sum_{\uGam^\tE \in \cD(\Lam\|\,i; \tE)}\;\prod_{\Gam^\tE \in \uGam^\tE} 
\rW(\Gam^\tE)  \prod_s \Xi\left({_3\rI_s(\Gam^\tE)}\,\big\|\,i\right).  
\label{4.21}\eeq
Here we used the fact that $\Xi\left({_3\rI_s(\Gam^\tE)}\,\big\|\,\iota^{\tE}(\Gam^\tE)\right) = \Xi\left({_3\rI_s(\Gam^\tE)}\,\big\|\,i\right)$ for all external {\BCs} $\Gam^\tE \in \uGam^\tE$. Iterating \eqref{4.21} we obtain the desired representation
\beq
\Xi  (\Lam\|\,i)=\sum_{\uGam}\;\prod_{\Gam \in \uGam} \rW(\Gam),  \label{4.22}
\eeq
where the sum is taken over the collections of {\BCs} $\uGam$ such that for any {\BC} $\Gam$ from this collection $\rB(\Gam) \subseteq {^2\Lam}$, $\iota^\tE(\Gam) = i$ and all $\rB(\Gam)$ are mutually disjoint. 

Representation \eqref{4.22} gives rise to a host of facts and constructions. Viz., the Peierls bound 
\eqref{2.44} takes a simple form
\beq\label{4.23}
\mu_{\Lam}\left(\cG (\uGam^{\tE})\| \,i \right)\leq \prod\limits_{\Gam^{\tE}\in\uGam^{\tE}}
W(\Gam^{\tE}).
\eeq
As was said, we will use \eqref{4.23} in Section~5, beginning with Eqn \eqref{5.24}.  

\bigskip
{\bf Proof of Theorem 1.1(I)} Assertion(I) of Theorem~1.1 follows in a standard way from the convergence of polymer expansion for the partition functions $\Xi  (\Lam\|\,i)$ in \eqref{4.22}. In turn, this convergence is implied by Theorem~6.1 as soon as  condition \eqref{A.03} is verified. 

Extending the definition in \eqref{3.12}, we set
\beq a(\Gam) = \ups \left(\rB(\Gam)\right) \log\,(1 +\rv^{0.9})\label{4.24}\eeq
for any {\BC} $\Gam$ (large or small). The inequality
\beq
\sum_{\Gam:\; \rB(\Gam) \supseteq \Ups(0)}\;\; \rW(\Gam)(1+\rv^{0.8})^{ \ups(\rB(\Gam))} \le \log\,(1+\rv^{0.9}),
\label{4.25} \eeq 
can be verified, after plugging in the definition \eqref{4.20} and estimate \eqref{4.19}, by using the same enumerating arguments as in the proof of \eqref{3.13}. The desired bound \eqref{A.03} follows. 

The polymer expansion Theorem~6.1 implies existence, for $i\in\St$, of a pure phase $\mu (\,\cdot\,\|\,i)$ (cf. \eqref{1.20}) which is shift-invariant, ergodic and has an exponential decay of correlations. For measure $\mu (\,\cdot\,\|\,i)$, $i\in\St$, the probability that the type of $\Ups(0)$ is $i$ tends to 1 as $z \to \infty$. Consequently, all pure phases named in Theorem~1.1(I) are different (and, due to the Lemma 3.3, symmetric). This completes the proof of assertion (I) in Theorem 1.1.\quad\quad $\blacksquare$

\bigskip

\section{Unstable boundary condition}

In Section 5 we prove Theorem 1.2. and assertion (II) of Theorem 1.1. Our approach is based 
on papers \cite{LM, Z}. In our opinion, constructions performed here yield simplified versions of 
those used in  \cite{LM, Z}. Although the results of this section are claimed for $q=4$, we keep 
using a general $q$ as it would clarify the nature of the bounds used. See, e.g., \eqref {5.06}, 
\eqref {5.07}, \eqref {5.09} and so on.
\bigskip

{\bf 5.1. Proof of Theorem 1.2 (I).} Recall, we assume that a single stable type is $1$. We will verify 
that $\mu(\,\cdot\,\|\,1) = \lim\limits_{L\to\infty}\mu_{\Lam_L}(\,\cdot\,\|\,1)$ is the only limit Gibbs/DLR 
state of the model. In the course of the proof we develop an argument that is later used in the proof 
of the remaining parts in Theorem~1.2 and of assertion (II) of Theorem~1.1. 

Given $L>0$, we work with a cubic box $\Lam_{L}$ of size $L$; cf. \eqref{1.05}. Next, given an 
admissible particle configuration $\ubY=(\bY_1,\ldots ,\bY_q) \in\cA$, we consider the distribution  
$\mu_{\Lam_{L}} \left(\;\cdot \;| \ubY^{\Lam_{L}^\cmp}\right)$ defined in Eqn \eqref{1.11}. It is instructive 
to consider a {\BL}  $\Gam^\rU$
for the boundary condition $\ubY^{\Lam_{L}^\cmp}$: this is a generalization of the similar concept 
introduced in Section~4 for the boundary condition containing only particles of a given unstable type. 
Namely,  $\Gam^\rU$ is the connected component (or connected components) of the unit cells not 
in a stable phase adjacent to $\Lam_L^\cmp$. As before, the meaning of the boundary layer is to 
describe a transition from  particles of unstable types present in $\ubY^{\Lam_{L}^\cmp}$ to  particles 
of the stable type occurring in box $\Lam_{L}$. Details are given below.  

Our aim is to show that, with probability tending to 1 as $L\to\infty$, the boundary layer does not 
penetrate into the smaller concentric cubic box $\Lam_{L/2}$. If this is the case for a particle 
configuration $\ubX \in \cA\left(\Lam_L|\,\ubY^{\Lam_L^\cmp}\right)$ then there exists a box 
$\Lam^\prime=\Lam^\prime(\ubX^{\Lam_L})$ such that $\Lam_{L/2} \subseteq \Lam^\prime 
\subseteq \Lam_{L}$ and $\ubX \in \cA (\Lam' \|\,1)$.  From the polymer expansions constructed 
for $\mu_{\Lam_{L}}(\; \cdot \; \|\,1)$ in Section~4 we know that the measure $\mu_{\Lam_{L}}
(\; \cdot \; \|\,1)$ forgets the boundary conditions exponentially fast. Thus, $\displaystyle{\lim_{L \to \infty}}
 \mu_{\Lam_{L}}\left(\;\cdot \;| \ubY^{\Lam_{L}^\cmp}\right) = \displaystyle{\lim_{L \to \infty}} 
\mu_{\Lam'}(\; \cdot \; \|\,1)$ which yields assertion (II) of Theorem 1.2.

We will now give a semi-formal review of the ideas, and we take some liberties during this stage of 
presentation. In particular, saying that a quantity behaves like $e^{-L}$ we mean that 
$e^{-L}$ is an upper bound for the quantity. Also we omit various positive constants, e.g. writing 
$e^{-L}$ instead of $e^{-cL}$, as constants are not essential for the arguments. We also 
use the notation $\Delta_L$ for the annulus $\Lam_{L+\oa } \setminus \Lam_L$ of thickness $\oa$
where $\oa$ is the largest hard-core diameter; cf. Eqn \eqref{1.17}. 

Graphically our ideas are presented by Figure 5. Here the base of the {\BL} is shown as the complement to the area occupied by the light grey color. The frames (i) and (ii) demonstrate two possibilities for the {\BL} to penetrate into $\Lam_{L/2}$; we will check that each of these possibilities occurs with probability tending to $0$ as 
$L\to\infty$. 

In frame (i) there is an annulus $\Delta_{L^\prime}$, with $3L/4 < L^\prime < L$, with a relatively low number of 
unstable unit cells inside it. When we estimate the probability of such a {\BL}, the unstable 
unit cells inside $\Delta_{L^\prime}$ serve as pinpoints  for connected components of empty cells (depicted, as
always, in a white color). At least one of these components is stretched over a long distance, at least $3L/4-L/2=L/4$,
which carries a small probability.  Furthermore, a relatively small amount of pinpoints guarantees that the 
entropy-type contribution to the probability coming from relatively short components is beaten by this small 
probability. This makes the overall  probability of such a {\BL} negligible as $L\to\infty$. Cf. Eqn  \eqref {5.15}. 

In frame (ii) no such annulus $\Delta_{L^\prime}$ can be found, and therefore a considerable part of $\Lam_L\setminus L_{3L/4}$ is covered by the {\BL}, although the amount of empty unit cells is not that large. This again makes probability of this type of {\BLs} small. Cf. Eqn \eqref{5.31}.

\begin{center}
\includegraphics[scale=1.05]{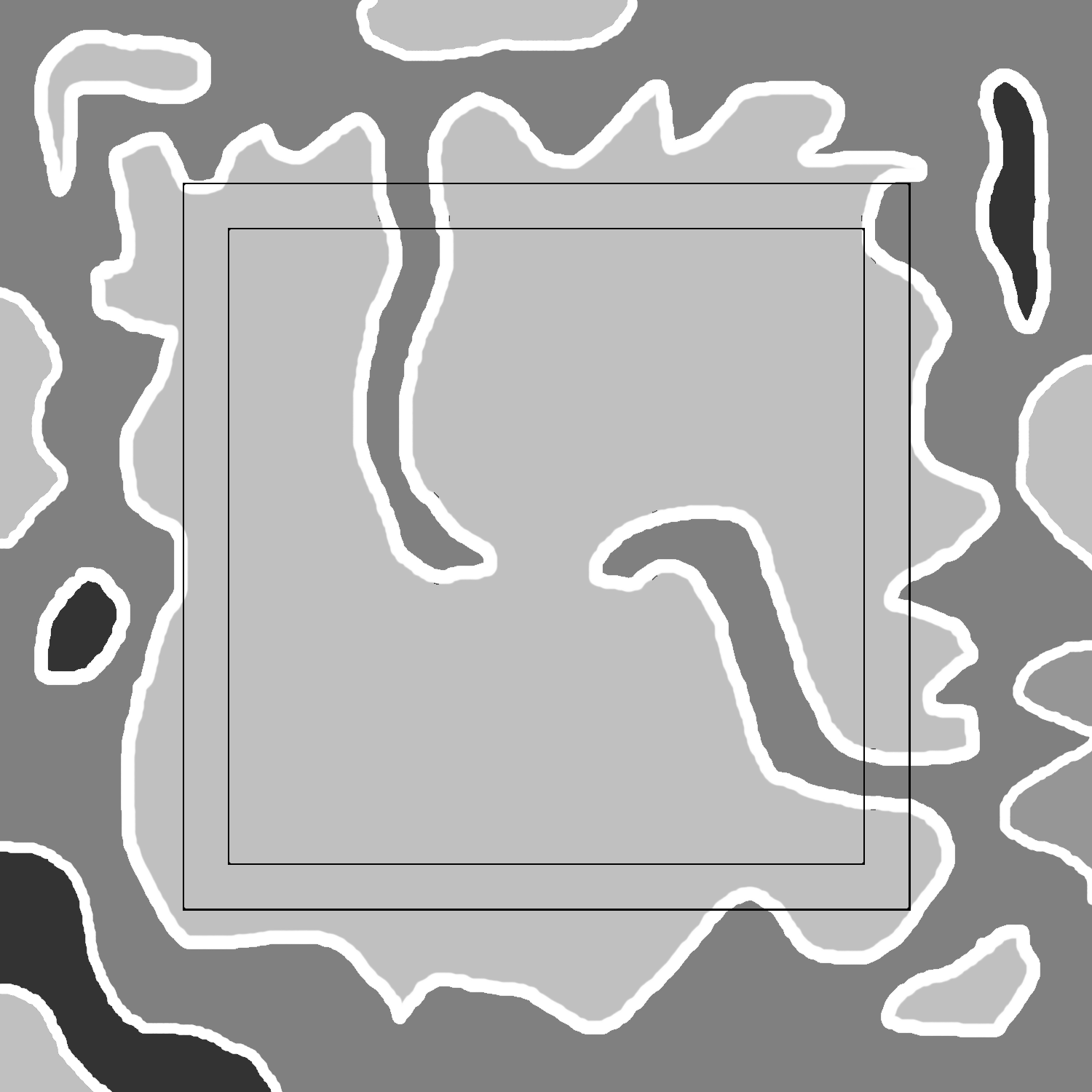}\;
\includegraphics[scale=1.05]{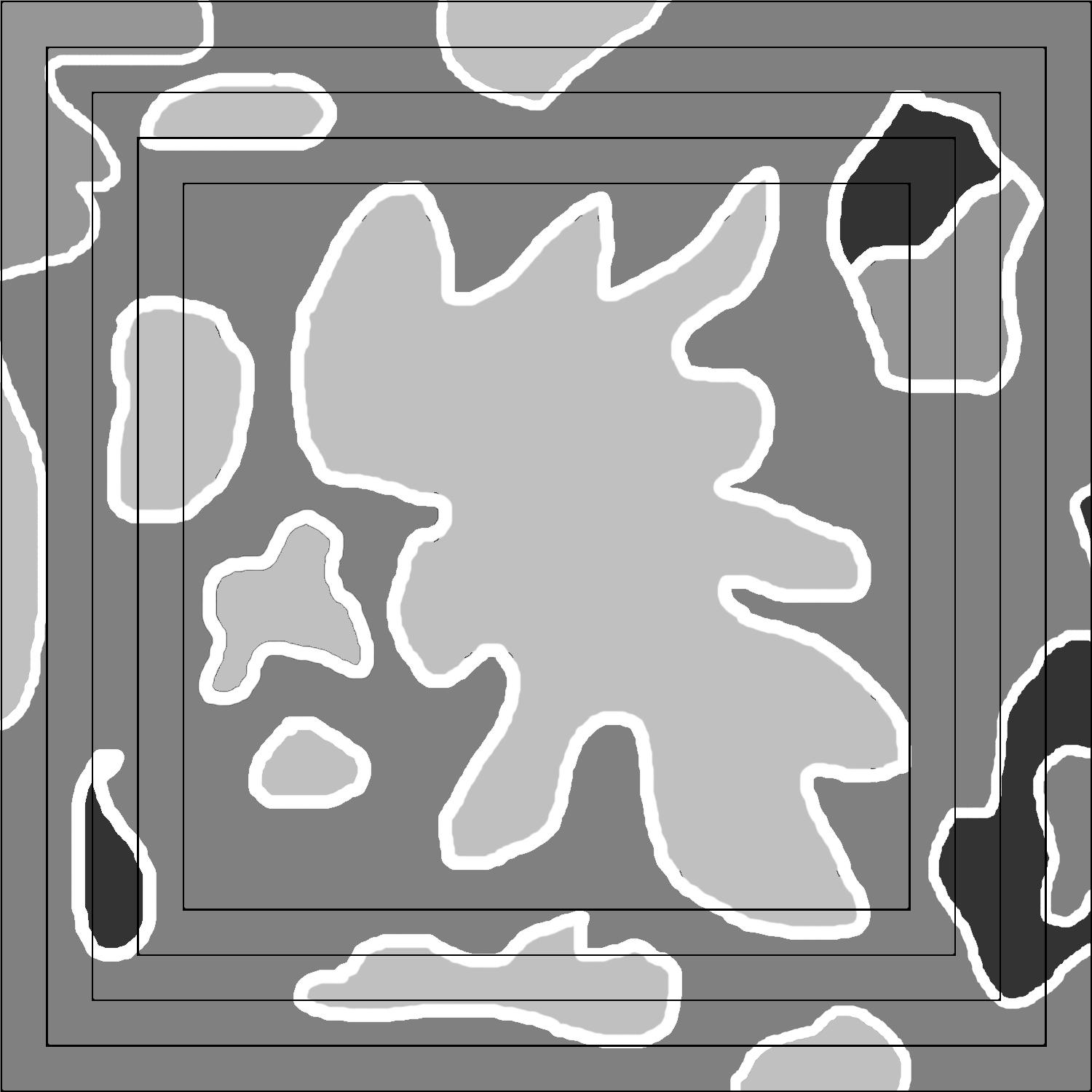}
\cl{{\bf (i)}\qquad\qquad\qquad\qquad\qquad\qquad\qquad {\bf (ii)}}
\cl{\bf Figure 5.}
\end{center}

\bigskip
More precisely, if the boundary condition $\ubY^{\Lam_{L^\prime}^\cmp}$, where $3L/4 < L^\prime < L$, contains inside $\Delta_{L^\prime}$ 
less than ${\sqrt L}$ unit cells   which are not in  
phase $1$ then the probability for the boundary layer to penetrate inside $\Lam_{L/2}$ decays like $e^{-L}$. This 
property can be verified based on the following observations {\bf (1)--(5)}:
\begin{description}   \setlength{\itemsep}{0pt} \setlength{\parskip}{0pt} 
\item[(1)\;] Only $\ubY^{\Delta_{L^\prime}}$ affects $\mu_{\Lam_{L^\prime}}\left(\;\cdot \;| \ubY^{\Lam_{L^\prime}^\cmp}\right)$. 
\item[(2)\;] Each unstable unit cell inside $\Delta_{L^\prime}$ induces a {\LBC} belonging to the boundary layer. (These induced {\LBCs} are not necessarily different for different unstable unit cells). 
\item[(3)\;] To penetrate into $\Lam_{L/2}$ the {\BL} should have at least $L/2$ unit cells in the base of at least 
one of the above {\LBCs}. 
\item[(4)\;] The $\mu_{\Lam_{L^\prime}}\left(\;\cdot \;| \ubY^{\Lam_{L^\prime}^\cmp}\right)$-probability of the collection of large \basic contours induced by unstable unit cells inside $\Delta_{L^\prime}$ is not larger than the $\mu_{\Lam_{L^\prime+\oa}}\left(\;\cdot \;\| 1\right)$-probability of the same collection multiplied by the factor of order $e^{\sqrt{L}}$. This statement is true because there are at most $\sqrt{L}$ unstable unit cells inside $\Delta_{L^\prime}$. 
\item[(5)\;] Due to the presence of a long {\LBC} in the collection the corresponding $\mu_{\Lam_{L^\prime+\oa}}\left(\;\cdot \;\| 1\right)$-probability of the collection decays as $2^{\sqrt{L}}e^{-L}$. Here the first factor estimates from above the contribution of all large \basic contours in the collection which do not reach $\Lam_{L/2}$. The second factor is the contribution of the large \basic contour which reaches $\Lam_{L/2}$. For large $L$ the factor of order $e^{-L}$ dominates the factors of order $2^{\sqrt{L}}$ which gives us the desired estimate. 
\end{description}
This argument covers the picture in frame (i) of Figure 5.

To handle boundary conditions which contain more than $\sqrt{L}$ unstable unit cells inside $\Delta_L$ we split $\Lam_L \setminus \Lam_{3L/4}$ into annuli of thickness $\oa$: $\Delta_{3L/4}$, $\Delta_{3L/4+\oa}, \ldots, \Delta_{L-\oa}$. First, we apply the preceding argument to $\Lam_{L^\prime}=\Lam_{3L/4}$ assuming that the boundary condition $\ubZ^{\Lam_{3L/4}^\cmp}$ contains at most $\sqrt{L}$ unstable unit cells. Then we proceed by induction, allowing the next annulus to have more unstable unit cells than the preceding one, for the price of doubling the estimate available for the preceding annulus. We select $e^{n/\sqrt{L}}$ as the maximal allowed amount of unstable unit cells in the $n$-th annulus. Note that this quantity remains smaller than $\sqrt{L}$ as soon as $n < {1\over 2}\sqrt{L} \log L$, and it becomes larger than the volume of the corresponding annulus for $n > 2 d\sqrt{L} \log L$. Therefore, we need at most $2 d\sqrt{L} \log L$ induction steps to reproduce the desired estimate for an arbitrary boundary condition. As each induction step worsens the estimate by the factor 2, at the end of the induction we accumulate the factor of $2^{2 d\sqrt{L} \log L}$. This factor  is still multiplied by the initial estimate $e^{-L}$. Thus, for large $L$ we have the desired resulting estimate $2^{2 d\sqrt{L} \log L} e^{-L} \le e^{-L/2}$.

The induction step itself is performed in the following way. Fix the configuration in the annulus number $n$ such that this configuration contains at most $e^{n/\sqrt{L}}$ unstable unit cells. This configuration serves as a boundary condition for $\Lam_{3L/4+n\oa}$. For any configuration inside $\Lam_{3L/4+n\oa}$ there are only two possibilities {\bf (1)} and {\bf (2)}:

\begin{description} \setlength{\itemsep}{0pt} \setlength{\parskip}{0pt} 
\item[(1)\;]
One of the layers indexed by $n' < n$ has less than $e^{n'/\sqrt{L}}$ unstable unit cells.  In this case by induction we already know that the probability for the boundary layer to penetrate inside $\Lam_{L/2}$ is small. 
\item[(2)\;]
The annuli indexed by $n' < n$ have in total at least $\sum\limits_{n^\prime= 1}^{n-1} e^{n^\prime/\sqrt{L}} > \diy \sqrt{L} e^{n/\sqrt{L}} /2$ unstable unit cells. The $\mu_{\Lam_{3L/4+n\oa}}\left(\;\cdot \;\| 1 \right)$-probability of this event behaves like $\exp(-\sqrt{L} e^{n/\sqrt{L}} /2)$, as follows from Lemma~4.1. The corresponding $\mu_{\Lam_{3L/4+n\oa}}\left(\;\cdot \;| \ubZ^{\Delta_{3L/4+n\oa}}\right)$-probability of the same event is at most $\exp(e^{n/\sqrt{L}})$ times larger. For large enough $L$ we obtain:
\beq
\exp\left(-\sqrt{L} e^{n/\sqrt{L}} /2+e^{n/\sqrt{L}}\right) < \exp\left(-\sqrt{L} e^{n/\sqrt{L}} /4\right).
\label{5.01}\eeq 
\end{description}

\noindent Adding the last estimate to the estimates obtained in the previous induction steps we at most double them. This completes the induction step and the argument covering the picture in frame (ii) of Figure 5.

\bigskip
We turn now to formal definitions and proofs. The event $\cA\left(\Lam_L|\,\ubY^{\Lam_L^\cmp}\right)$ defined in Eqn \eqref{1.09} is  written in terms of the restriction $\ubX^{\Lam_L}$ of the configuration $\ubX$. This restriction is a collection of finite sets $\bX_j^{\Lam_L}=\bX_j\cap\Lam_L$, $1\leq j\leq q$ belonging to $\Lam_L$. With a certain abuse of notation, we write $\ubX^\Lam\in\cA\left(\Lam_L |\,\ubY^{\Lam_L^\cmp}\right)$ meaning that $\ubX\in\cA\left(\Lam_L |\,\ubY^{\Lam_L^\cmp}\right)$. In the argument below we need to embed the sets $\bX_j^{\Lam_L}$ in a larger box, $\Lam^\prime\supset\Lam_L$. For this purpose, the notation $\ubX^{\Lam_L}\vee{\uvnth}^{\Lam^\prime\setminus\Lam_L}$ will be employed. A similar meaning is assigned to symbol $\ubX^{\Lam^\prime}\vee{\uvnth}^{\Lam_L\setminus\Lam^\prime}$ when box  
$\Lam^\prime\subset\Lam_L$.

Further,  the events $\cA (\Lam )$ and $\cA (\Lam \|\,i)$ 
in Eqns \eqref{1.07} and \eqref{1.13} are also written in terms of the restriction 
$\ubX^\Lam$. Consequently, with the same degree of the notational abuse we write 
$\ubX^{\Lam_L}\vee{\uvnth}^{\Lam^\prime\setminus\Lam_L}\in\cA (\Lam^\prime\|\,1)$ when 
$\Lam^\prime\supset\Lam$,
and $\ubX^{\Lam^\prime}\vee{\uvnth}^{\Lam_L\setminus\Lam^\prime}\in\cA (\Lam\|\,1)$ 
when $\Lam^\prime\subset\Lam$, by following a similar meaning. Under these agreements, it becomes obvious that
\beq\label{5.02}
\ubX^{\Lam_{L}}\in\cA (\Lam_{L} | \ubY^{\Lam_{L}^\cmp}) \implies \ubX^{\Lam_L}\vee
{\uvnth}^{\Delta_L}\in \cA (\Lam_{L+\oa} \| 1)
\eeq 
and, for $L>\oa$, 
\beq\label{5.03}
\ubX^{\Lam_{L-\oa}}\in\cA (\Lam_{L-\oa} \| 1) \implies \ubX^{\Lam_{L-\oa}}\vee {\uvnth}^{\Delta_{L-\oa}}\in\cA\left(\Lam_{L} | \ubY^{\Lam_{L}^\cmp}\right).
\eeq 
Extending these manipulations to events, we obtain that
\beq\label{5.04}
\cB\subseteq\cA\left(\Lam |\,\ubY^{\Lam_L^\cmp}\right) \implies  \cB\vee{\uvnth}^{\Delta_L}\subset\cA (\Lam_{L+\oa} \|\,1)
\eeq 
and 
\beq\label{5.05}
\cB\subseteq\cA (\Lam_{L-\oa} \|\,1) \implies  \cB\vee
{\uvnth}^{\Delta_{L-\oa}}\subset\cA\left(\Lam_{L} |\,\ubY^{\Lam_{L}^\cmp}\right).
\eeq 
Consequently, for an event $\cB\subseteq\cA\left(\Lam |\,\ubY^{\Lam_L^\cmp}\right)$ we have
\beq\label {5.06}\begin{array}{cl}
\diy\mu_{\Lam_{L}}\left(\cB\;| \ubY^{\Lam_{L}^\cmp}\right)  
&\diy < \mu_{\Lam_{L+\oa}}\left(\cB\vee{\uvnth}^{\Delta_L}\|\,1\right)\,
\frac{{\bbP}(\cA (\Lam_{L+\oa}\|\,1))}{{\bbP}(\cA (\Lam_{L-\oa}\|\,1))} \\  \;&\\
\;&\diy <\mu_{\Lam_{L+\oa}}\left(\cB\vee{\uvnth}^{\Delta_L}\|\,1\right) 
\Big(\frac{q}{\rv}\Big)^{\ups(\Lam_{L+\oa} \setminus \Lam_{L-\oa})}
\end{array}\eeq   
In fact, we will need a more precise version of \eqref {5.06}. 

\def\uLam{{_*\Lam}} 
\def\oLam{{^*\!\Lam}}

\bigskip
{\bf Definition 5.1.} We construct the set $\uLam_L =\uLam_L(\ubY^{\Delta_L};1)$ by removing from $\Lam_{L}$ the cubes of size $2\oa + 1$ centered at unit cells that are not in phase $1$ in $\ubY^{\Delta_L}$. Similarly, consider the set $\oLam_L =\oLam_L(\ubY^{\Delta_L};1)$ obtained by adding to $\Lam_L$ the cubes of size $2\oa + 1$ centered at unit cells that are not in phase $1$ in $\ubY^{\Delta_L}$ and then intersecting the result with $\Lam_{L + \oa}$. $\qquad\blacktriangle$

\begin{center}
\includegraphics[scale=1.05]{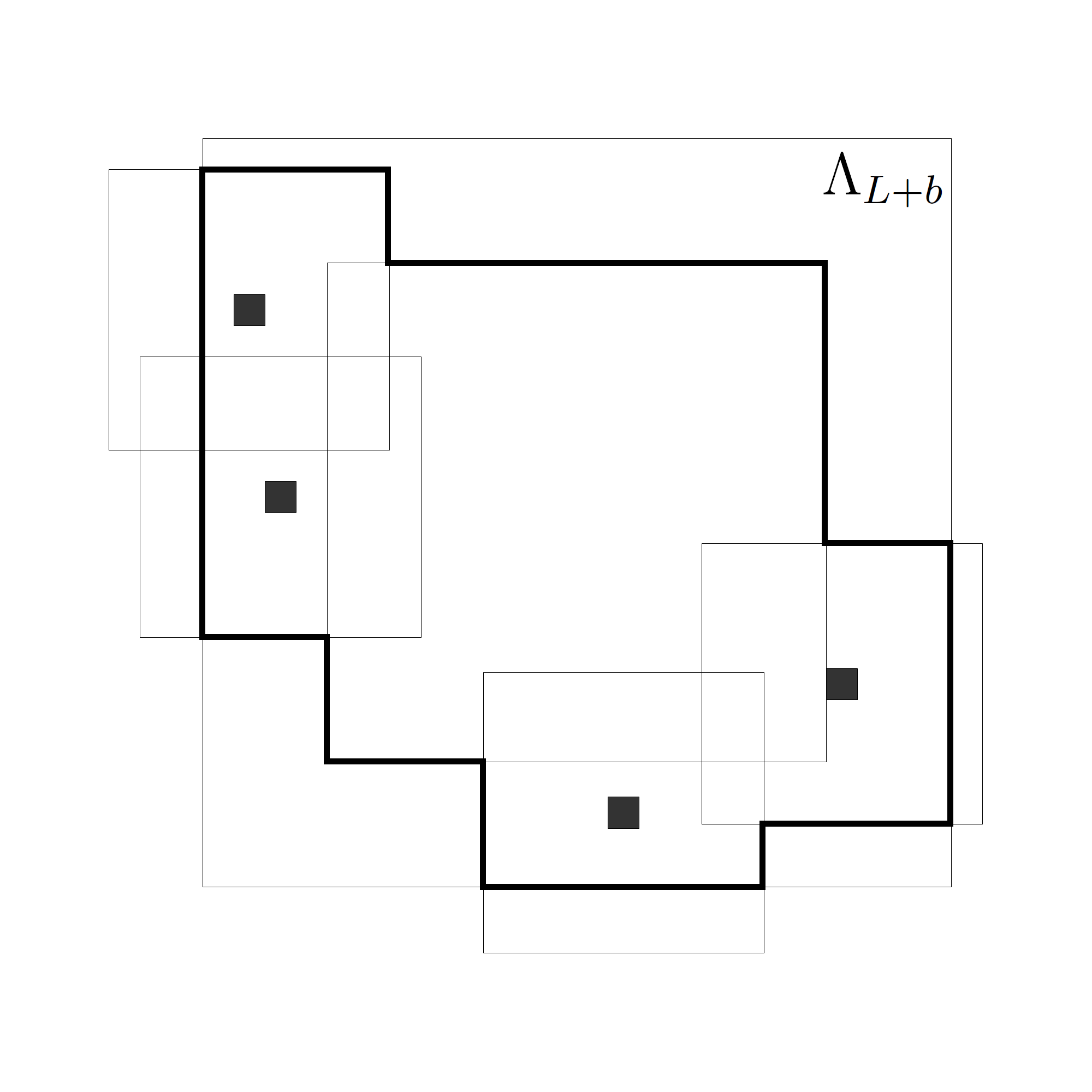}
\includegraphics[scale=1.05]{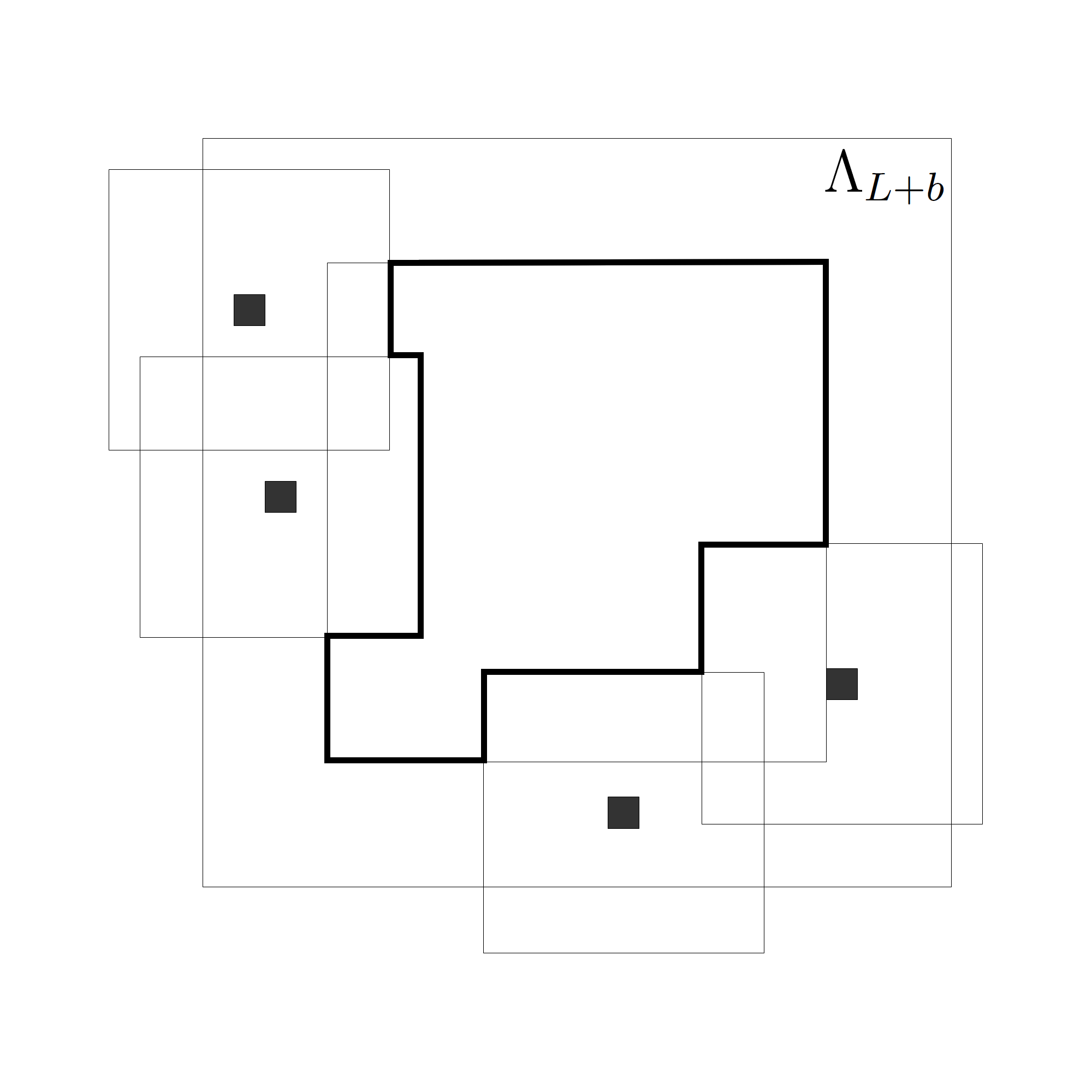}
\cl{\quad\quad{\bf (i)}\quad$\uLam_L$\qquad\qquad\qquad\qquad\qquad\qquad\qquad {\bf (ii)}\quad$\oLam_L$}
\cl{ }
\cl{\bf Figure 6. Sets $\uLam_L$ and $\oLam_L$}
\end{center}

\bigskip 
Assume that $\ubY^{\Delta_L}$ contains $\leq n$ unit cells that are not in phase $1$. 
Then, for an event $\cB\subseteq\cA\left(\Lam |\,\ubY^{\Lam_L^\cmp}\right)$,
\beq\label {5.07}\begin{array}{cl}
\diy\mu_{\Lam_{L}}\left(\cB\,| \ubY^{\Lam_{L}^\cmp}\right) &\diy
< \mu_{\oLam_L}\left(\cB\vee{\uvnth}^{\oLam_L\setminus\Lam_L}  \;\|\,1\right) 
\frac{{\bbP}(\cA (\oLam_L\|\,1))}{{\bbP}(\cA (\uLam_L\|\,1))}\\ \;&\;\\ \;&\diy
< \mu_{\oLam_L}\left(\cB\vee{\uvnth}^{\oLam_L\setminus\Lam_L}  \;\|\,1\right) \Big(\frac{q}{\rv}\Big)^{n\,(2\oa +1)^{\,d}}.
\end{array}\eeq   

Using inequalities \eqref {5.06} and \eqref {5.07}, it is possible to obtain an upper bound for $\mu_{\Lam_{L}}\left(\cB\,| \ubY^{\Lam_{L}^\cmp}\right)$ from a similar bound for $\mu_{\oLam_L}\left(\cB\vee{\uvnth}^{\oLam_L\setminus\Lam_L}  \;\|\,1\right)$. Since type $1$ is stable we can estimate probabilities of events in $\cA (\cdot \|1)$ by using the machinery developed in Sections 3 and 4. 

Our next step is a formal definition of the boundary layer in $\Lam_{L}$ for the case of an arbitrary boundary condition $\ubY^{\Lam_{L}^\cmp}$, cf. Definition~4.1. 

\bigskip
{\bf Definition 5.2.}  Consider a cubic box $\Lam_L$ and a particle configurations $\ubY\in\cA$. Given
a particle configuration $\ubX\in\cA
\left(\Lam |\,\ubY^{\Lam_L^\cmp}\right)$, consider  the set ${\rm U}$ of the unit cells $\Ups\subset\Lam_L$ which are not in phase 1 in $\ubX^{\Lam_L}$. A {\it base} of the {\BL} (in $\ubX^{\Lam_L}$) under boundary condition $\ubY^{\Lam_{L}^\cmp}$ is the union of those connected components of the set  ${\rm U}$ which are adjacent to unstable unit cells in $\Lam_L^\cmp$.  As in Definition 4.1, we refer to a {\it {\BL} $\Gam^\rU=\Gam^\rU(\ubX^{\Lam_L})$ under boundary condition $\ubY^{\Lam_{L}^\cmp}$} by considering its base, $\rB (\Gam^\rU)$, together with the types attached to unit cells $\Ups\in\rB (\Gam^\rU )$. $\qquad\blacktriangle$

\bigskip
Equivalently, the {\BL} can be defined in the following way. First, we construct the set $\oLam_L=\oLam_L(\ubY^{\Delta_L};1)$ as described in Definition 5.1 and pass from a configuration $\ubX\in\cA\left(\Lam |\,\ubY^{\Lam_L^\cmp}\right)$ to its restriction $\ubX^{\oLam_L}\in\cA\left(\oLam_L \|\,1\right)$. Due to the definition of $\oLam_L$, all unit cells of $\oLam_{L} \setminus \Lam_{L}$ are empty and therefore belong to the bases of some external contours $\Gam^{\tE}_k$. Among these external {\BCs} there are {\LBCs} which we denote by $\Gam^{\rL, \tE}_l$. Some of the external {\LBCs} $\Gam^{\rL, \tE}_l$ have interior components $\rI_t(\Gam^{\rL, \tE}_l)$ with unstable types $\iota_t(\Gam^{\rL, \tE}_l)$. Any such component gives rise to a {\BL} $\Gam^\rU_{l,t}$, with base
$\rB(\Gam^\rU_{l,t})$  as described by Definition~4.1. The full collection of {\BLs} $\Gam^\rU_{l,t}$, for varying $l$ and $t$,  together with all external contours $\Gam^{\tE}_k$, forms the {\BL} $\Gam^\rU$ under boundary condition $\ubY^{\Lam_{L}^\cmp}$. Consequently, the base $\rB(\Gam^{\rU}) = \left(\operatornamewithlimits{\cup}\limits_{l,t} \rB(\Gam^\rU_{l,t})\right) \cup \left(\operatornamewithlimits{\cup}\limits_k \big(\rB(\Gam^{\tE}_k) \cap \Lam \big) \right)$. As in Definition~4.1 $\rB(\Gam^\rU) \subseteq \Lam_L$. $\qquad\blacktriangle$ 

\begin{center}
\includegraphics[scale=1.25]{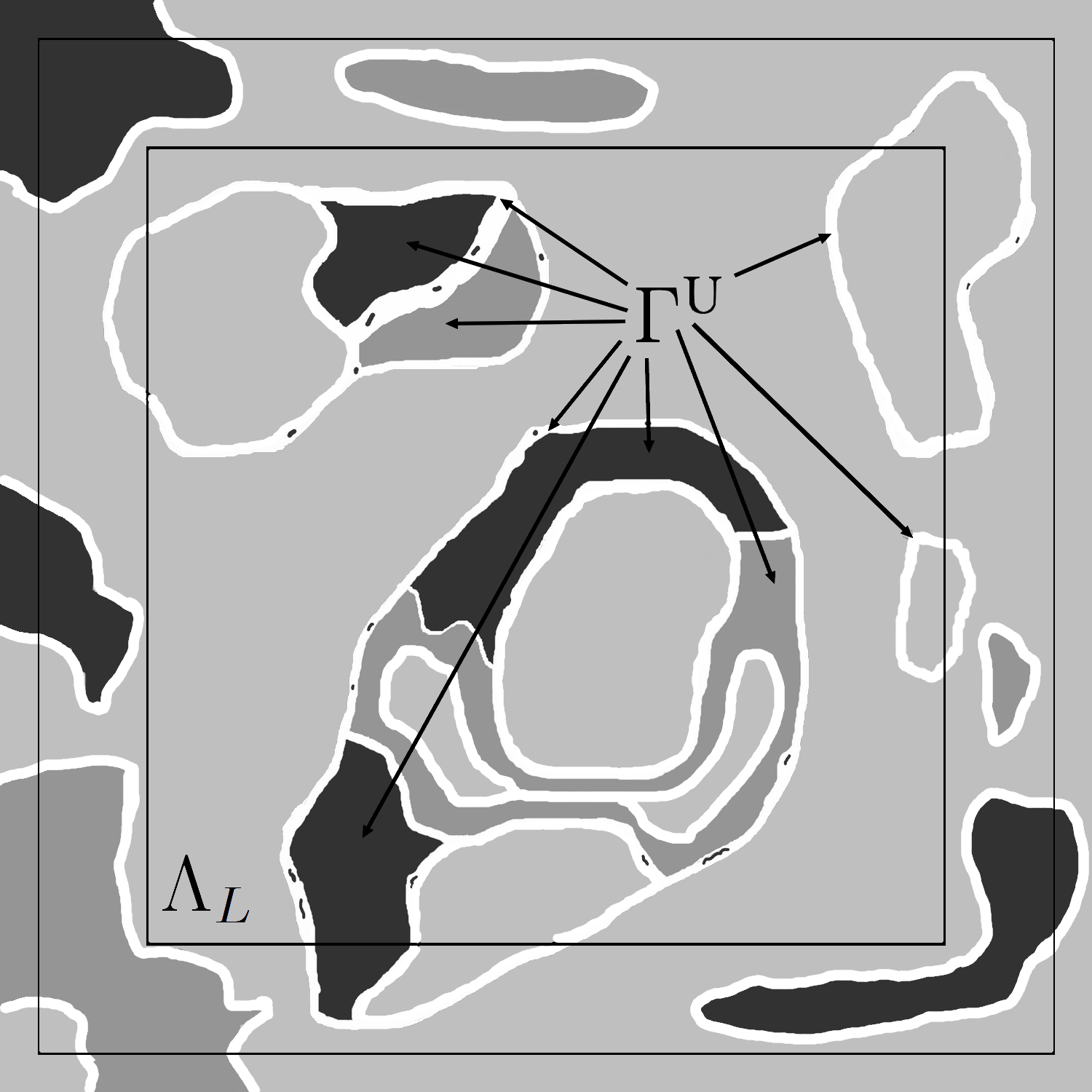}
\vskip 5 truemm

\cl{\bf Figure 7. Boundary layer under a generic boundary condition }
\end{center}
\vskip 5 truemm

\noindent In Figure 7, the contours have been drawn in configuration $\ubX^{\Lam_L}\vee\ubY^{\Lam_L^\cmp}$. The base $\rB(\Gam^\rU)$ of the boundary layer is the part of $\Lam_L$ which is not colored light grey. There are four {\BCs} in the collection $\{\Gam^{\tE}_k\}$. Their bases are shown as connected components of white color inside $\Lam_L$. 
Among these {\BCs} one is small and three are large.  These {\LBCs} constitute the collection $\{\Gam^{\rL, \tE}_l\}$. Five connected components of dark colors constitute the collection $\{\rB(\Gam^\rU_{l,t})\}$. The remaining {\BCs} of the configuration $\ubX^{\Lam_{L}}$ are not shown. They are situated inside light grey areas in $\Lam_L$.  

\bigskip 
We are interested in estimating of $\mu_{\Lam_{L}}\left(\cdot\,| \ubY^{\Lam_{L}^\cmp}\right)$-probability of the event 
	\beq\label {5.08}\cB_0=\Big\{\ubX^{\Lam_L}\in\cA\left(\Lam_L | \ubY^{\Lam_{L}^\cmp}\right):\;\rB(\Gam^{\rU}) \cap \Lam_{L/2} \not= \emptyset\Big\}.\eeq
Note that event $\cB_0$ occurs only if among $\Gam^{\rL, \tE}_l$ there exists at least one {\LBC} $\Gam^{\rL, \tE}_*$ for which either $\Lam_{L/2} \subset \rI(\Gam^{\rL, \tE}_*)$ or $\rB(\Gam^{\rL, \tE}_*) \cap \Lam_{L/2} \not= \emptyset$. In both cases $\ups(\rB(\Gam^{\rL, \tE}_*)) > L/2$ (regardless of dimension $d$). 

Assume for now that $\ubY^{\Delta_L}$ has at most $\sqrt{L}$ unit cells not in the phase $1$. Then, according to \eqref {5.07},
\beq\label {5.09}\begin{array}{rl}
\diy\mu_{\Lam_{L}}\left(\cB_0\,| \ubY^{\Lam_{L}^\cmp}\right) 
&\diy < \mu_{\oLam_L}\left(\cB_0\vee{\uvnth}^{\oLam_L\setminus\Lam_L}  \;\|\,1\right) \Big(\frac{q}{\rv}\Big)^{\sqrt{L}\,(2\oa +1)^{\,d}}.
\end{array}\eeq   
The upper bound for $\mu_{\oLam_L}\left(\cB_0\vee{\uvnth}^{\oLam_L\setminus\Lam_L}  \;\|\,1\right)$ can be obtained from the following considerations: 
\begin{description} \setlength{\itemsep}{0pt} \setlength{\parskip}{0pt}
\item[(1)\;] As follows from  Peierls estimate \eqref{4.23}, the probability to have all {\LBCs} of $\Gam^{\rL, \tE}_l$ among entire 
collection of external {\BCs} in $\oLam_{L}$ does not exceed the product of renormalized statistical weights $\prod\limits_l \rW(\Gam^{\rL, \tE}_l)$.
\item[(2)\;] For any {\LBC} $\Gam^{\rL, \tE}_l$ there exists a unit cell $\Ups \subset\oLam_L \setminus \Lam$ such that $\Ups \subset \rB(\Gam^{\rL, \tE}_l)$.  This unit cell $\Ups$ can be used as a pinpoint in summation over bases $\rB(\Gam^{\rL, \tE}_*) \supset \Ups$ as in \eqref{4.25}.  
\item[(3)\;] The product $\prod\limits_l \rW(\Gam^{\rL, \tE}_l)$ includes a {\LBC} $\Gam^{\rL, \tE}_*$ with $\ups(\rB(\Gam^{\rL, \tE}_*)) > L/2$. Therefore $\rW(\Gam^{\rL, \tE}_*) \le 2\left(\rv/(1 - \rv)\right)^{L/5^d 2}$ as follows from \eqref{4.19} and \eqref{2.31}. There are at most $\sqrt{L}$ possibilities to select, among unstable unit cells in $\Delta_L$, a unit cell $\Ups$ giving rise to $\rB(\Gam^{\rL, \tE}_*)$.
\item[(4)\;] The remaining part of the product $\prod\limits_l \rW(\Gam^{\rL, \tE}_l)$ corresponds to {\LBCs} `originated' from remaining at most $\sqrt{L}-1$ unstable unit cells in $\Delta_L$. The sum of factors $\prod\limits_l \rW(\Gam^{\rL, \tE}_l)/\rW(\Gam^{\rL, \tE}_*)$ over all possibilities to `draw' these {\LBCs} is less than $(1+\rv^{0.9})^{\sqrt{L}-1}$, as follows from \eqref{4.25}. 
\end{description}
Combining {\bf (1)--(4)}, we conclude that
\beq\label {5.10}\begin{array}{rl}
\diy\mu_{\Lam_{L}}\left(\cB_0\,| \ubY^{\Lam_{L}^\cmp}\right) 
&\diy < \Big(\frac{q}{\rv}\Big)^{\sqrt{L}\,(2\oa+1 )^{\,d}}  2\left({\rv \over 1 - \rv}\right)^{{L / (5^d 2)}} \; \sqrt{L}\; \Big(1+\rv^{0.9}\Big)^{{\sqrt L}-1} \\ 
&\diy \le \rv^{L/(5^d 3)}
\end{array}\eeq   
provided $L$ is large enough.

\def\Lamn{\Lam_{(n)}}
\def\Deltan{\Delta_{(n)}}
\def\oLamn{\oLam_{(n)}}
Next, set $\Lamn=\Lam_{3L/4 + n\oa}$. Consider the annuli $\Deltan = \Lam_{3L/4 + n\oa + \oa} \setminus \Lam_{3L/4 + n\oa}$ indexed by an integer $n$, $1\leq n\leq 2d \sqrt{L} \log L$, and 
introduce a threshold value
\beq G(n) = \max(1,e^{n/\sqrt{L}})\label {5.11}\eeq 
as an upper bound for the amount of unstable unit cells in the corresponding annulus. For $n \ge 2d \sqrt{L} \log L$ we have
\beq\label {5.12} 
G(n) > (3L/4 + n\oa + \oa)^d - (3L/4 + n\oa)^d
\eeq
so that the above threshold is not really a limitation. 

Consider the part of the boundary layer $\Gam^\rU$ which lies outside $\Lamn$, i.e.  consider the restriction of $\Gam^{\rU}$ to $\rB(\Gam^{\rU}) \setminus \Lamn$. We denote by $\rB^{(n)}(\Gam^{\rU})$ the part of $\rB(\Gam^{\rU}) \setminus \Lamn$ formed by connected components of $\rB(\Gam^{\rU}) \setminus \Lamn$ adjacent to  $\Lam_L^\cmp$. Further, define the event
\beq\label {5.13}
\cB_n=\Big\{\ubX^{\Lam_L}\in\cA\left(\Lam_L | \ubY^{\Lam_{L}^\cmp}\right):\;\ups\left(\rB^{(n)}(\Gam^\rU) \cap\Deltan\right)\le G(n)\Big\}.
\eeq
In other words, event $\cB_n$ occurs when $\Deltan$ contains at most $G(n)$ unstable unit cells from $\rB^{(n)}(\Gam^{\rU})$.

For $n <{\frac{1}{2}}\sqrt{L} \log L$ we have
\beq\label {5.14} 
G(n) < \sqrt{L}.\eeq
Thus, adjusting the argument leading to \eqref {5.10} we obtain
\beq\label {5.15}\begin{array}{rl}
\diy\mu_{\Lamn}\left(\cB_0\,| \ubZ^{\Lamn^\cmp}\right) 
&\le \rv^{(3L/4 +\oa n)/(5^d 3)}
\end{array}\eeq
provided $L$ is large enough and $\ubZ^{\Lamn^\cmp}$ contains at most $G(n)$ unstable unit cells inside the annulus $\Deltan$. The difference between \eqref {5.15} and \eqref {5.10} is that the {\LBC} $\Gam^{\rL, \tE}_*$ has $\ups(\rB(\Gam^{\rL, \tE}_*)) > (3L/4 + \oa n)/2$. Set:
\beq\label {5.16}\begin{array}{l}
\diy
\mu_{\Lam_L}\left(\cB_0\,| \cB_n,\ubY^{\Lam_L^\cmp} \right)\diy := 
\frac{\mu_{\Lam_L}\left(\cB_0\,\cap\, \cB_n \,| \ubY^{\Lam_L^\cmp}\right)}
{\mu_{\Lam_L}\left(\cB_n \,| \ubY^{\Lam_L^\cmp}\right)}\\
\end{array}\eeq
By using the DLR property together with \eqref {5.15}, we obtain that for $n <{\frac{1}{2}}\sqrt{L} \log L$,
\beq\label {5.17}\begin{array}{l}
\diy
\mu_{\Lam_L}\left(\cB_0\,| \cB_n,\ubY^{\Lam_L^\cmp} \right)\diy 
=\frac{\int\limits_{\cB_n }\mu_{\Lamn}\left(\cB_0\,| \ubZ\right) \rd\mu_{\Lam_L}\left(\ubZ \,| \ubY^{\Lam_L^\cmp}\right)}
{\int\limits_{\cB_n}\rd\mu_{\Lam_L}\left(\ubZ \,| \ubY^{\Lam_L^\cmp}\right)} \le \rv^{(3L/4 +\oa n)/(5^d 3)}
\end{array}\eeq
where $\ubZ =\ubZ^{\Lamn^\cmp}$. Writing, for brevity, $\mu_{\Lam_L}\left(\cB_0\,| \cB_{n}
\right)$ instead of\\ $\mu_{\Lam_L}\left(\cB_0\,| \cB_{n},\ubY^{\Lam_L^\cmp} \right)$, we can estimate:
\beq\sum_{n'=1}^{{1 \over 2}\sqrt{L} \log L}\mu_{\Lam_L}\left(\cB_0\,| \cB_{n^\prime} 
\right)\leq \rv^{3L/(5^d 24)}\label {5.18}\eeq
provided that $L$ is large enough.

On the other hand, for $n \ge \frac{1}{2} \sqrt{L} \log L$, with $\cB^*_n:=\operatornamewithlimits{\cap}\limits_{n'=1}^{n-1} \cB_{n^\prime}^\cmp$, we have:
\beq\label {5.19}\begin{array}{rl}
\diy\mu_{\Lam_L}\left(\cB_0\,| \cB_n \right) 
&\le \sum\limits_{n' = 1}^{n-1} \mu_{\Lam_L}\left(\cB_0\,| \cB_{n'} \right) \mu_{\Lam_L}\left(\cB_{n'}\,| \cB_n \right) \\
&+ \diy\mu_{\Lam_L}\left(\cB_0\,|\cB^*_n\right) \mu_{\Lam_L}\left(\cB^*_n\,| \cB_n \right) \\
&\le \sum\limits_{n' = 1}^{n-1} \mu_{\Lam_L}\left(\cB_0\,| \cB_{n'} \right) +  \mu_{\Lam_L}\left(
\cB^*_n \,| \cB_n \right).
\end{array}\eeq
Owing to \eqref {5.07} and a calculation similar to \eqref {5.17},
\beq\label   {5.20}\begin{array}{rl}
\diy\mu_{\Lam_L}\left(\cB^*_n \,| \cB_n \right) &\diy \le \sup\,\left[ \mu_{\Lamn}\left(\cB^*_n \,|\ubZ \right):\;{\ubZ\in\cB_n}\right] \\
\end{array}\eeq
where again $\ubZ = \ubZ^{\Lamn^\cmp}$. The subsequent inequalities \eqref {5.22}--\eqref {5.30} 
aim at estimating the probability $ \mu_{\Lamn}\left(\cB^*_n \,|\ubZ \right)$ uniformly in 
${\ubZ\in\cB_n}$.

The event $\cB^*_n$ is that inside $\Lamn \setminus \Lam_{3L/4}$ the {\BL} contains at least 
\beq M=M(n):=\sum\limits_{n^\prime = 1}^{n-1} G(n^\prime )\label {5.21}\eeq
unstable unit cells. The probability of this event can be estimated in the following way:
\beq\label {5.22}\begin{array}{rl}
\diy\mu_{\Lamn}\left(\cB^*_n \,|\ubZ^{\Lamn^\cmp} \right) &\diy \le  \left(\frac{q}{\rv}\right)^{G(n)}    \mu_{\oLamn}\left(\cB^*_n \,\| \,1 \right);\\\end{array}\eeq
cf. Eqn \eqref{5.09}. Recall, box $\oLamn$ obtained from cube $\Lamn$ is $\ubZ^{\Lamn^\cmp}$-dependent
(see Definition 5.1). Further, we write 
\beq\label {5.23}\begin{array}{rl}
\diy\mu_{\oLamn}\left(\cB^*_n \,\| \,1 \right)=&
\diy{\sum\limits_{\uGam^{\rL, \tE}}}^{\sharp,\oLamn} \frac{\mu_{\oLamn} \left(\cG(\uGam^{\rL, \tE})  \| \,1 \right)}
{\prod\limits_{\Gam^{\rL,\tE} \in \uGam^{\rL,\tE}} \prod\limits_{t} \Xi\left({_3\rI_t(\Gam^{\rL,\tE})}\,\big\|\,\iota_t(\Gam^{\rL,\tE})\right)} \\ \\
&\times \diy{\sum\limits_{\Gam^{\rU}}}^{\uGam^{\rL, \tE},M} \rw(\Gam^{\rU})
\prod\limits_{k,s} \Xi\left({_3\rI_{k,s}(\Gam^{\rU}})\,\Big\|\,\iota_{k,s}\left(\Gam^\rU \right)\right).
\end{array}\eeq
Here the sum $\;{\sum\limits_{\uGam^{\rL, \tE}}}^{\sharp,\,\oLamn}$ is taken over collections $\uGam^{\rL, \tE}$ of external {\LBCs}  $\Gam^{\rL,\tE}$ with base $\rB (\uGam^{\rL,\tE})=\operatornamewithlimits{\cup}\limits_{\Gam^{\rL,\tE}\in\uGam^{\rL,\tE}}\rB (\Gam^{\rL,\tE})$ such that each $\rB (\Gam^{\rL,\tE})$ contains a unit cell unstable in $\ubZ^{\Deltan}$. The event $\cG (\uGam^{\rL, \tE})$ is defined in \eqref{2.43}. Next, the sum $\;{\sum\limits_{\Gam^{\rU}}}^{\uGam^{\rL, \tE},\,M}$ is taken over {\BLs} $\Gam^{\rU}$ in $\operatornamewithlimits{\cup}\limits_{\Gam^{\rL,\tE}\in \uGam^{\rL,\tE}}\operatornamewithlimits{\cup}\limits_t\; {_3\rI_t(\Gam^{\rL,\tE})}$ such that $\ups (\rB (\Gam^{\rU}))\geq M$. Note that for each $\Gam^{\rL,\tE}$ only components $_3\rI_t(\Gam^{\rL,\tE})$ having unstable $\iota_t(\Gam^{\rL,\tE})$ contribute to the double union above (and correspondingly to the double product in \eqref{5.23}). The statistical weight $\rw(\Gam^\rU)$ is defined by \eqref{4.02} and  the sum over $\Gam^\rU$ in \eqref{5.23} originates from \eqref{4.03}. To be more precise,  Eqn \eqref{4.02} defines $\rw(\Gam^\rU(_3\rI_t(\Gam^{\rL,\tE})))$ for the {\BL} $\Gam^\rU(_3\rI_t(\Gam^{\rL,\tE}))$ inside $_3\rI_t(\Gam^{\rL,\tE}))$ and $\rw(\Gam^\rU)$ is obtained as the double product $\prod\limits_{\Gam^{\rL,\tE} \in \uGam^{\rL,\tE}} \prod\limits_{t}\rw(\Gam^\rU(_3\rI_t(\Gam^{\rL,\tE})))$. Finally, in the product $\prod\limits_{k,s}$ the {\BL} $\Gam^\rU$ is treated as a single object, i.e. indexes $k$ and $s$ are defined according to item {\bf (i)} below Definition~4.1 and are not partitioned by index $t$.

By virtue of the Peierls bound \eqref{4.23}, the probability of event  $\cG (\uGam^{\rL, \tE})$ can be estimated by the product of renormalized statistical weights:
\beq\label{5.24}
\mu_{\oLamn}\left(\cG (\uGam^{\rL, \tE})\| \,1 \right)\leq \prod\limits_{\Gam^{\rL, \tE}\in\uGam^{\rL, \tE}}
W(\Gam^{\rL, \tE}).
\eeq
Hence, with the help of \eqref{4.20} and \eqref{4.06}, the RHS in \eqref {5.23} is less than or equal to
\beq{\sum\limits_{\uGam^{\rL, \tE}}}^{\sharp,\oLamn} \prod\limits_{\Gam^{\rL, \tE}\in\uGam^{\rL, \tE}}\rw (\Gam^{\rL, \tE})
\;{\sum\limits_{\Gam^{\rU}}}^{\uGam^{\rL, \tE},M} \rW (\Gam^{\rU}),
\label{5.25}\eeq
which can be seen by multiplying the fraction in \eqref{5.23} by the double product $\prod\limits_{\Gam^{\rL,\tE} \in \uGam^{\rL,\tE}} \prod\limits_{t} \Xi\left({_3\rI_t(\Gam^{\rL,\tE})}\,\big\|\,1\right)$ and dividing the sum ${\sum\limits_{\Gam^{\rU}}}^{\uGam^{\rL, \tE},M}$ by the same quantity; cf. \eqref{4.04}.

Now, referring to \eqref{4.08}, we conclude that the expression \eqref{5.25} does not exceed
\beq
\exp\,\left(-M\rv^{\oa^{\,2d}}\right)
{\sum\limits_{\uGam^{\rL, \tE}}}^{\sharp,\oLamn} \prod\limits_{\Gam^{\rL, \tE}\in\uGam^{\rL, \tE}}\rw (\Gam^{\rL, \tE})
\;{\sum\limits_{\Gam^{\rU}}}^{\uGam^{\rL, \tE},M} \rW (\Gam^{\rU})e^{a(\Gam^{\rU})}
\label {5.26}\eeq
since $a(\Gam^{\rU})\geq \ups (\rB (\Gam^{\rU}))\rv^{\oa^{\,2d}}$.

\noindent
Next, we apply the bound \eqref{4.09}:  
\beq
{\sum\limits_{\Gam^{\rU}}}^{\uGam^{\rL, \tE},M} \rW (\Gam^{\rU})e^{a(\Gam^{\rU})}
\leq\prod\limits_{\Gam^{\rL, \tE}\in\uGam^{\rL, \tE}}\Big( 1 + \rv^{\,0.5\,\oa^{\,2d}}  \Big)^{\ups(\rB (\Gam^{\rL, \tE}))}.
\label {5.26.1}\eeq

\noindent
Therefore, expression \eqref {5.26} is less than or equal to
\beq
\exp\,\left(-M\rv^{\,\,\oa^{\,2d}}\right)
{\sum\limits_{\uGam^{\rL, \tE}}}^{\sharp,\oLamn} \prod\limits_{\Gam^{\rL, \tE}\in\uGam^{\rL, \tE}}\rw (\Gam^{\rL, \tE})
\,\Big( 1 + \rv^{\,0.5 \,\oa^{\,2d}}  \Big)^{\ups(\rB (\Gam^{\rL, \tE}))}.
\label {5.27}\eeq
Further,  as follows from \eqref{4.19}, statistical weights $\rw(\Gam )$ and  $W(\Gam )$ obey
\beq
\rw(\Gam^{\rL, \tE})\leq W(\Gam^{\rL, \tE})\Big( 1 + \rv^{\,0.5\,\oa^{\,2d}}  \Big)^{\ups(\rB (\Gam^{\rL, \tE}))}
\label {5.27.1}\eeq
because $\sum\limits_t\ups\left(\partial\left({_3}\rI_t(\Gam )\right)\right)\leq\ups(\rB (\Gam ))$. This yields that \eqref {5.27} does not exceed
\beq\exp\,\left(-M\rv^{\,\oa^{\,2d}}\right)
{\sum\limits_{\uGam^{\rL, \tE}}}^{\sharp,\oLamn} \prod\limits_{\Gam^{\rL, \tE}\in\uGam^{\rL, \tE}}\rW (\Gam^{\rL, \tE})
\,\Big( 1 + \rv^{\,0.5\,\oa^{\,2d}}  \Big)^{2\ups(\rB (\Gam^{\rL, \tE}))}\label {5.28}\eeq
which in turn is not bigger than
\beq\begin{array}{l} 
\exp\,\left(-M\rv^{\,\oa^{\,2d}}\right)\left(1+\rv^{\,0.9}\right)^{G(n)}
\end{array}\label{5.29} \eeq
Here we used \eqref{4.25} and the fact that we have at most $G(n)$ reference unit cells 
inside $\oLamn\setminus
\Lamn$, and each base $\rB (\Gam^{\rL, \tE})$ contains at least one such reference unit cell. Note that the 
quantities \eqref{5.29}  do not depend on particle configuration $\ubZ$.

All in all, for the probability in the LHS of \eqref   {5.20} is estimated as follows:
\beq \mu_{\Lam_L}\left(\cB^*_n \,| \cB_n \right)\leq
\left(\frac{q}{\rv}\right)^{G(n)}\exp\,\left(-\sum\limits_{n^\prime =1}^{n-1}G(n^\prime )\rv^{\,\oa^{\,2d}}\right)\left(1+\rv^{\,0.9}\right)^{G(n)}.\label{5.30} \eeq
Observe that $\forall$ $n\geq 1$, as follows from the definition \eqref {5.11},
$$\sum\limits_{n^\prime =1}^{n-1}G(n^\prime )\geq\frac{1}{2}{\sqrt L}\,G(n).$$
For $L$ large enough this yields
$$ \mu_{\Lam_L}\left(\cB^*_n \,| \cB_n \right)\leq \exp\,\left(-\frac{1}{4}G(n){\sqrt L}\,\rv^{\,\oa^{\,2d}}\right)$$
which, for $n\geq\frac{1}{2}{\sqrt L}\ln\,L$, leads to
\beq\mu_{\Lam_L}\left(\cB^*_n \,| \cB_n \right)\leq \exp\,\left(-\frac{1}{4}L\,\rv^{\,\oa^{\,2d}}\right)\,.
\label{5.31} \eeq

Therefore, with the help of \eqref {5.18}, we obtain a recursive bound for the probability in the 
LHS of \eqref {5.19}: for $n\geq{\frac{1}{2}}{\sqrt L}\ln\,L$, 
\beq\begin{array}{rl}
\mu_{\Lam_L}\left(\cB_0 \,| \cB_n \right)&\leq 
\rv^{3L/(5^d 24)}\\
&+ \diy\sum_{n'={1 \over 2}\sqrt{L} \log L}^{n-1}
 \mu_{\Lam_L}\left(\cB_0\,| \cB_{n'} \right)+ \exp\,\left(-\diy\frac{1}{4}L\,\rv^{\,\oa^{\,2d}}\right).
\end{array}\label{5.32} \eeq
This yields that for $n\geq{\frac{1}{2}}{\sqrt L}\ln\,L$,
\beq\label{5.33}  \mu_{\Lam_L}\left(\cB_0 \,| \cB_n \right)\\
\leq 2^{n}\left[\rv^{3L/(5^d 24)}
+ \exp\,\left(-\frac{1}{4}L\,\rv^{\,\oa^{\,2d}}\right)\right].\eeq

Finally, observe that for $n=2d{\sqrt L} \log L$ and $L$ large enough,
by virtue of \eqref {5.12}, $\cB_n$ is the full event.  Therefore, for $L$ large enough,
\beq\begin{array}{rl}
\mu_{\Lam_L}(\cB_0\,|\,\ubY^{\Lam_L^\cmp}  ) &=\mu_{\Lam_L}\left(\cB_0\,| \cB_{n},\ubY^{\Lam_L^\cmp} \right) \\ \\
&\leq \diy 
 2^{2d{\sqrt L} \ln L}\left[\rv^{3L/(5^d 24)}
+ \exp\,\left(-\frac{1}{4}L\,\rv^{\,\oa^{\,2d}}\right)\right] \\
&\leq \diy \exp\,\left(-\frac{1}{8}L\,\rv^{\,\oa^{\,2d}}\right).
\end{array}\label{5.34} \eeq 

\def\s{\sigma}

Denote by $\cB_0^\cmp$ the complement of $\cB_0$ and consider a cylindrical event
$\cB$ localized in $\Lam_{L/4}$. Then, for $L$ large enough, a standard consequence of polymer expansion is
\beq\left|\frac{\mu_{\Lam_L}(\cB\cap\cB_0^\cmp |\ubY^{\Lam_L^\cmp})}{\mu_{\Lam_L}(\cB_0^\cmp |\ubY^{\Lam_L^\cmp})} 
- \mu (\cB\|\,1) \right| \leq e^{-cL} \label{5.35} \eeq
where $c\in (0,\infty )$ is a constant. Finally, 
\beq\begin{array}{rl}
\left| \mu_{\Lam_L}(\cB|\ubY^{\Lam_L^\cmp} ) - \mu(\cB\|1) \right| &\le
\left|\diy\frac{\mu_{\Lam_L}(\cB\cap\cB_0 |\ubY^{\Lam_L^\cmp} )}{\mu_{\Lam_L}(\cB_0 |\ubY^{\Lam_L^\cmp} )} 
-\mu(\cB\|1) \right|\mu_{\Lam_L}(\cB_0|\ubY^{\Lam_L^\cmp} ) \\ \\
&+\left|\diy \frac{\mu_{\Lam_L}(\cB\cap\cB_0^\cmp |\ubY^{\Lam_L^\cmp} )}{\mu_{\Lam_L}(\cB_0^\cmp |\ubY^{\Lam_L^\cmp} )}  
- \mu(\cB\|1) \right|\mu_{\Lam_L}(\cB_0^\cmp |\ubY^{\Lam_L^\cmp}) \\
&\le \diy e^{-cL}+ \exp\,\left(-\frac{1}{8}L\,\rv^{\,\oa^{\,2d}}\right). \label{5.36} \end{array}\eeq
The RHS in \eqref {5.36}  tends to $0$ as $L\to\infty$. This completes the proof of Theorem 1 (I).

\bigskip\def\ts{{\tt s}}
{\bf 5.2. Proof of Assertions (IIa), (III) and (IV) of Theorem 1.2.} Here we suppose that $\St=\{1,2\}$ (and hence, unstable types are $3$ and $4$). A common property in cases (IIa) and (IV) is a complete symmetry between stable types $1$ and $2$. Namely, for any particle configuration $\ubX$ with restrictions $\ubX^{\Lam_L}$
and $\ubX^{\Lam_L^\cmp}$ we define the corresponding {\it factorized} particle configurations $\ubX^{\tF}$,
$\ubX^{\Lam_L,\tF}$ and $\ubX^{\Lam_L^\cmp,\tF}$ by replacing each particle of a stable type with a particle of a factorised type $\ts$. (Here $\ts$ stands for  stable.) In general, more than one original particle configuration is mapped to the same factorized particle configuration. To be more specific, if in the factorized particle configuration the number of $D(1,2)$-connected components of $\ts$-type particles equals $k$ then the number of original particle configurations mapped to this factorized configuration is $2^k$. (Recall, in this sub-section $\sh\St=2$.) Assigning to the factorized particle configurations additional statistical weights $2^k$, we obtain an equivalent model which 
is a kind of the FK representation for the original model (cf. \cite{GHM, CCK}). 

Figure~8 is a factorized version of Figure~2. We assume that two lighter grey colors in Figure~2 correspond to stable types 1 and 2. In Figure~8 both of them are drawn as a single light grey color which corresponds to the type~$\ts$.

\begin{center}
\includegraphics[scale=1]{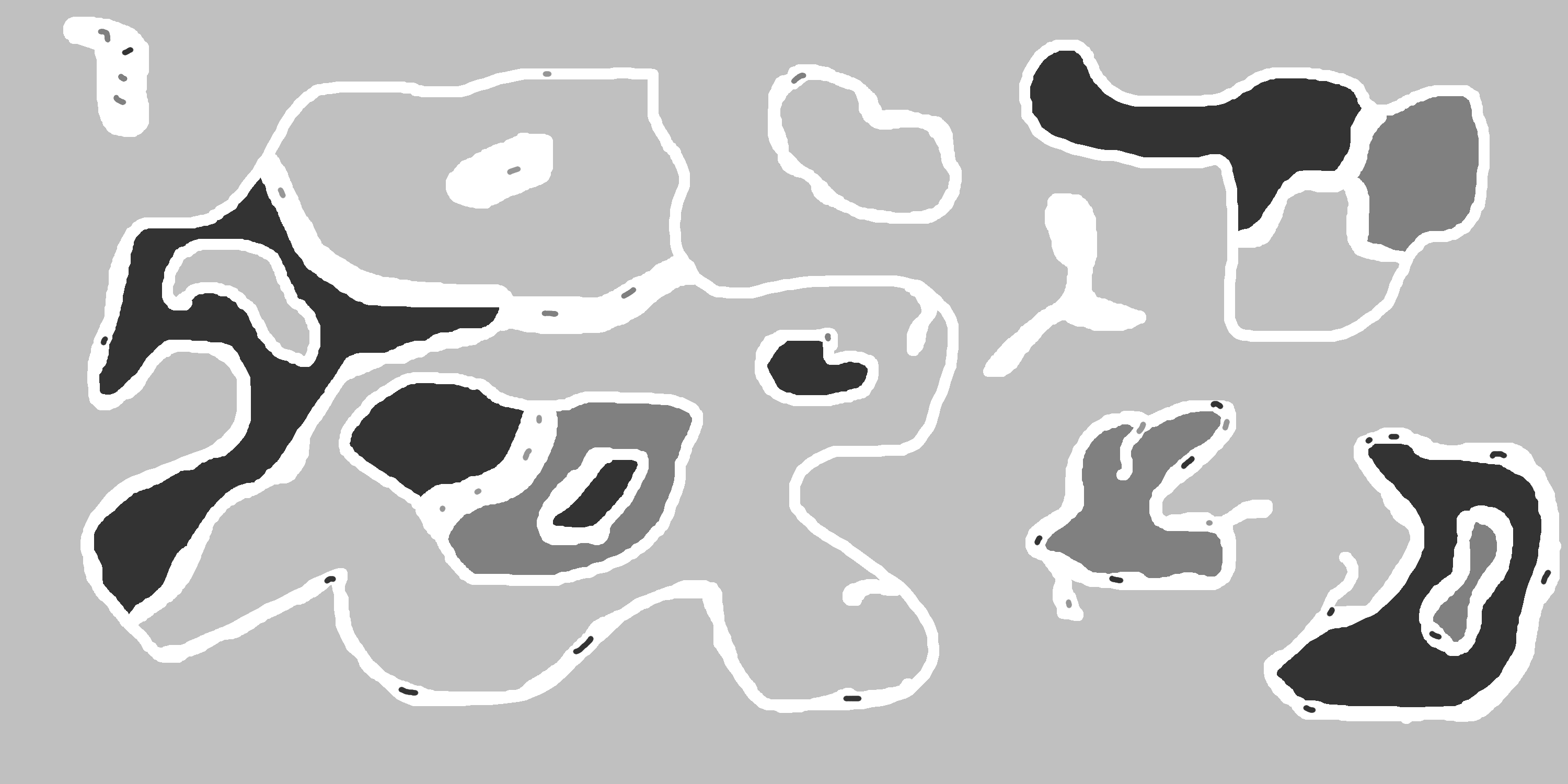}\quad
\cl{\bf Figure 8. Factorized configuration}
\end{center}

\bigskip
All constructions from the previous sections can be repeated for the factorized model with minor modifications which we list below. 

In the original model the measure $\bbP$  was defined for particles of the type 1,2,3 and 4. In the factorized model this measure $\bbP$ is defined for particles of the type 3, 4 and~$\ts$.

The original objects carrying the notation $(\;\cdot\;\| i)$ with $i=1,2$ have a direct analogues $(\;\cdot\;\|\ts)$ in the factorized model. To define in the factorized model the statistical weight of a {\BC} $\Gam^{\tF}$ we need to integrate over the generating configurations $\ubX^{\tF} \in \cA (\Gam^{\tF},\Lam)$ the statistical weight $2^{k(\ubX^{\tF})-1}$. Here $k(\ubX^{\tF})$ is the number of $D(1,2)$-connected component of $\ts$-type particles in $\ubX^{\tF}$; cf. \eqref{2.30}. We use the power $k-1$ rather than $k$ because the external $D(1,2)$-connected component (i.e. the one intersecting $\rE(\Gam^{\tF})$) inherits its original type from the boundary condition or from the enclosing {\BC}. Consequently, the analogue of the upper bound in \eqref{2.31} has an additional factor $2^{\ups(\rB(\Gam^{\tF}))}$.

Similarly, in analogues of bounds \eqref {5.06} and \eqref {5.07} we replace $\left(\diy\frac{q}{\rv} \right)$ with $\left(\diy\frac{2q}{\rv} \right)$ estimating from above the maximal number of $D(1,2)$-connected components of $\ts$-type particles in the boundary condition $\ubY^{\Lam_L^\cmp, \tF}$. (Recall, we work with $q=4$.)

The definition of the event $\cB_0$ (see \eqref {5.08}) remains valid for the factorized model. Moreover, repeating the argument \eqref {5.15} -- \eqref {5.34}  from Section~5.1 we conclude that the probabilities of the complement event $\cB_0^\cmp$ under factorized measures $\mu_{\Lam_L}(\,\cdot\,\|\,3)$, $\mu_{\Lam_L}(\,\cdot\,\|\,4)$ and $\mu_{\Lam_L}(\,\cdot\,)$ (the latter is with the empty boundary condition) tend to $1$ as $L\to\infty$. Note that the probability laws for random configurations $\bX^{\Lam_L}_3$ and $\bX^{\Lam_L}_4$ under measures $\mu_{\Lam_L}(\,\cdot\,\|j)$, $j=3,4$, and $\mu_{\Lam}(\,\cdot\,)$ are the same in both factorized and original models. If in the factorized model $\ubX^{\Lam_{L},\tF} \in \cB_0^\cmp$ then this factorized configuration contains an annulus enclosing $\Lam_{L/2}$ and formed by unit cells of type $\ts$. Due to the symmetry between stable types $1$ and $2$, the corresponding original configuration $\ubX^{\Lam_{L}} \in \cB_0^\cmp$ contains, inside this annulus particles of specific type $i \in \St$ with the probability which is $1/2$ of the probability of the factorized configuration. Moving back from probabilities of the factorized configurations $\ubX^{\Lam_{L},\tF}$ to probabilities of the original configurations $\ubX^{\Lam_{L}}$, we see that, for unstable types $j=3,4$,
\beq\label{5.37}
\lim\limits_{L\to\infty}\mu_{\Lam_L}(\,\cdot\,\|j) = \lim_{L\to\infty}\mu_{\Lam}(\,\cdot\,)
={\diy\frac{1}{2}}\big[\mu (\,\cdot\,\| 1)+\mu (\,\cdot\,\| 2)\big].
\eeq 
This yields Assertion (IIa) of Theorem 1.2 and Assertion (IV) in the case where $\St=\{1,2\}$. 

Note that the technique from Section 5.1 which we applied to the factorized model implies the  uniqueness of the limit Gibbs state in the factorized model. This uniqueness does not imply, of course, the uniqueness in the original model. In particular, the not translation periodic limit Gibbs states which can exist in the original model can't be detected by the factorized model as the sigma-algebra of the factorized model is not fine enough.

Modifications of the above argument covering Assertion (III) (with $\sh\St =3$) are straightforward. The same is true of Assertion (IV) in case $\sh\St =3$. For the proof in case $\sh\St=4$, see Section 5.4.  

\bigskip \def\tu{{\tt u}}
{\bf 5.3. Proof of Assertions (IIb) of Theorem 1.2.} For the definiteness we again assume that the stable types are $1$ and $2$, now with $D(4,2) = D(3, 1) < D(3,2) = D(4,1)$. In this case types $1$ and $2$ are not symmetric with respect to a given unstable type, e.g. type $3$. Nevertheless, the pair of types $(1,3)$ is symmetric relative to the pair $(2,4)$, in the sense that if we simultaneously replace type $1$ with $2$ and type $3$ with $4$ then an admissible configuration remains admissible. Based on this symmetry, we use the ideas of Section~5.2 and work with factorized configurations. 

The factorized model now contains two factorized particle types: $\ts$ and~$\tu$. The factorized particle type $\ts$ replaces stable types $1$ and $2$ while the factorized particle type $\tu$ replaces unstable types $3$ and $4$. We say that two factorized particles are connected if the distance between them is less than or equal to: $D(1,2)$ for two $\ts$-type particles, $D(3,4)$ for two $\tu$-type particles, and $D(1,4)$ for two particles of different factorized type. Each connected component of factorized particles has two implementations by original particles: ($\ts=1, \tu=3$) and ($\ts=2, \tu=4$). Therefore, if a factorized particle configuration $\ubX^{\tF} \in \cA$ contains $k$ connected components of the above type then the corresponding statistical weight is $2^{k(\ubX^{\tF})}$. Repeating the argument from Section~5.2, we conclude that $\mu_{\Lam_L}\left(\cB_0^\cmp\, \|\,\tu\right)$ tends to $1$ as $L\to\infty$. Here the meaning of the notation $(\cdot \|\tu)$ is straightforward.

The factorized particle configuration belonging to the event $\cB_0^\cmp$ can be implemented by the original configurations in 4 different ways. The difference between implementations is in the choice of the unstable particle type which is present in the boundary condition for $\Lam_L$ and the choice of the stable particle type placed in the annulus enclosing $\Lam_{L/2}$. Each specific choice is a pair of unstable and stable particle types, e.g. $(3, 1)$. 

It turns out that the $\mu_{\Lam_L}(\cdot \|\,\tu)$-probability of pairs  $(3,2)$  and  $(4, 1)$ tends to $0$ as $L \to \infty$. Indeed, any {\BL} configuration implementing the pair $(3, 2)$ can be mapped into a {\BL} configuration implementing the pair $(3,1)$ such that the integral of the statistical weights of all particle configurations implementing the pair $(3, 2)$ which are mapped to the same configuration implementing the pair $(3,1)$ is at least $\rv^{-L/5^d}$ times smaller than the statistical weight of the configuration they are mapped to. The $1\leftrightarrow 2, 3\leftrightarrow 4$ symmetry defines a similar map for pairs $(4,1)$ and $(4,2)$.

Recall that the {\BL} is a collection of {\LBCs}. Then the aforementioned map is constructed by erasing in this collection of the most external {\LBC} satisfying the following conditions: it encloses $\Lam_{L/2}$ and implements a transition from the type $3$ outside to the type $4$ inside. If such  {\LBC} is not present in the {\BL} then the map is constructed by replacing the most internal {\LBC} satisfying the following conditions: it encloses $\Lam_{L/2}$ and implements a transition from the type $3$ outside to the type $2$ inside. The replacing {\LBC} is chosen to be the smallest {\LBC} implementing the transition from the type $3$ outside to the type $1$ inside and having the same interior as the original {\LBC}. Here the smallest {\LBC} $\Gam$ means the {\LBC} having minimal possible $\ups(\rO(\Gam))$. 

\begin{center}
\includegraphics[scale=0.7]{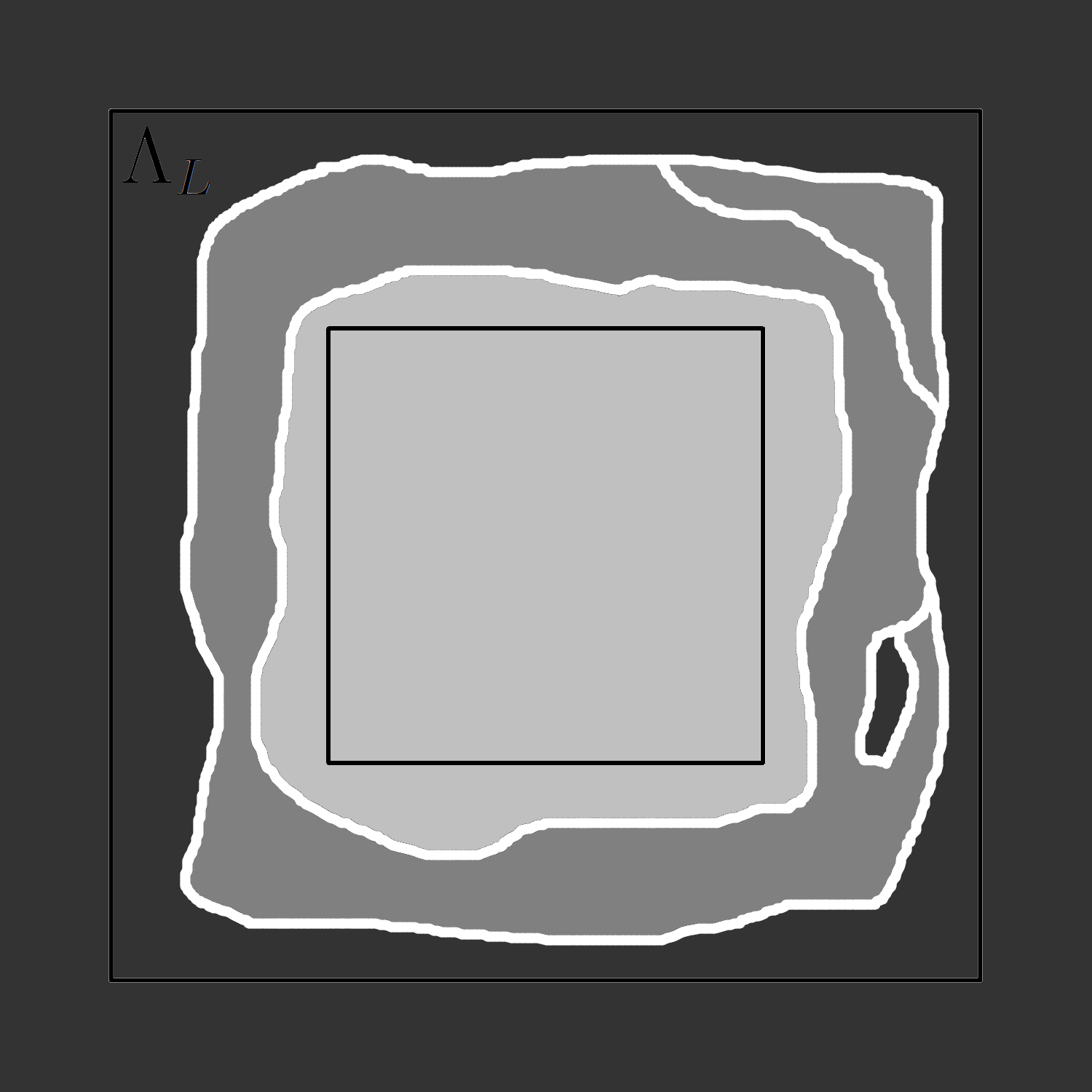}
\includegraphics[scale=0.7]{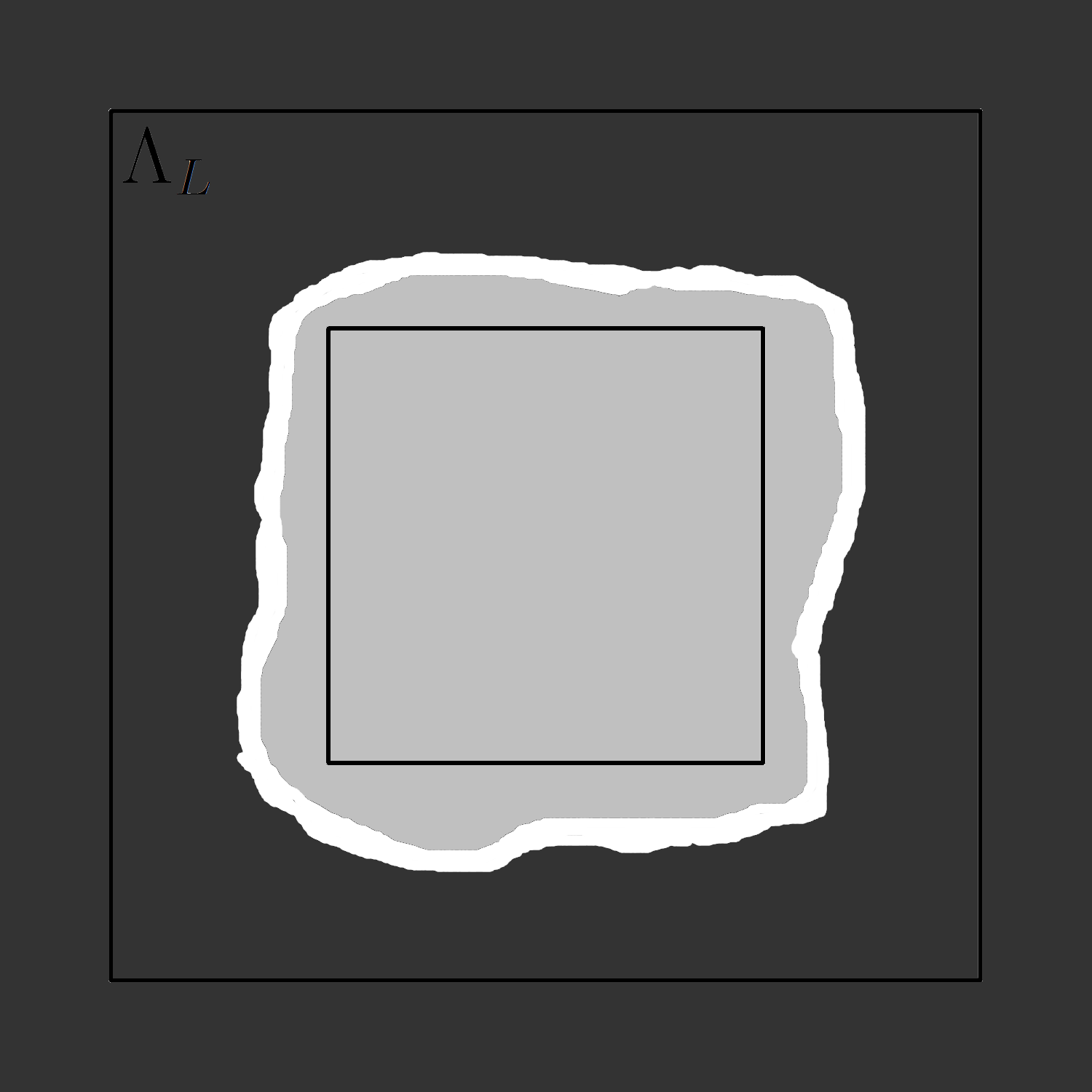}
\includegraphics[scale=0.7]{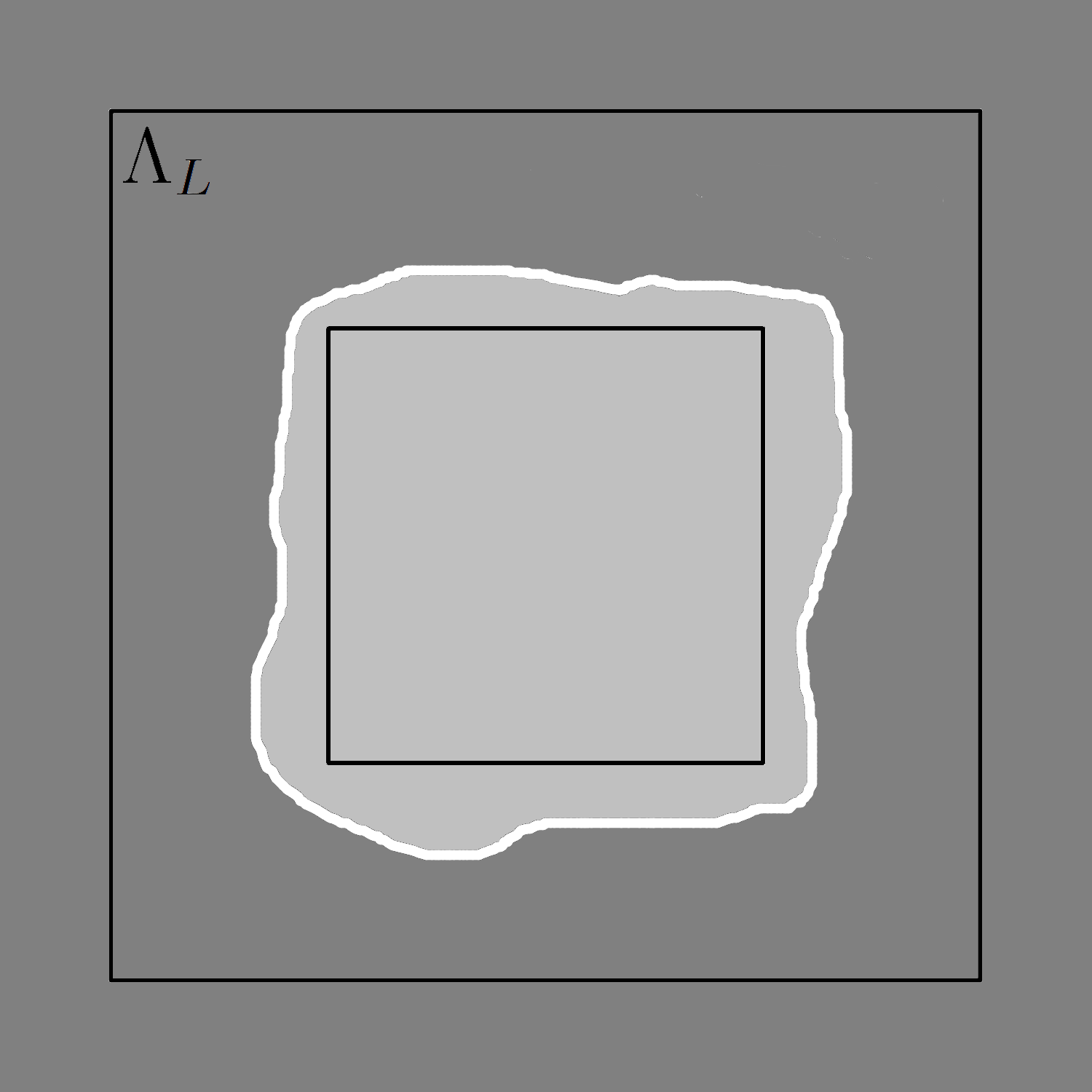}
\cl{{\bf (i)}\quad\quad\quad\quad\quad\quad\quad\quad\quad\quad{\bf (ii)}\quad\quad\quad\quad\quad\quad\quad\quad\quad\quad{\bf (iii)}}
\cl{ }
\cl{\bf Figure 9. The {\BL} map}
\end{center}

Figure~9 shows two types of the original {\BLs} (frames (i) and (ii)) and the resulting {\BL} (frame (iii)).

\bigskip
The desired estimate of the statistical weights is a consequence of the facts that: the map decreases the number of empty unit cells in the boundary layer by at least $2d(L/2)^{d-1}$ unit cells, all removed empty unit cells are 5-connected and this connected set encloses $\Lam_{L/2}$.

Due to the symmetry, the surviving pairs $(3,1)$  and  $(4,2)$ have the same  $\mu_{\Lam_L}(\cdot \|\,\tu)$-probability. This yields Assertion (IIb). 
\bigskip

{\bf 5.4. Proof of Assertion (IV) of Theorem 1.2.} In Sections 5.2 and~5.3 we showed how to construct 
factorized models for all cases except for the case ${\sh\St} = 4$. Assume for the definiteness that $D(1,2)=D(3,4)=\ua$ and $D(1,4) = \oa$ as in the previous section. Then the model is symmetric under the same transformation $1\leftrightarrow 2, 3\leftrightarrow 4$ and the factorized model is constructed as in the previous section by using two factorized types $\ts'$ and $\ts''$. The factorized type $\ts'$ combines particles of the types 1 and 2 while the factorized type $\ts''$ combines particles of the types 3 and 4. The types $\ts'$ and $\ts''$ are symmetric again such we can perform yet another factorization and define a model in terms of a single stable phase $\ts$. As in previous Sections, a factorized configuration $\ubX^{\Lam_{L},\tF} \in \cB_0^\cmp$ contains and annulus enclosing $\Lam_{L/2}$ and formed by unit cells of type $\ts$. Due to the symmetry, the corresponding original configuration $\ubX^{\Lam_{L}}  \in \cB_0^\cmp$ contains, inside this annulus particles of specific type $i$ with the probability which is  $1/4$ of the probability of the factorized configuration. Moving back from probabilities of the factorized configurations $\ubX^{\Lam_{L},\tF}$ to probabilities of the original configurations $\ubX^{\Lam_{L}}$,  we arrive at Assertion (IV) in case where $\sh\St =4$. $\qquad\blacksquare$

\bigskip
Note that in the model consisting of a single particle type $\ts$ the probability of $\cB_0$ can easily be estimated without the machinery from previous sections. In this case the event $\cB_0$ simply means that there exists a contour $\Gam^{\tF}$ adjacent to $\Lam_L^\cmp$ and intersecting $\Lam_{L/2}$. Due to the Peierls bound, the probability of such contour is less than $\rv^{\ups(\rO(\Gam^{\tF}))}$. The amount of contours originating from a given unit cell $\Ups \in \p \Lam_L$ and having given value of $\ups(\rO(\Gam^{\tF}))$ is less than $c^{\ups(\rO(\Gam^{\tF}))}$. There is at most $2d L^{d-1}$ possibilities to chose the originating unit cell $\Ups$. Thus, the probability of $\cB_0$ does not exceed $2d L^{d-1} (c\rv)^{\ups(\rO(\Gam^{\tF}))}$.
 
\bigskip
{\bf 5.5. Proof of Assertion (II) of Theorem 1.1.} In this section we follow ideas from \cite{Z} (cf. Lemma  in Section~3.2 of \cite{Z}). 

Given a unit cell $\Ups$ denote by $\cB_{\Ups}$ the event formed by particle configurations $\ubX \in \cA$ for which there is no bounded box $\Lam' \supset \Ups$ and a particle type $i$ (stable or unstable) such that $\ubX^{\Lam'} \in \cA (\Lam'\|\,i)$. It is not hard to see that 
\beq\begin{array}{l}\cB_{\Ups}=\{\ubX \in\cA:\;\Ups \subset \rB(\Gam )\;\hbox{ for some}\\ 
\qquad\qquad \hbox{{\BC} $\Gam\in\uGam (\ubX)$ with 
unbounded base $\rB(\Gam )$}\}.\end{array}\label{5.38}\eeq

Similarly, consider a box $\Lam$ and an admissible particle configuration $\ubY\in\cA$. Then for a given unit cell $\Ups \subset \Lam$, we define the event $\cB_{\Ups, \Lam, \ubY} \in \cA$ consisting of particle configurations $\ubX$ such that $\ubX^{\Lam^\cmp}=\ubY^{\Lam^\cmp}$ and there is no bounded box $\Lam': \;\Ups \subset \Lam' \subseteq \Lam$ and a particle type $i$ (stable or unstable) such that $\ubX^{\Lam'} \in \cA (\Lam'\|\,i)$. Equivalently,
\beq\begin{array}{l}
\cB_{\Ups,\Lam,\ubY}=\{\ubX \in\cA:\;\ubX^{\Lam^\cmp}=\ubY^{\Lam^\cmp},\;\Ups \subset \rB(\Gam )\;\hbox{ for}\\ 
\qquad\qquad\qquad \hbox{some {\BC} $\Gam\in\uGam(\ubX)$ such that $\rB(\Gam )\cap\Lam^\cmp\neq\varnothing$}\}.\end{array}\label{5.39} \eeq
Obviously, 
for $\ubY \not\in \cB_{\Ups}$,
\beq
\operatornamewithlimits\cap\limits_{\Lam:\;\Lam\supset\Ups} \;\cB_{\Ups, \Lam, \ubY}=\varnothing.\label {5.41} \eeq
Furthermore,  
\beq\cB_{\Ups} = \operatornamewithlimits\cap\limits_{\Lam:\;\Lam\supset\Ups}\cB_{\Ups, \Lam}
\hbox{ where }\;\cB_{\Ups, \Lam} = \operatornamewithlimits{\cup}\limits_{\ubY \in \cA} \cB_{\Ups, \Lam, \ubY}.
\label{5.42} \eeq

Let $\mu(\cdot)$ be a translation periodic DLR measure. Suppose that
\beq
\mu(\cB_{\Ups}) > 0
\label{5.43} \eeq
for some $\Ups$. Due to translation periodicity of $\mu$ and owing to \eqref {5.42} ,
\beq
\sum_{\Ups \subset \Lam_L} \mu(\cB_{\Ups, \Lam_L}) > \sum_{\Ups \subset \Lam_L} \mu(\cB_{\Ups}) > \e \ups(\Lam_L) = \e L^d
\label{5.44} \eeq
where $\e > 0$. 

The sum in the LHS of \eqref {5.44}  gives an expectation of the amount of unit cells inside $\Lam_L$ which 
belong to bases of so-called interface (or open ) {\BCs} in $\Lam$. An {\it interface {\BC}} in $\Lam$ is a {\BC} $\Gam$ that has more than one connected component of the intersection $\rE(\Gam) \cap \Lam$. 
(Recall, $\rE(\Gam)$ stands for the exterior of $\Gam$; see Definition 2.3.) Note that for an interface {\BC} $\Gam$, base $\rB(\Gam)$ is always adjacent to $\Lam^\cmp$. 

\begin{center}
\includegraphics[scale=1]{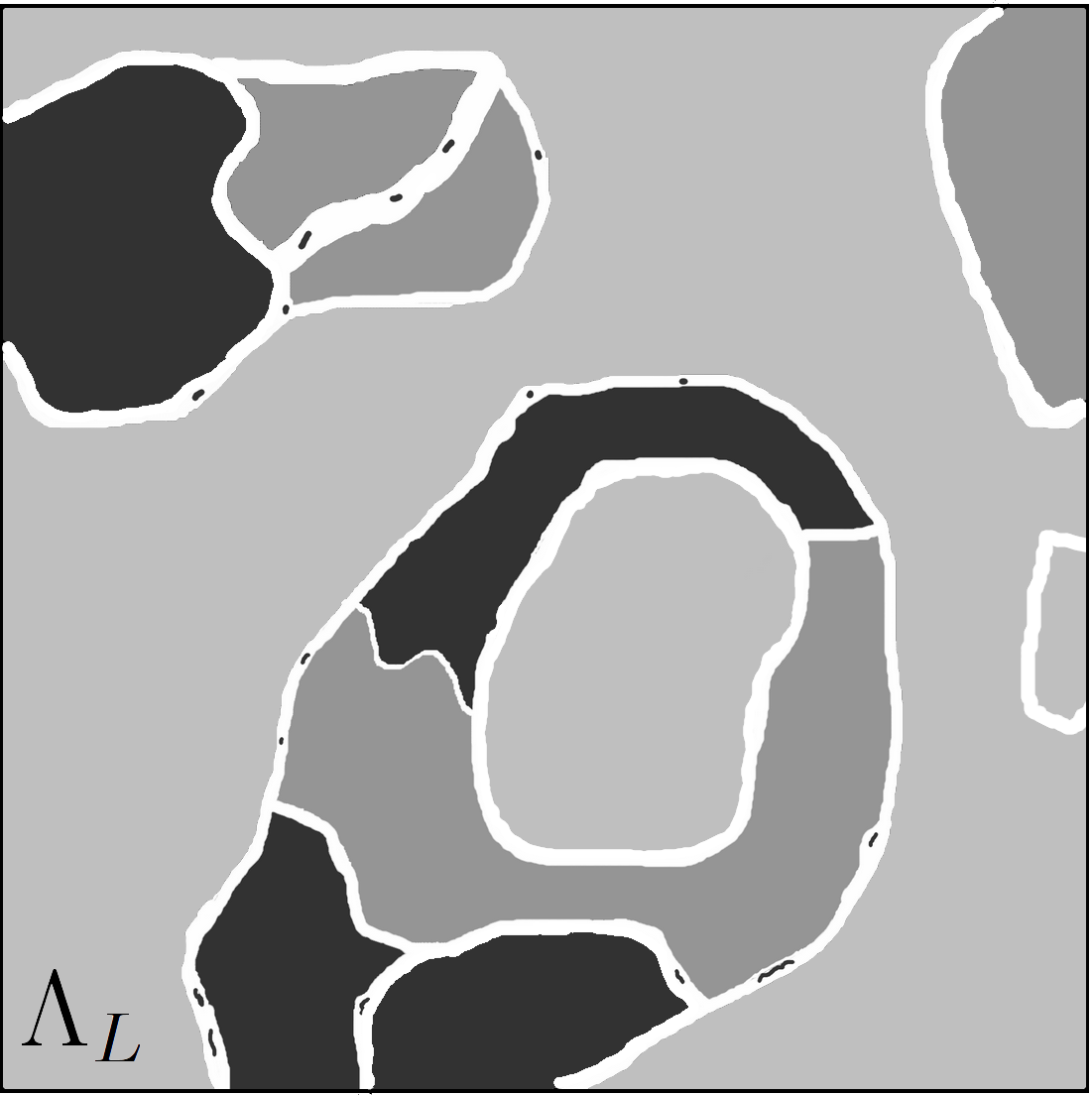}
\cl{\bf Figure 10. Interface {\BCs}}
\end{center}

Consider the event
\beq \cB_{M, \Lam_L}=\left\{\ubX\in\cA:\;{\sum}^{{\rm{I}},\Lam}\ups(\rB(\Gam)) = M\right\}\label{5.45}\eeq
where the sum $\sum^{{\rm{I}},\Lam}$ is taken over all interface {\BC}s $\Gam\in\uGam (\ubX)$ in $\Lam$. By 
virtue of the DLR property, the expectation
\beq
\sum_{M\geq 0} M\, \mu(\cB_{M, \Lam_L}) = \int_{\cA} \mu(d \ubY) \sum_{M\geq 0}
M\,\mu_{\Lam}(\cB_{M, \Lam_L} | \ubY^{\Lam^\cmp}).
\label{5.46}\eeq
According to \eqref {5.06}, 
\beq\label{5.47}
\diy\mu_{\Lam}(\cB_{M, \Lam_L} | \ubY^{\Lam^\cmp}) < \mu_{\Lam_{L+\oa}}\left(\cB_{M, \Lam_L} \vee{\uvnth}^{\Delta_L}\|\,1\right) 
\Big(\frac{q}{\rv}\Big)^{\ups(\Lam_{L+\oa} \setminus \Lam_{L-\oa})};
\eeq   
here we assume, without loss of generality, that type $1$ is stable. The probability $\mu_{\Lam_{L+\oa}}\left(\cB_{M, \Lam_L} \vee{\uvnth}^{\Delta_L}\|\,1\right) $ can be upper-bounded by using \eqref{2.31} and polymer expansions from Section~4. The resulting inequality is
\beq\label{5.48}
\mu_{\Lam_{L+\oa}}\left(\cB_{M, \Lam_L} \vee{\uvnth}^{\Delta_L}\|\,1\right) \le \left(\frac{\rv}{1-\rv}\right)^{0.5 M /5^d}
\eeq   
as, together with ${\uvnth}^{\Delta_L}$, the interface {\BCs} form a single {\BC} inside $\Lam_{L+\oa}$ containing at least $M$ unit cells in its base. Partitioning the sum $\sum\limits_{M\geq 0}$ in the RHS of \eqref  {5.46} into $\sum\limits_{0\leq M\leq L^{d - 1/2}}$ and $\sum\limits_{M>L^{d - 1/2}}$ and plugging into the later sum estimates \eqref   {5.47} and \eqref {5.48} we conclude that for $L$ large enough, 
\beq\sum_{\Ups \subset \Lam_L} \mu(\cB_{\Ups, \Lam_L}) = \sum_{M=0}^{\infty} M\, \mu(\cB_{M, \Lam_L}) 
\le 2L^{d - 1/2}.\label{5.49}\eeq
But \eqref {5.49} contradicts \eqref {5.44}, and so the assumption \eqref {5.43}  is false. 

The negation of \eqref {5.43}  means that with $\mu$-probability $1$ a particle configuration $\ubX \in \cA$ does not have unbounded {\BCs} in $\uGam (\ubX)$. Thus, for any given $L>1$, with $\mu$-probability $1$ there exists a box $\Lam \supset \Lam_L$ such that the restriction $\ubX^{\,{}^2\Lam \setminus  \Lam }$ contains at least one particle of the same type in each unit cell $\Ups\subset\,{}^2\Lam \setminus  \Lam$.  Therefore, the $\mu$-probability of the event that $\Lam_{L/2}$ is enclosed by an annulus of unit cells in a stable phase tends to $1$ as $L \to \infty$ as was shown in Sections~5.1--5.3. This implies Assertion (II) of Theorem 1.1.


\def\slr{\Bigg/}
\def\sll{\Bigg\backslash}

\def\qed{{\vrule height6pt width4pt depth0pt}}
\def\hd{\frac{d}{2}}
\def\T{\!{{} \atop \scriptstyle{T}}}
\def\a{\alpha}
\def\b{\beta}
\def\o{\omega}
\def\e{\varepsilon}
\def\d{\delta}
\def\G{\Gamma}
\def\g{\gamma}
\def\L{\Lambda}
\def\Th{\Theta}
\def\th{\theta}
\def\cl{\{L_j\}}
\def\sL{{\cal L}}
\def\W{{M}}
\def\pg{[\th_i^{\a_i}]}
\def\p{\partial}
\def\f1{\frac{2}{3}}
\def\se{\sqrt{\e} }

\bigskip\bigskip
\section{Appendix: the Polymer expansion theorem}

Consider a finite or countable set $\Th$ the elements of which are called
(abstract) contours and denoted $\th,\th'$, ets. Fix some anti-reflexive
and symmetric relation $\sim$ on $\Th\times\Th$ (with $\th\not\sim\th$ and
$\th\sim\th'$ equivalent to $\th'\sim\th$).  A pair $\th,\th' \in
\Th\times\Th$ is
called incompatible ($\th\not\sim\th'$) if it {\bf does not} belong to the relation
and compatible ($\th\sim\th'$) in the opposite case. (In our context, 
two contours are compatible when they are mutually external and have the 
same external type.)
A collection $\{\th_j\}$ is called a compatible collection of
contours if any two its elements are compatible.  Every contour $\th$ is
assigned a (generally speaking) complex-valued statistical weight
denoted by $w(\th)$, and for any finite $\L\subseteq\Th$ an (abstract)
partition function is defined as
\beq \label{A.01} Z(\L)=\sum\limits_{\{\th_j\}\subseteq\L}
\prod_j w(\th_j),\eeq   
 where the sum is extended to all compatible collections of contours
$\th_i\in\L$.  The empty collection is compatible by definition, and it is
included in $Z(\L)$ with statistical weight $1$.

A polymer $\Pi=[\pg ]$ is an (unordered) finite collection of different
contours $\th_i \in \Th$ taken with positive integer multiplicities
$\a_i$, such that for every pair $\th',\ \th'' \in \Pi $ there exists a
sequence $\th'=\th_{i_1},\ \th_{i_2},\ldots, \th_{i_s}=\th'' \in \Pi$ with
$\th_{i_j}\not\sim\th_{i_{j+1}},\ j=1,2,\ldots,s-1$. The notation
$\Pi\subseteq \L$ means that $\th_i \in \L$ for every $\th_i \in \Pi$.

With every polymer $\Pi$ we associate an (abstract) graph $G(\Pi)$ which
consists of
$\sum\limits_i \a_i$ vertices labeled by the contours from $\Pi$ and
edges joining every two vertices labeled by incompatible contours. As
follows from the definition of $\Pi$, graph $G(\Pi)$ is connected. We
denote by $r(\Pi)$ the quantity
\beq \label{A.02} r(\Pi)=\prod_i (\a_i!)^{-1} \sum\limits_{G' \subseteq G(\Pi)} (-1)^{\sh\cE(G')}
\eeq   
(the M{o}ebius-type inversion coefficient). Here the sum is taken over all 
connected subgraphs $G'$ of $G(\Pi)$
containing all $\sum\limits_i \a_i$ vertices, and $\sh\cE(G')$ denotes
the number of edges in $G'$. For any $\th \in \Pi$ we denote by
$\a(\th,\Pi)$ the multiplicity of contour $\th$ in polymer $\Pi$.

The Polymer expansion theorem (Theorem 6.1 below) is a modification of 
assertions from \cite{KP} (see also \cite{Se}) which has been established in 
\cite{MS2}.

\bigskip

{\bf Theorem 6.1.} {\sl Suppose that there exists a function $a(\th):\ \Th
\mapsto {\bbR}^{+}$ such that for any contour $\th$
\beq \label{A.03}
\sum\limits_{\th':\ \th'\not\sim\th} |w(\th')| e^{a(\th')} \le
a(\th). \eeq   
Then, for any finite $\L$,
\beq\log Z(\L)=\sum\limits_{\Pi\subseteq\L} w(\Pi),\label{A.04} \eeq  
where the statistical weight of a polymer $\Pi=\pg$ equals
\beq  w(\Pi)=r(\Pi) \prod_i w(\th_i)^{\a_i}. \label{A.05} \eeq  
Moreover, the series \eqref{A.04} for
$\log Z(\L)$ absolutely converges in view of the bound
\beq \label{A.06} \sum\limits_{\Pi:\ \Pi\ni\th} \a(\th,\Pi) |w(\Pi)| \le
|w(\th)| e^{a(\th)}, \eeq  
which holds true for any contour $\th$.}


\subsection*{Acknowledgments}
This work has been conducted under Grant 2011/51845-5 provided by 
the FAPESP, Grant 2011.5.764.45.0 provided by The Reitoria of the 
Universidade de S\~{a}o Paulo.  YS and IS 
express their gratitude to the IME, Universidade de S\~{a}o Paulo,
Brazil, for the warm hospitality. YS
expresses his gratitude to the ICMC, Universidade de S\~{a}o Paulo,
Brazil, for the warm hospitality.

\end{document}